\documentclass[a4paper,fleqn,usenatbib]{mnras}
\pdfoutput=1
\usepackage{newtxtext,newtxmath}
\usepackage[T1]{fontenc}
\usepackage{ae,aecompl}

\usepackage{graphicx}
\usepackage{amsmath,amssymb}
\usepackage{gensymb}
\usepackage{hyperref}

\newcommand{\ciii}{\hbox{C\,{\sc iii}}}
\newcommand{\civ}{\hbox{C\,{\sc iv}}}
\newcommand{\oiii}{\hbox{O\,{\sc iii}}}
\newcommand{\oii}{\hbox{O\,{\sc ii}}}
\newcommand{\cii}{\hbox{C\,{\sc ii}}}
\newcommand{\nii}{\hbox{N\,{\sc ii}}}

\newcommand{\nev}{\hbox{Ne\,{\sc v}}}
\newcommand{\sii}{\hbox{S\,{\sc ii}}}

\newcommand{\alii}{\hbox{Al\,{\sc ii}}}
\newcommand{\hii}{\hbox{H\,{\sc ii}}}

\newcommand{\heii}{\hbox{He\,{\sc ii}}}
\newcommand{\hei}{\hbox{He\,{\sc i}}}
\newcommand{\feii}{\hbox{Fe\,{\sc ii}}}

\providecommand{\e}[1]{\ensuremath{\times 10^{#1}}}
\newcommand{\unit}[1]{\ensuremath{\, \mathrm{#1}}}

\newcommand{\hst}{\textit{HST}}
\newcommand{\hstcos}{\textit{HST}/COS}

\newcommand{\jwst}{\textit{JWST}}

\newcommand{\ott}{\ensuremath{\mathrm{O}_{32}}}
\newcommand{\rtt}{\ensuremath{\mathrm{R}_{23}}}

\newcommand{\cloudy}{\textsc{cloudy}}

\newcommand{\beagle}{\textsc{beagle}}
\newcommand{\tauV}{\hbox{$\hat{\tau}_V$}}

\newcommand{\Us}{\hbox{$U_\mathrm{S}$}}
\newcommand{\xid}{\hbox{$\xi_\mathrm{d}$}}
\newcommand{\nH}{\hbox{$n_{\mathrm{H}}$}}

\newcommand{\Hii}{\mbox{H\,{\sc ii}}}
\newcommand{\Zism}{\hbox{$Z_\mathrm{ISM}$}}
\newcommand{\mup}{\hbox{$m_{\rm{up}}$}}

\title[HeII Emitters via COS]{Ultraviolet spectra of extreme nearby star-forming regions --- approaching a local reference sample for \textit{JWST}}
\author[P. Senchyna et al.]{
    Peter Senchyna$^{1}$\footnotemark[1], 
    Daniel P. Stark$^{1}$, 
    Alba Vidal-Garc\'ia,$^{2}$
    Jacopo Chevallard$^{3}$, \newauthor
    St\'{e}phane Charlot$^{2}$,
    Ramesh Mainali$^{1}$, 
    Tucker Jones$^{4,5,\dagger}$, \newauthor
    Aida Wofford$^{6}$,
    Anna Feltre$^{7}$
    and Julia Gutkin$^{2}$
    \vspace{0.1in}\\
    $^{1}$ Steward Observatory, University of Arizona, 933 N Cherry Ave, Tucson, AZ 85721 USA \\  
    $^{2}$ Sorbonne Universit\'{e}s, UPMC-CNRS, UMR7095, Institut d'Astrophysique de Paris, F-75014 Paris, France \\
    $^{3}$ Scientific Support Office, Directorate of Science and Robotic Exploration, ESA/ESTEC, Keplerlaan 1, 2201 AZ Noordwijk, The Netherlands \\
    $^{4}$ Department of Physics, University of California Davis, 1 Shields Avenue, Davis, CA 95616, USA \\
    $^{5}$ Institute for Astronomy, University of Hawaii, 2680 Woodlawn Drive, Honolulu, HI 96822, USA \\
    $^{6}$ Instituto de Astronom\'{i}a, UNAM, Ensenada, CP 22860, Baja California, Mexico \\
    $^{7}$ Centre de Recherche Astrophysique de Lyon, Universit\'{e} Lyon 1, 9 Avenue Charles Andr\'{e}, F-69561 Saint Genis Laval Cedex, France \\
    $^{\dagger}$ Hubble Fellow \\
}

\date{Accepted XXX. Received YYY; in original form ZZZ}

\pubyear{2017}

\begin{document}
\label{firstpage}
\pagerange{\pageref{firstpage}--\pageref{lastpage}}
\maketitle

\begin{abstract}
    Nearby dwarf galaxies provide a unique laboratory in which to test stellar population models below $Z_\odot/2$.
    Such tests are particularly important for interpreting the surprising high-ionization UV line emission detected at $z>6$ in recent years.
    We present \hstcos{} ultraviolet spectra of ten nearby metal-poor star-forming galaxies selected to show \heii{} emission in SDSS optical spectra.
    The targets span nearly a dex in gas-phase oxygen abundance ($7.8<12+\log\mathrm{O/H}<8.5$) and present uniformly large specific star formation rates (sSFR $\sim 10^2$ $\mathrm{Gyr}^{-1}$).
    The UV spectra confirm that metal-poor stellar populations can power extreme nebular emission in high-ionization UV lines, reaching \ciii{}] equivalent widths comparable to those seen in systems at $z\sim 6-7$.
    Our data reveal a marked transition in UV spectral properties with decreasing metallicity, with systems below $12+\log\mathrm{O/H}\lesssim 8.0$ ($Z/Z_\odot \lesssim 1/5$) presenting minimal stellar wind features and prominent nebular emission in \heii{} and \civ{}.
    This is consistent with nearly an order of magnitude increase in ionizing photon production beyond the $\mathrm{He^+}$-ionizing edge relative to H-ionizing flux as metallicity decreases below a fifth solar, well in excess of standard stellar population synthesis predictions.
    Our results suggest that often neglected sources of energetic radiation such as stripped binary products and very massive O-stars produce a sharper change in the ionizing spectrum with decreasing metallicity than expected.
    Consequently, nebular emission in \civ{} and \heii{} powered by these stars may provide useful metallicity constraints in the reionization era.
\end{abstract}

\begin{keywords}
    galaxies: evolution -- galaxies: stellar content -- stars: massive -- ultraviolet: galaxies
\end{keywords}

\footnotetext[1]{E-mail: senchp@email.arizona.edu}

\section{Introduction}

The first deep spectra of galaxies at $z>6$ present a striking contrast to the properties of typical star-forming galaxies at lower redshift.
Ground-based spectroscopy probing the rest frame ultraviolet (rest-UV) of gravitationally lensed Lyman-$\alpha$ emitters at $z>6$ has revealed nebular emission in transitions of highly-ionized species including $\mathrm{C}^{2+}$, $\mathrm{O}^{2+}$, and $\mathrm{C}^{3+}$ \citep{Stark2015,Stark2015a,Stark2017,Mainali2017}.
The observed equivalent width of \ciii{}] and \oiii{}] exceed that seen in typical UV-selected star-forming galaxies at $z\sim 2-3$ by an order of magnitude, and nebular \civ{} is rarely seen in emission at all in these or local star-forming samples.
Studies with ALMA targeting [\cii{}] and [\oiii{}] emission and with \textit{Spitzer} probing rest-optical nebular line excesses in broadband photometry at $z>6$ paint a similar picture, suggesting that extreme radiation fields are more common in star-forming galaxies at these early times (see \citealt{Stark2016} for a review).

The rest-UV properties of these objects are not entirely without precedent in lower-$z$ samples.
For instance, nebular \civ{} and \heii{} emission (requiring flux beyond $\sim 50$ eV) is seen in some lensed star-forming dwarf galaxies at $z\sim 2-3$ \citep{Erb2010,Christensen2012,Stark2014,Vanzella2016,Vanzella2017a}.
This suggests that the additional ionizing flux may be provided by low-metallicity stars.
Rest-optical spectroscopy of galaxies at $z\sim 2-3$ has also revealed differences with respect to models calibrated at near-solar metallicity.
In particular, offsets in diagnostic line ratios have been interpreted as due to some combination of higher nitrogen abundance and harder ionizing radiation fields at these redshifts \citep[e.g.][]{Kewley2013a, Steidel2014, Shapley2015, Sanders2016, Kashino2017, Strom2017, Kojima2017}.

Indeed, these surprising detections at high-$z$ were presaged by spectroscopy of low-metallicity dwarf galaxies in the nearby universe.
Detections of nebular \heii{} $\lambda 4686$ emission in nearby stellar-photoionized \hii{} regions date back to at least 1985, and its origin remains mysterious \citep[e.g.][]{Garnett1991,Thuan2005,Brinchmann2008,Shirazi2012,Kehrig2015}.
The nebular \heii{} $\lambda 1640$ and $4686$ lines are emitted in the cascading recombination of $\mathrm{He}^{++}$, produced by ionizing photons beyond 54.4 eV.
Due to strong absorption in the atmospheres and winds of massive stars, even very hot stellar models considered in standard population synthesis prescriptions generally predict very few photons in this energy range.
The necessary ionizing flux has thus been attributed variously to very massive stars, high-mass X-ray binaries, and fast radiative shocks.

Ultraviolet spectra of nearby star-forming regions are far less ubiquitous than optical spectra, but previous UV work also hints at some commonalities with extreme high-$z$ galaxies.
The International Ultraviolet Explorer satellite \citep[IUE; e.g.][and references therein]{Kinney1993,Giavalisco1996,Heckman1998} as well as the Goddard High Resolution Spectrograph (GHRS) and Faint Object Spectrograph (FOS) previously onboard the \textit{Hubble Space Telescope} \citep[\textit{HST}; ][]{Garnett1995,Leitherer2011} enabled detailed study of rest-UV stellar and nebular features in local star-forming galaxies.
Data from these instruments reveal that the [\ciii{}], \ciii{}] $\lambda\lambda 1907,1909$ semi-forbidden doublet (hereafter \ciii{}] doublet) reaches extremely high equivalent widths in a handful of systems below half-solar metallicity, but does not appear to follow a monotonic trend with gas-phase oxygen abundance or Ly$\alpha$ equivalent width \citep[e.g.][]{Bayliss2014,Rigby2015}.
Further interpretation of this data is complicated by the limited resolution and sensitivity of these UV spectrographs and by the small number of metal-poor objects with archival coverage.

Observations of both nearby massive stars and peculiar galaxies in the distant universe have motivated a great deal of work in both stellar modeling and stellar population synthesis.
Models of the atmospheres and winds of massive stars have advanced considerably in the past decades, incorporating non-LTE effects and full hydrodynamical modeling of winds, with substantial effects on the predicted emergent ionizing flux \citep[e.g.][]{Kudritzki1987,Pauldrach2001,Smith2002,Todt2015}.
The incorporation of physics such as stellar rotation \citep[e.g.][]{Maeder2000,Levesque2012,Szecsi2015} and binary mass transfer \citep[e.g.][]{Eldridge2008,deMink2014,Gotberg2017} has significant effects on the evolution of individual stars, and both tend to enhance the ionizing flux produced by composite stellar populations \citep[e.g. Figure 2 of][]{Wofford2016}.
Much progress has also been made in self-consistently predicting the emergent stellar continuum and nebular emission of galaxies and linking spectral observations to underlying physical parameters of interest \citep[e.g.][Charlot \& Bruzual 2017 in-prep]{Charlot2001,Eldridge2009,Chevallard2016,Gutkin2016,Leja2017,Byler2017,Vidal-Garcia2017}.
However, the calibration of all components of these models is most challenging at the low metallicities expected in the reionization era.

The relatively-common detection of high-ionization emission at $z>6$ and in star-forming dwarf galaxy populations nearby suggests substantial evolution in the ionizing spectra of stellar populations with metallicity.
Empirical constraints on stellar models at low metallicities are difficult to obtain.
Individual stars can only be resolved reliably within the Local Group, and thus direct calibration of massive star models extend down only to approximately $Z/Z_\odot \sim 1/5$, or $12+\log\mathrm{O/H} \sim 8.0$ \citep[see e.g.][]{Massey2003,Garcia2014,Bouret2015,Crowther2016}.
Comprehensive tests of stellar population models at the highest masses and lowest metallicities requires high-sSFR populations outside the Local Group, and detailed spectroscopy of weak nebular lines and stellar features in integrated light spectra of individual systems is generally only possible at low-redshift \citep[see e.g.][]{Leitherer2011,Wofford2014}.
The metal-poor dwarf galaxy population nearby (within a few hundred Mpc) is thus a critical laboratory in which to test stellar population synthesis models at sub-SMC metallicities.

However, locating and studying local galaxies with UV emission comparable to that seen in $z>6$ systems has proved challenging.
In particular, no nearby star-forming galaxies with nebular \civ{} approaching the $\sim 20-40$ \AA{} observed at $z\sim 6-7$ \citep{Stark2015a,Mainali2017} have yet been identified.
Previous local UV spectroscopic samples have focused on Ly$\alpha$, the UV slope, and C/O as probed via the \ciii{}] and \oiii{}] doublets \citep[e.g.][]{Giavalisco1996,Garnett1995,Leitherer2011,Berg2016}.
While sufficient to reveal that \ciii{}] equivalent widths do reach $\sim 20$ \AA{} in some metal-poor galaxies \citep{Rigby2015}, these observations lack the coverage and resolution necessary to study the most extreme UV lines.
Measuring \civ{} and \heii{} especially at low equivalent width requires sufficient resolution to disentangle nebular emission from broad stellar wind lines and interstellar absorption; and thus all but the strongest nebular emission is difficult to constrain with FOS and COS low-resolution gratings ($R\equiv \lambda / \Delta \lambda \lesssim 2000$, unbinned).
Archival samples are also biased towards relatively high metallicities; the FOS/GHRS atlas compiled by \citet{Leitherer2011} contains only six galaxies below $12+\log\mathrm{O/H} \lesssim 8.0$.
In order to study nebular line production and constrain the associated stellar populations, we require large samples of metal-poor galaxies (necessarily probing to fainter objects) with moderate resolution spectral coverage.
The Cosmic Origins Spectrograph (COS) installed by \hst{} Servicing Mission 4 now provides the sensitivity and medium-resolution gratings ($R >16000$, unbinned) necessary to begin this work in-earnest.

We have undertaken a campaign to investigate the stellar populations and physical conditions which power high-ionization nebular emission locally.
In Cycle 23 (GO: 14168, PI: Stark)  we obtained \hstcos{} UV spectra of ten star-forming galaxies selected to have \heii{} $\lambda 4686$ emission in optical SDSS spectra \citep{Shirazi2012}.
This emission is indicative of the hard $\gtrsim50$ eV radiation necessary to power UV high-ionization lines.
Grating settings were selected to constrain the full suite of high-ionization UV lines detected at high-$z$: \civ{}, \heii{}, \oiii{}], and \ciii{}].
The galaxies were chosen to span nearly a dex in gas-phase metallicity $7.6 \lesssim 12+\log\mathrm{O/H} \lesssim 8.4$ (roughly $1/10 < Z/Z_\odot < Z/2$) in order to explore the metallicity dependence of the UV spectra.
We discuss the sample and data in more detail in Section 2.
In Section 3, we present the \hstcos{} UV spectra in the context of the optical measurements.
We describe first results from stellar population synthesis fits to the full UV spectra in Section 4, to be continued in a follow-up paper (Chevallard et al.\ 2017, in-prep).
We then discuss implications for high-redshift observations and stellar population synthesis at low-metallicities in Section 5, and conclude in Section 6.

We assume a solar oxygen abundance of $12+\log_{10}\left([\text{O/H}]_\odot\right)  = 8.69$ \citep{Asplund2009}.
For distance calculations and related quantities, we adopt a flat cosmology with $H_0 = 70$ \unit{km \, s^{-1} \, Mpc^{-1}}.

\section{Sample Selection and Data}

\subsection{SDSS Galaxies With \heii{} Emission}
\label{sec:selection}

To find a set of metal-poor galaxies with hard ionizing stellar spectra, we rely on the detection of diagnostic lines in SDSS optical spectra.
\citet[][SB2012 hereafter]{Shirazi2012} searched the SDSS DR7 spectral database \citep{Abazajian2009, York2000, Ahn2012} for nebular \heii{} $\lambda 4686$ emission.
This helium recombination line (technically a blended multiplet) is indicative of a very hard ionizing continuum, as the energy required to strip $\text{He}^+$ of its one electron is 54.4 eV ($\lambda \simeq 228$ \AA{}).

The SDSS DR7 spectroscopic sample consists of all objects targeted in SDSS I and II.
Ignoring the stars and supernovae targeted as part of SDSS-II (SEGUE and the SDSS Supernova Survey), this sample is equivalent to the SDSS Legacy survey.
The SDSS Legacy Survey selected galaxies, quasars, and luminous red galaxies for spectroscopic follow-up by making a variety of photometric cuts \citep{Strauss2002,Richards2002,Eisenstein2001}.
In particular, the main galaxy sample is estimated to be complete to $>$99\% for all galaxies in the SDSS footprint with $r$-band Petrosian magnitudes $r\leq 17.77$ and half-light surface brightnesses $\mu_{r,50} \leq 24.5$ $\text{mag}/\text{arcsec}^2$ \citep{Strauss2002}.
In total, the sample contains $\sim 1.5$ million spectra collected over $\sim 8000$ \unit{deg^2}.
While the SDSS spectroscopic sample is not complete to all dwarf galaxy morphologies \citep[see, for instance,][]{James2015}, our selection goal was to identify metal-poor objects sufficiently bright and compact for \hstcos{} follow-up.

In order to separate AGN from predominantly star-forming galaxies, SB2012 employed a range of line ratio diagnostics.
\citeauthor{Shirazi2012} required a $\text{S/N} > 3$ detection of H$\beta$, [\oiii{}] $\lambda 5007$, H$\alpha$, and [\nii{}] $\lambda 6584$ so as to enable a traditional \citet*[][BPT]{Baldwin1981} diagram analysis.
In addition, nebular line diagnostics incorporating \heii{} were used to place firmer constraints on the hardness of the ionizing spectrum and minimize contamination from AGN.
This analysis identified 2865 \heii{} detections with linewidths comparable to the strong forbidden and recombination lines; and 189 with line ratios indicating a predominantly stellar ionizing spectrum.

The final SB2012 star-forming subsample is overwhelmingly nearby, with the vast majority located at $z<0.1$.
The objects span the metallicity range $7.5 < 12+\log_{10}(\mathrm{O/H}) < 9.5$.
They range in character from \hii{} regions embedded in larger galaxies to isolated blue compact dwarfs.

We selected ten of these star-forming \heii{}-emitters to target with \hstcos{} UV spectroscopy in the \hst{} program GO 14168 (PI: Stark). 
Our primary goal was identifying moderately metal-poor ($12+\log_{10}(\text{O/H})\sim 7.7-8.2$, i.e. $Z/Z_{\odot} \sim 1/8 - 1/3$) objects with intense radiation fields.
Thus, we selected targets spread evenly throughout this range.
We utilized the gas phase metallicity estimates produced by SB2012 who used the grid of photoionization models described by \citet{Charlot2001} to fit the measured emission line fluxes.

In addition, we selected targets such that approximately half showed signs of Wolf-Rayet (WR) stars in the optical according to SB2012.
These stellar wind signatures are the broad blue and red wind emission bumps located near $4650$ and $5808$ \AA{} respectively, consisting of blended \heii{} $\lambda 4686$, \civ{} $\lambda \lambda 5801-12$, and various metal lines \citep[e.g.][]{Crowther2007}.
While Wolf-Rayet stars have been suggested as the most likely stellar population to produce the hard ionizing spectrum required for nebular \heii{} emission, previous studies have found an unclear association between this nebular emission and the WR wind bumps at low metallicity \citep[e.g.][]{Shirazi2012, Brinchmann2008, Guseva2000}.

\begin{figure*}
    \includegraphics[width=\textwidth]{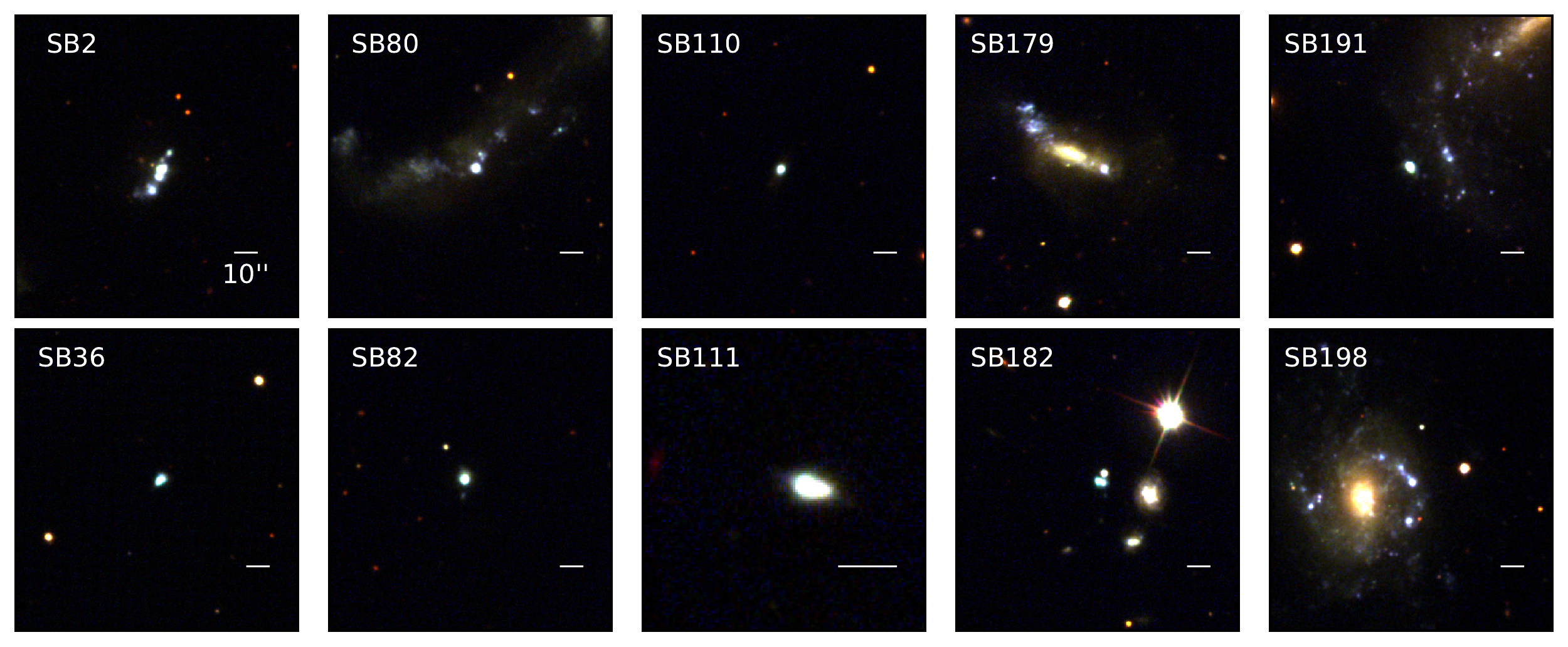}
    \caption{
        SDSS $u,g,r$ montage images centred on our targets.
        The scalebar is fixed at 10\arcsec{} in length (SB111 is zoomed-in to avoid the edge of the SDSS frame).
        The objects span a range of environments, from isolated dwarfs to \hii{} regions embedded in larger galaxies.
    }
    \label{fig:sdss_montage}
\end{figure*}

The targets selected for \hstcos{} observation (see Table~\ref{tab:basicprop}) reside in a variety of environments, from isolated dwarf galaxies to \hii{} regions embedded in larger disk systems; and range in distance from $\sim 10-200$ Mpc.
The final sample of ten is listed in Table~\ref{tab:basicprop}, and SDSS cutouts for each are plotted in Fig.~\ref{fig:sdss_montage}.
To estimate distances, we adopt the local velocity flow model described by \citet{Tonry2000} with $H_0=70 \unit{km}\unit{s}^{-1}\unit{Mpc}^{-1}$ and check the literature for more robust measurements.
The uncertainty in redshift-only distance estimates is dominated by the random motion of galaxies and by systematic error arising from group assignment.
The inverse model relating observed recessional velocity to distance does not have a unique solution for objects near the Virgo cluster.
In cases where the sky position and redshift are consistent with Virgo, and no other literature distances or group assignments are available, we assume a distance of 16.5 Mpc \citep{Mei2007}.
A case-by-case distance analysis is presented in Appendix~\ref{app:distances}.
Uncertainty in the process of distance assignment in this work translates mainly into potential systematic uncertainty in inferred stellar masses and absolute star formation rates.

Since this work is primarily concerned with understanding the UV spectra of young star-forming regions, we focus our analysis on the \hstcos{} aperture.
At these distances, the projected \hstcos{} aperture radius corresponds to physical scales ranging from 60 -- 1200 pc; but the star formation rate and stellar mass surface densities within this aperture span a much smaller dynamic range (see Section~\ref{sec:measurements}).
The bulk properties of the galaxies in which several of our objects are embedded likely have negligible direct impact on the emergent spectra of the star-forming regions within the \hstcos{} aperture, and thus are beyond the scope of this paper.

Selecting systems by high-ionization line emission such as in \heii{} may result in a bias towards systems with unusual IMF sampling.
As found in Section~\ref{sec:photres}, the masses and star formation rates ($\gtrsim 10^{4.7} M_\odot$ and $\gtrsim 10^{-2} M_\odot/\mathrm{yr}$) of our sample are above the range where stochastic sampling of the IMF has been previously inferred to significantly impact UV and H$\alpha$ emission \citep[e.g.][]{Lee2009}.
However, the impact of stochastic sampling on stellar wind features and the $>54.4$ eV continuum is less clear.
We do not expect IMF sampling to have a significant impact on our interpretation of trends in our sample beyond potentially introducing scatter, but future work considering these effects will be important especially for detailed comparisons with stellar population synthesis models (Vidal-Garc\'{i}a et al., in prep.).

\subsection{\hstcos{}}

\begin{figure*}
	\includegraphics[width=\textwidth]{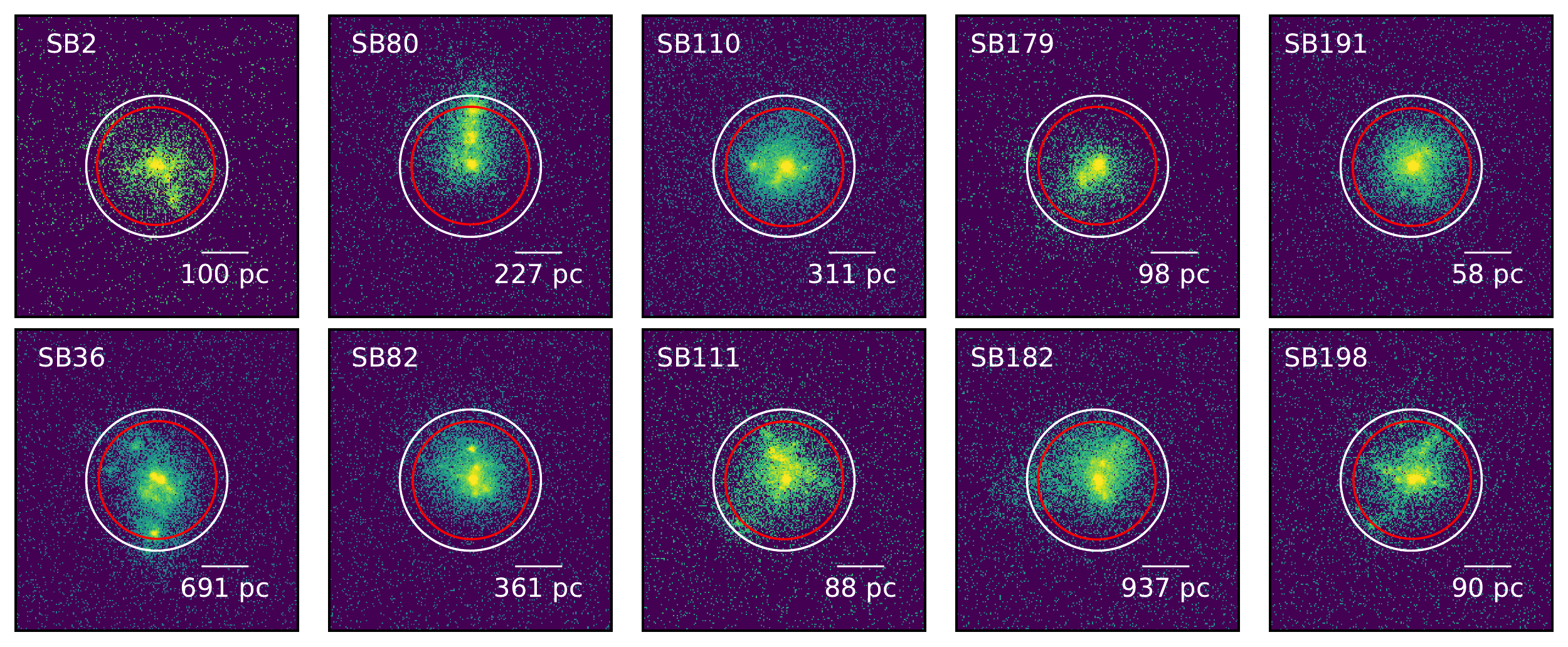}
    \caption{
        \hstcos{} NUV/MIRRORA target acquisition images.
        The white circle in each represents the SDSS 3\arcsec{} fiber aperture centred on the SDSS coordinates of the target; and the red circle represents the COS 2.5\arcsec{} aperture after centring on the flux centroid of the image.
        The 1\arcsec{} scalebar is labelled with an estimate of the comoving distance corresponding to this angle at the redshift of each target.
    }
    \label{fig:tacq}
\end{figure*}

The \hstcos{} observations were performed in the NUV at $\sim 1910$ \AA{} (with the G185M grism) and the FUV targeting $1450-1700$ \AA{} (G160M) with the 2.5\arcsec{} diameter Primary Science Aperture (PSA).
These grisms provide an optimal balance between spectral resolution and wavelength coverage for our program.
The observations are described in Table~\ref{tab:basicprop}.
Wavelength settings were chosen for each target to provide rest-frame coverage of [\ciii{}] 1907, \ciii{}] 1909 \AA{}; \oiii{}] $\lambda \lambda 1661, 1666$; \heii{} $1640$; and \civ $\lambda \lambda 1548, 1550$.
The targets were acquired in ACQ/IMAGE mode with MIRRORA and 43-95 second exposures (adjusted for the GALEX/NUV flux of each target).
The target acquisition images are displayed in Fig.~\ref{fig:tacq}.

\begin{table*}
	\centering
	\caption{
        Basic properties and \hstcos{} exposure times for our ten targets.
    }
	\label{tab:basicprop}
\begin{tabular}{lccccccc}
\hline
SBID  & RA & Dec & Wolf-Rayet class & Distance & 2.5\arcsec{} & $u$, $i$ & NUV/G185M, FUV/G160M\\ 
      & (J2000) & (J2000) & from SDSS (SB2012) & (Mpc) & comoving kpc & AB mag & exposure (s)\\ 
\hline
2 & 9:44:01.87 & -0:38:32.2 & Non-WR & 19 & 0.23 & 18.2, 18.1 & 2168, 2577 \\ 
36 & 10:24:29.25 & 5:24:51.0 & Non-WR & 141 & 1.71 & 18.1, 18.0 & 2136, 2609 \\ 
80 & 9:42:56.74 & 9:28:16.2 & WR & 46 & 0.56 & 17.9, 18.1 & 2132, 2608 \\ 
82 & 11:55:28.34 & 57:39:52.0 & Non-WR & 76 & 0.92 & 18.0, 18.0 & 2344, 2853 \\ 
110 & 9:42:52.78 & 35:47:26.0 & Non-WR & 63 & 0.76 & 18.2, 18.4 & 2156, 2665 \\ 
111 & 12:30:48.60 & 12:02:42.8 & WR & 16 & 0.20 & 18.5, 18.4 & 2132, 2613 \\ 
179 & 11:29:14.15 & 20:34:52.0 & WR & 25 & 0.30 & 18.3, 18.5 & 2160, 2588 \\ 
182 & 11:48:27.34 & 25:46:11.8 & Non-WR & 191 & 2.31 & 18.3, 17.9 & 2112, 2625 \\ 
191 & 12:15:18.60 & 20:38:26.7 & WR & 10 & 0.12 & 17.7, 18.2 & 2156, 2616 \\ 
198 & 12:22:25.79 & 4:34:04.8 & Non-WR & 16 & 0.20 & 18.3, 18.8 & 2120, 2605 \\ 
\hline
\end{tabular}
\end{table*}

The data were taken in TIME-TAG mode with FP-POS=ALL and FLASH=YES.
These settings allow a flat image to be constructed and minimize the impact of fixed-pattern noise in the detectors.
The data were reduced using \texttt{CALCOS} 3.1.7 (2016-02-03) and the latest calibration files (as of 2016-07-01) downloaded using STSDAS.

Extraction was performed using the default parameters for each detector.
The NUV extraction is performed using the BOXCAR algorithm by default; we confirmed that all target flux visible in the 2D NUV corrtag spectra was within the default extraction box.
As our targets are not point sources, we checked to ensure that the FUV TWOZONE extraction (optimized for point sources) was collecting all available target flux.
To do so, we re-ran the FUV extractions with the BOXCAR algorithm after checking to ensure that the box width captured the entirety of the target trace in the 2D corrtag spectra.
The final 1D spectra extracted using these two methods were virtually indistinguishable, so we chose to use the preferred TWOZONE output for the FUV data.

The G160M and G185M gratings have dispersions of 12.23 and 34 m\AA{}/pixel; or (since the FUV XDL and NUV MAMA have 6 and 3 pixels per resolution element) 73.4 and 102 m\AA{}/ resolution element.
We checked the approximate attained resolution of the spectra by fitting narrow Milky Way absorption features.
As these are much more common in the FUV/G160M wavelength range, most of our constraints come from this grism.
These fits yield line $\sigma$ (FWHM) of approximately 0.254 \AA{} (0.598 \AA{}) on average, corresponding to approximately 3.5 (8.1) resolution elements in the FUV and 2.5 (5.9) resolution elements in the NUV, respectively.
The one-dimensional spectra are binned via boxcar averaging over the length of this resolution element (or a multiple thereof) to achieve higher S/N per pixel.

\subsection{MMT}

The redshifts of our sample shift [\oii{}] $\lambda 3727$ into the SDSS spectral range for only two galaxies.
We obtained supplementary optical spectra for the other eight targets with the Blue Channel spectrograph on the MMT.
These observations were made with the 1.5\arcsec{}x180\arcsec{} slit and either the 800 or 300 lines/mm grating.
The resulting 800 lpm (300 lpm, respectively) spectra have a dispersion of 0.75 (1.96) \AA{}/pixel, a spectral FWHM resolution measured from the [\oiii{}] lines of approximately 3.3 (6.9) \AA{}, a spatial scale of $\sim$ 0.6\arcsec{}pixel along the slit, and all cover the wavelength range 3200-5200 \AA{}.
Data for SB2 and 80 were taken on the night of January 20th, 2016 with the 800 lpm grating at airmass $\lesssim 1.5$ and with (guider) seeing ranging from $0.9''$ to $1.5''$.
The targets SB 179, 110, 82, 191, 198, and 111  were observed in the second half of the night of January 8th, 2017 with the 300 lpm grating at airmass $\lesssim 1.2$; seeing reported by the guider remained near $1''$.
Arcs were obtained for each target with the HeAr and HgCd lamp combinations separately to cover the full observed wavelength range.
Standard stars LB 227, GD 108, and Feige 66 were observed at the beginning, middle, and end of the night (resp.) on January 20th 2016; and Feige 34 was observed on January 8th 2016.
Each target was observed at the parallactic angle to minimize slit loss.
The observations are summarized in Table~\ref{tab:mmt_obs}.
The data was reduced with standard longslit techniques in IRAF\footnote{IRAF is distributed by the National Optical Astronomy Observatories, which are operated by the Association of Universities for Research in Astronomy, Inc., under cooperative agreement with the National Science Foundation.}, and strong line fluxes were compared with the SDSS measurements to obtain an effective aperture correction for each target.

\begin{table}
	\centering
	\caption{Log of MMT observations from the nights of January 20 2016 and January 8 2017.}
	\label{tab:mmt_obs}
	\begin{tabular}{lcc}
		\hline
        SBID & Airmass & Exposure (s)\\
		\hline
        \hline
        \multicolumn{3}{c}{January 20 2016} \\
        \hline
		80 & 1.2 & 4500 \\
		2 & 1.2 & 2833 \\
        \hline
        \multicolumn{3}{c}{January 8 2017} \\
        \hline
        179 & 1.1 & 1200 \\
        110 & 1.1 & 1200 \\
        82 & 1.2 & 1200 \\
        191 & 1.0 & 900 \\
        198 & 1.1 & 1200 \\
        111 & 1.0 & 1500 \\
		\hline
	\end{tabular}
\end{table}

\subsection{ESI}

We also obtained data with the Echellette Spectrograph and Imager \citep[ESI, ][]{Sheinis2002} on Keck II for our targets.
These data were taken with the 1\arcsec{}x20\arcsec{} slit on March 29, 2016 and January 20--21, 2017 under seeing ranging from $\sim$ 0.8 -- 1.2 arcseconds (see Table~\ref{tab:esi_obs}).
This yields spectra covering the wavelength range 3900-10900 \AA{} with approximately 11.5 km/s/pixel dispersion and 0.154 arcseconds/pixel along the slit.

We reduced the ESI spectra using the \texttt{ESIRedux} code\footnote{\url{http://www2.keck.hawaii.edu/inst/esi/ESIRedux/}}.
Dome flats and HgNeXeCuAr lamp exposures were used to perform flat and wavelength calibrations.
The object continuum was used to trace the echelle orders in each exposure, and the spectra were boxcar-subtracted with a radius of 20 pixels to ensure all object flux was captured.
Since an archival sensitivity curve was used to perform first-order flux corrections, the final one-dimensional spectra are flux-calibrated only in a relative sense.
These spectra yield an extremely high-resolution view of optical lines, which is critical in crowded line complexes such as that near \heii{} $\lambda 4686$.

\begin{table}
	\centering
	\caption{Log of ESI observations.}
	\label{tab:esi_obs}
	\begin{tabular}{lcc}
		\hline
        \hline
        SBID & Airmass & Exposure (s)\\
		\hline
        \multicolumn{3}{c}{March 29, 2016}\\
        \hline
		110 & 1.1 & 8100 \\
		82 & 1.3 & 9000 \\
		191 & 1.1 & 5400 \\
		198 & 1.4 & 4200 \\
        \hline
        \multicolumn{3}{c}{January 20, 2017}\\
        \hline
        2 & 1.6 & 7200 \\
        182 & 1.4 & 9000 \\
        111 & 1.1 & 8100 \\
        \hline
        \multicolumn{3}{c}{January 21, 2017}\\
		\hline
        80 & 1.3 & 6400\\
        179 & 1.1 & 5400\\
        36 & 1.2 & 9600\\
        \hline
	\end{tabular}
\end{table}

\subsection{SDSS}

Our targets were selected using optical fiber spectra and imaging originally released in SDSS DR7.
We obtained reduced spectra and imaging frames for our objects from the SDSS archive.
The spectra cover approximately the wavelength range 3800-9200 \AA{} at a spectral resolution $R \sim 1800$.
The fiber diameter is approximately 3\arcsec{} on-sky \cite{York2000}.
Imaging from SDSS covers 5 filters denoted $u$, $g$, $r$, $i$, and $z$, with central wavelengths ranging from 3551 to 8932 \AA{}.

\subsection{Line Measurement and Photometry}
\label{sec:linefit}

We use custom fitting software to measure line features in the one-dimensional spectra.
For isolated emission and absorption lines, we adopt a model consisting of a linear function describing the local continuum plus a Gaussian described by a mean, total area, and standard deviation.
For close doublets and lines with potentially two different velocity components, we add a second Gaussian to this model (in the former case, with linewidths enforced to be similar to the first).
We define the Bayesian posterior likelihood function as the product of a $\chi^2$ likelihood describing the data-model difference with a flat prior over physically-plausible ranges of the model parameters.
To infer parameter values given our model and data, we explore the resulting posterior distribution using the affine-invariant Markov chain Monte Carlo (MCMC) ensemble sampler \texttt{emcee} \citep{Foreman-Mackey2013}.
After removing burn-in, we adopt the $16-50-84$ percentiles from the sampler distribution as our parameter value and error estimates.
These fits are initiated with parameters centred on a guess at the feature to be fit; we adjust burn-in and check the results visually to ensure the chain has converged on a reasonable model and that the resulting likelihood functions are not significantly multimodal.

The spectroscopic apertures of COS and SDSS are roughly circular and of-order the same size (2.5\arcsec{} and 3\arcsec{} in diameter, respectively); thus aperture photometry is appropriate for analysis of the broadband properties of the targets.
We measure flux in the SDSS images using a 3\arcsec{}-diameter circular aperture and estimate the local sky background within 1\arcmin{} after sigma-clipping (low $10\sigma$, high $3\sigma$).

\subsection{Dust corrections}
\label{sec:extinction}

Correction of attenuation due to dust absorption and scattering is of paramount concern in the UV.
The choice of attenuation model can have a significant effect on parameters inferred from UV spectral features \citep[see e.g.][]{Wofford2014}.
For fitting photometry and UV spectral features in this paper, we adopt the 2-component model presented by \citet{Charlot2000}, as described in Section~\ref{sec:beagle}.

For the purposes of correcting optical emission lines for dust attenuation in Section~\ref{sec:measurements}, we adopt a simplified approach.
In particular, we utilize extinction curves rather than an attenuation model, as scattering should have a minimal effect in the optical over the small field of view probed by the SDSS aperture.
First, we correct for Galactic dust extinction towards each object.
Galactic extinction maps and curves are relatively well-determined; in particular, we use the maps of \citet{Schlafly2011} to determine E(B-V) and assume the $R_\text{V}=3.1$ extinction curve of \citet{Fitzpatrick1999}.
To estimate the residual intrinsic reddening of the nebular emission in each galaxy, we measure the Balmer decrement relative to the Case B recombination value of H$\alpha$/H$\beta$ computed with \texttt{PyNeb} \citep{Luridiana2015} for the $T_e$ and $n_e$ values computed from [\oiii{}] with this software as described in Section~\ref{sec:optgas} --- we repeat the process of deriving $n_e$, $T_e$, and $\mathrm{E(B-V)}$ until these quantities converge (which occurs within three iterations). 
The final adopted values of H$\alpha$/H$\beta$ range from 2.78 to 2.86.
We adopt the SMC bar average extinction curve measured by \citet{Gordon2003}.
The results for optical lines are essentially unchanged if we instead use the Galactic diffuse average curve of \citet{Fitzpatrick1999}, since these curves diverge significantly only in the UV.
These systems are dominated by young stars, such that correction of the Balmer lines due to underlying stellar absorption is negligible; the ESI data reveal underlying absorption in H$\beta$ for only one system (SB 111) at $4\%$ of the total line flux.

For each object, we check that the extinction derived via the Balmer decrement is consistent with Case B predictions for the other Balmer lines accessible in the SDSS spectra (H$\gamma$, H$\delta$).
The agreement is good to within a few percent and consistent with the measurement errors in the line ratios, except in the case of SB 2.
The reddening inferred from H$\alpha$/H$\beta$ yields Case B predictions for H$\delta$ and H$\gamma$ which are very inconsistent with the observed values even accounting for reasonable variation in $T_e$.
In addition, the H$\alpha$ line profile for SB 2 shows an asymmetric redward extension which accounts for about half the total flux in the line and does not appear in the other strong nebular lines or in the ESI spectrum of this source.
Ignoring H$\alpha$ and instead using the ratio of H$\beta$/H$\gamma$ to de-redden the optical yields an intrinsic $\mathrm{E(B-V)}$ in much better agreement with the other objects (c.\ f.\ $\sim 0.5$ using H$\alpha$/H$\beta$) and good agreement with H$\delta$.
For the purposes of measuring extinction and other properties of the nebular gas in this object, we ignore the raw H$\alpha$ flux and instead predict it where-necessary from H$\beta$/H$\gamma$ and Case B.

\section{Results}
\label{sec:measurements}

The UV spectra of star-forming systems are sensitive to both gas conditions as well as the winds and ionizing spectra of massive stars.
In this section, we first present measurements and physical parameters derived from optical photometry and spectra.
We then explore the \hstcos{} UV spectra in the context of these other measurements.
Finally, we present the Keck/ESI optical spectra, which provide a close look at \heii{} $\lambda 4686$ and the Wolf-Rayet stars in these systems.

\subsection{Bulk Stellar Population Constraints from Photometry}
\label{sec:photres}

The optical broadband photometry provides constraints on the total stellar mass and star formation activity.
We fit the SDSS photometry using the Bayesian spectral analysis code \beagle{}\footnote{http://www.jacopochevallard.org/beagle/} \citep{Chevallard2016}; this code and the parameter space explored by the models are described in-detail in Section~\ref{sec:beagle}.
We adopt a constant star formation history for our analysis as this provides an adequate fit to the measurements.
Incorporating an older stellar population by fitting an exponentially-delayed plus burst star formation model increases the inferred stellar mass by $0.4 \pm 0.2$ dex.
Since \beagle{} incorporates nebular line predictions from \cloudy{}, photometric band contamination from strong optical lines is naturally and self-consistently modeled.
We experiment with excluding the $g$ and $r$ bands (contaminated by the strong lines H$\beta$, [\oiii{}]$4959,5007$, and H$\alpha$) and with including aperture-corrected GALEX magnitudes \citep[using T-PHOT:][]{Merlin2015}, but neither have a significant effect on the derived masses or specific star formation rates.
For simplicity and to avoid systematic error due to aperture effects, we present results from fitting the full set of SDSS photometry only (bands $u$, $g$, $r$, $i$, and $z$) which probe the stellar continuum and strong optical emission lines from $\sim 3500 - 9000$ \AA{}.
A visual inspection confirms that the $u$, $i$, and $z$ bands closely approximate the continuum in the SDSS spectra.
The results are displayed in Table~\ref{tab:sedres}.

\begin{table}
	\centering
	\caption{
        Parameter estimates derived from broadband SED fitting with \beagle{} (see Section~\ref{sec:beagle} for details).
        These measurements indicate that the systems are low-mass (within the spectroscopic and photometric aperture: $\log_{10}(M/M_\odot) \sim 4.7$ -- $7.5$) and dominated by recent star formation, yielding extremely large specific star formation rates (median sSFR $100$ \unit{Gyr^{-1}} from photometry only).
        Note that the SFR indicated here is measured over the last 10 Myr.
    }
    \label{tab:sedres}
\begin{tabular}{lccc}
\hline
SBID & $\log_{10}(M/M_{\odot})$ & $\log_{10}(\mathrm{SFR}/(M_{\odot}/\mathrm{yr}))$ & \tauV{}\\ 
\hline
2 & $5.1 \pm 0.0$ & $-1.88 \pm 0.04$ & $0.57 \pm 0.10$ \\ 
36 & $7.5 \pm 0.2$ & $-0.25 \pm 0.08$ & $0.08 \pm 0.08$ \\ 
80 & $6.1 \pm 0.1$ & $-0.94 \pm 0.06$ & $0.33 \pm 0.13$ \\ 
82 & $6.4 \pm 0.1$ & $-0.58 \pm 0.06$ & $0.25 \pm 0.06$ \\ 
110 & $6.4 \pm 0.1$ & $-0.91 \pm 0.05$ & $0.02 \pm 0.02$ \\ 
111 & $5.6 \pm 0.2$ & $-2.20 \pm 0.10$ & $0.19 \pm 0.08$ \\ 
179 & $5.2 \pm 0.1$ & $-1.77 \pm 0.05$ & $0.20 \pm 0.07$ \\ 
182 & $7.3 \pm 0.1$ & $0.23 \pm 0.04$ & $0.27 \pm 0.05$ \\ 
191 & $4.9 \pm 0.1$ & $-2.11 \pm 0.07$ & $0.37 \pm 0.07$ \\ 
198 & $4.7 \pm 0.3$ & $-2.27 \pm 0.12$ & $0.03 \pm 0.04$ \\ 
\hline
\end{tabular}
\end{table}

The photometric data suggest that the systems are dominated by recently-formed stars.
We infer total stellar masses spanning 2 orders of magnitude, from $10^{5}$ to $10^{7}$ $M_\odot$.
The lowest-mass objects $\log_{10} M_\star/M_\odot \lesssim 5.5$ are predominantly giant \hii{} regions embedded in nearby disk systems (SB 198, 191, 179; see Fig.~\ref{fig:sdss_montage}).
Adopting the \hstcos{} spectroscopic aperture as a rough measure of the size of the star-forming region probed, the implied stellar mass surface densities span approximately a half-dex around $10^{6.5} M_\odot / \unit{kpc}^2$.
They are generally unreddened, with $V$-band optical depth to dust $\tauV{} \lesssim 0.6$ (corresponding approximately to $A_V \lesssim 0.6$).
The derived specific star formation rates (sSFRs) are uniformly high, $10-100$ \unit{Gyr^{-1}}, implying these systems have undergone intense recent star formation.
These sSFRs are comparable to those measured for large samples of local extreme emission-line galaxies such as the green peas \citep{Cardamone2009, Izotov2011a}, though our objects are lower in mass than the median for such samples \citep[$\sim 10^9 M_\odot$ total with $10^7 M_\odot$ in a recent burst;][]{Izotov2011a}.
The sSFRs in our sample are comparable to those inferred for photometric samples at $z\sim 7$ \citep{Schaerer2010,Stark2013,Salmon2015}, as well as those inferred from photoionization modeling of systems at $z\sim 6-7$ with high-ionization UV line detections \citep{Stark2015,Stark2016}.
Very young local star-forming galaxies such as these are likely to have moderately metal-poor gas and extreme nebular line emission reflecting the presence of numerous massive stars.

\subsection{Gas Conditions from Optical Spectra}
\label{sec:optgas}

Nebular lines in the optical spectra provide information about the ionization state, physical conditions, and composition of the gas.
We report diagnostic line ratios as well as several derived parameters in Tables~\ref{tab:optneb} and \ref{tab:optneb2}.
In these tables and in the rest of the paper (unless explicitly stated), all equivalent widths are measured in the rest-frame.
In these tables and in deriving quantities in this section, all line measurements are from SDSS with the exception of [\oii{}] $\lambda 3727$, which is measured in the MMT spectra where noted.
For measurements involving [\oii{}], we scale the MMT [\oii{}] flux by the median ratio of strong lines measured in both SDSS and MMT for aperture correction.
This correction factor varies by $<10\%$ between lines spanning $3900-5000$ \AA{} for each object; we add an additional 10\% uncertainty in-quadrature to the measured [\oii{}] $\lambda 3727$ uncertainty to account for this aperture correction.

We use the direct-$T_e$ method to measure gas-phase oxygen abundances.
In particular, we determine the electron temperature and density appropriate for singly and doubly ionized oxygen separately using the \texttt{getCrossTemDen} method provided by \texttt{PyNeb} \citep{Luridiana2015}.
For \oii{} and \oiii{}, we fit a temperature-sensitive line ratio ([\oii{}] $\lambda \lambda 3726, 3729$ / [\oii{}] $\lambda \lambda 7320, 7330$ and [\oiii{}] $\lambda 4363$ / [\oiii{}] $\lambda \lambda 4959, 5007$, respectively) alongside the density-sensitive [\sii{}] $\lambda 6731$ / [\sii{}] $\lambda 6716$ doublet.
We adopt the most up-to-date atomic and collisional data packaged with \texttt{PyNeb} for the three species involved, since adjustments to these quantities can have a significant impact on the gas properties derived using forbidden-line diagnostics \citep[e.g.][]{Sanders2016}.
We use collision strengths from \citet{Tayal2010} for [\sii{}], \citet{Kisielius2009} for [\oii{}], \citet{Storey2014} for [\oiii{}]; and atomic data from \citet{Fischer2004} for [\oii{}] and [\oiii{}] alongside \citet{Tayal2010} for [\sii{}].
With $T_e$ and $n_e$ measured directly for [\oii{}] and [\oiii{}] in-hand, we then compute total gas-phase oxygen abundance from the [\oii{}] $\lambda \lambda 3726, 3729$ and [\oiii{}] $\lambda \lambda 4959, 5007$ fluxes relative to H$\beta$ using the \texttt{getIonAbundance} method in \texttt{PyNeb}.
Uncertainties are propagated through this process by repeating the computation with resampled line fluxes from the line fit posterior distributions.
As described in Section~\ref{sec:extinction}, we compute the intrinsic Case B Balmer spectrum used to estimate extinction with these $T_e$, $n_e$ values, and iterate until convergence.
In addition, we infer star formation rates from the H$\alpha$ luminosities by converting to an ionizing photon luminosity above 13.6 eV assuming Case B using the \texttt{PyNeb}-derived $T_e$ and $n_e$, applying the ratio between ionizing photon luminosity and SFR derived from a $Z_\odot/5$ constant star formation history model at 100 Myr with a \citet{Chabrier2003} IMF produced with \beagle{} \citep{Chevallard2016}.
This is then compared to the stellar masses inferred from photometric SED fitting assuming the same IMF and a constant star formation history (see Table~\ref{tab:sedres}) to derive specific star formation rates.
We present total oxygen abundances, $T_e(\mathrm{\oiii{}})$, $n_e(\mathrm{\sii{}})$, and specific star formation rates derived in this manner in Table~\ref{tab:optneb2}.

The direct-$T_e$ method as-applied has the advantage of being independent from photoionization modeling and assumptions about the ionizing spectrum, but has some limitations.
For simplicity, we ignore contributions from $\mathrm{O}^{3+}$ in this direct metallicity computation as we lack access to any lines from this species.
\citet{Izotov2006} provides an approximate correction formula for this species based upon the $\mathrm{He^{2+}/He^{+}}$ ionic abundance; using the nebular \heii{} $\lambda 4686$ / \hei{} $\lambda 6678$ ratio measured in the high-resolution ESI data (dust-corrected, $\leq 2.2$) and the abundance formulae provided by \citet{Benjamin1999}, we find $\mathrm{He^{2+}/He^{+}} \leq 0.06$, and a negligible $\lesssim 0.01$ dex correction to our derived oxygen abundances.
Recently, \citet{Paalvast2017} described a systematic offset towards lower $T_e$-derived metallicities at very low star formation rates due to stochastic IMF sampling; but they predict this effect to be negligible ($\ll 0.1$ dex) at the star-formation rates probed here ($\gtrsim 10^{-2} M_\odot/\mathrm{yr}$).
Finally, this method implicitly assumes a simplified two-zone ionization structure; but note that the metallicities derived here are in reasonable agreement with those derived from full photoionization modeling of the strong UV and optical lines, as described in Section~\ref{sec:uvspectra}.

\begin{table*}
	\centering
	\caption{
        Optical nebular line measurements (flux and equivalent width, uncorrected for extinction) and derived extinction (galactic and residual Balmer decrement).
    }
	\label{tab:optneb}
\begin{tabular}{lccccccc}
\hline
SBID & H$\beta$ & H$\beta$ & [\oiii{}] 4959,5007 & [\oiii{}] 4959,5007 & [\oiii{}] 4363 & [\nii{}] 6584 & $\mathrm{E(B-V)}$\\ 
 & ($10^{-15}$ ergs/s/cm$^2$)& $W_0$ (\AA{})& ($10^{-15}$ ergs/s/cm$^2$)& $W_0$ (\AA{})& ($10^{-15}$ ergs/s/cm$^2$)& ($10^{-15}$ ergs/s/cm$^2$)& intrinsic (galactic)\\ 
\hline
2 & $67.4 \pm 2.1$ & $273 \pm 16$ & $516.6 \pm 38.3$ & $1844 \pm 253$ & $7.59 \pm 0.12$ & $2.68 \pm 0.07$& 0.15 (0.05)\\ 
36 & $26.6 \pm 0.4$ & $93 \pm 3$ & $189.1 \pm 6.0$ & $592 \pm 30$ & $2.57 \pm 0.10$ & $1.64 \pm 0.04$& 0.06 (0.02)\\ 
80 & $63.6 \pm 2.4$ & $243 \pm 17$ & $481.8 \pm 26.2$ & $1661 \pm 173$ & $3.30 \pm 0.11$ & $7.01 \pm 0.21$& 0.13 (0.02)\\ 
82 & $54.7 \pm 0.6$ & $178 \pm 4$ & $496.5 \pm 18.0$ & $1304 \pm 82$ & $7.10 \pm 0.14$ & $3.09 \pm 0.07$& 0.10 (0.03)\\ 
110 & $19.1 \pm 0.3$ & $86 \pm 2$ & $131.2 \pm 3.4$ & $542 \pm 24$ & $1.39 \pm 0.06$ & $2.19 \pm 0.07$& 0.06 (0.01)\\ 
111 & $21.7 \pm 0.6$ & $102 \pm 5$ & $144.5 \pm 6.5$ & $642 \pm 51$ & $2.42 \pm 0.11$ & $0.65 \pm 0.04$& 0.07 (0.02)\\ 
179 & $35.3 \pm 0.7$ & $196 \pm 8$ & $215.1 \pm 9.5$ & $1081 \pm 92$ & $1.15 \pm 0.06$ & $4.87 \pm 0.17$& 0.17 (0.02)\\ 
182 & $33.6 \pm 0.5$ & $151 \pm 4$ & $276.4 \pm 8.4$ & $1053 \pm 49$ & $3.45 \pm 0.08$ & $2.26 \pm 0.09$& 0.11 (0.02)\\ 
191 & $88.6 \pm 3.2$ & $393 \pm 23$ & $723.5 \pm 59.8$ & $2435 \pm 306$ & $4.04 \pm 0.10$ & $5.03 \pm 0.14$& 0.02 (0.03)\\ 
198 & $32.3 \pm 0.9$ & $189 \pm 11$ & $167.9 \pm 9.8$ & $871 \pm 96$ & $1.06 \pm 0.06$ & $5.02 \pm 0.24$& 0.10 (0.02)\\ 
\hline
\end{tabular}
        
\end{table*}

\begin{table*}
	\centering
	\caption{
        Optical nebular line ratios and derived gas physical conditions, all corrected for extinction.
        The \ott{} and \rtt{} measurements come from both SDSS [1] and MMT [2].
    }
	\label{tab:optneb2}
\begin{tabular}{lcccccccc}
\hline
SBID & $\mathrm{O}_{32}$ & $\mathrm{R}_{23}$ & [\oiii{}] 5007 / H$\beta$ & [\nii{}] 6584 / H$\alpha$ & $n_e(\mathrm{\sii{}})$ & $T_e(\mathrm{\oiii{}})$ & $12+\log_{10}(\mathrm{O/H})$ & $\mathrm{sSFR}/\mathrm{Gyr}^{-1}$\\ 
  &  &  &  &  & ($\mathrm{cm}^{-3}$) & ($10^4 \mathrm{K}$) & direct-$T_e$ & H$\alpha$\\ 
\hline
2& $7.5_{-0.6}^{+0.8}$ [2] & $8.2_{-1.1}^{+1.7}$& $5.59 \pm 0.52$& $0.0113 \pm 0.0005$ \textdagger & $186 _{ -32 } ^{ +33 }$ & $1.58 \pm 0.05$ & $7.81 \pm 0.07$ & $339 _{ -33 } ^{ +40 }$ \textdagger\\ 
36& $5.5_{-0.3}^{+0.3}$ [1] & $8.3_{-0.3}^{+0.3}$& $5.27 \pm 0.21$& $0.0200 \pm 0.0008$ & $119 _{ -34 } ^{ +37 }$ & $1.49 \pm 0.03$ & $7.92 \pm 0.04$ & $19 _{ -5 } ^{ +10 }$\\ 
80& $3.7_{-0.3}^{+0.3}$ [2] & $9.6_{-1.1}^{+1.2}$& $5.58 \pm 0.38$& $0.0324 \pm 0.0027$ & $168 _{ -39 } ^{ +44 }$ & $1.15 \pm 0.02$ & $8.24 \pm 0.06$ & $187 _{ -26 } ^{ +33 }$\\ 
82& $9.2_{-1.1}^{+1.3}$ [2] & $8.9_{-1.8}^{+3.0}$& $6.73 \pm 0.30$& $0.0174 \pm 0.0010$ & $158 _{ -39 } ^{ +47 }$ & $1.54 \pm 0.03$ & $7.91 \pm 0.04$ & $177 _{ -34 } ^{ +53 }$\\ 
110& $3.8_{-0.1}^{+0.1}$ [2] & $8.7_{-1.2}^{+1.6}$& $5.10 \pm 0.17$& $0.0377 \pm 0.0020$ & $70 _{ -33 } ^{ +36 }$ & $1.33 \pm 0.03$ & $8.17 \pm 0.08$ & $37 _{ -6 } ^{ +9 }$\\ 
111& $6.4_{-0.2}^{+0.2}$ [2] & $7.5_{-0.5}^{+0.8}$& $4.92 \pm 0.30$& $0.0097 \pm 0.0007$ & $22 _{ -16 } ^{ +26 }$ & $1.65 \pm 0.05$ & $7.81 \pm 0.08$ & $21 _{ -6 } ^{ +13 }$\\ 
179& $2.7_{-0.1}^{+0.1}$ [2] & $8.5_{-1.4}^{+1.9}$& $4.43 \pm 0.25$& $0.0389 \pm 0.0034$ & $117 _{ -35 } ^{ +34 }$ & $1.07 \pm 0.02$ & $8.35 \pm 0.07$ & $214 _{ -45 } ^{ +74 }$\\ 
182& $5.4_{-0.3}^{+0.4}$ [1] & $9.6_{-0.3}^{+0.3}$& $6.08 \pm 0.26$& $0.0208 \pm 0.0014$ & $125 _{ -34 } ^{ +37 }$ & $1.45 \pm 0.02$ & $8.01 \pm 0.04$ & $91 _{ -15 } ^{ +22 }$\\ 
191& $9.5_{-0.9}^{+0.7}$ [2] & $9.0_{-1.5}^{+1.8}$& $6.07 \pm 0.61$& $0.0187 \pm 0.0033$ & $95 _{ -62 } ^{ +81 }$ & $1.05 \pm 0.03$ & $8.30 \pm 0.07$ & $128 _{ -28 } ^{ +35 }$\\ 
198& $1.7_{-0.2}^{+0.1}$ [2] & $7.9_{-1.6}^{+3.2}$& $3.81 \pm 0.27$& $0.0475 \pm 0.0051$ & $42 _{ -23 } ^{ +29 }$ & $1.11 \pm 0.03$ & $8.48 \pm 0.11$ & $146 _{ -68 } ^{ +281 }$\\ 
\hline
\end{tabular}

\vspace{1mm}
{\footnotesize \textdagger Due to the complicated H$\alpha$ line profile for SB 2 (see Sec.~\ref{sec:extinction}), the dust-corrected H$\alpha$ flux is predicted by rescaling H$\beta$ according to Case B.}
\end{table*}

\begin{figure}
	\includegraphics[width=\columnwidth]{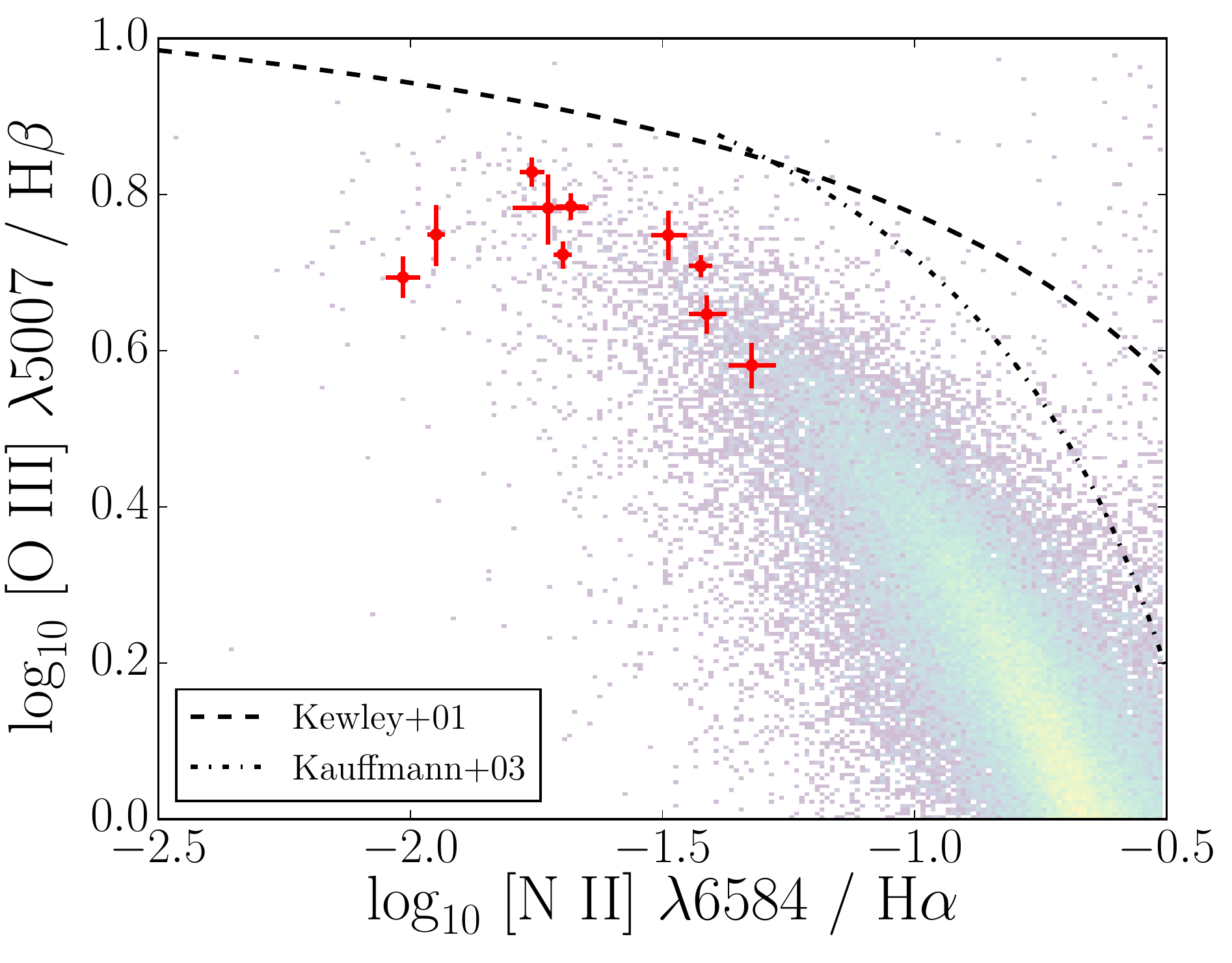}
    \caption{
        The relevant segment of the BPT diagram depicting our sources (red) and the log-histogram of SDSS galaxies with S/N$>3$ in each of the relevant lines (background density plot).
        The maximum starburst lines of \citet[][theoretical]{Kewley2001} and \citet[][empirically-modified]{Kauffmann2003} are displayed as dashed lines.
        Our galaxies lie in the extreme tail of the star-forming sequence, with very high [\oiii{}]/H$\beta$ ratios.
    }
    \label{fig:bpt}
\end{figure}

The optical line measurements confirm the extreme nature of these objects.
The strength of H$\alpha$ and [\oiii{}] are very different from typical nearby galaxies.
The star formation rates inferred from H$\alpha$ validate the very large specific star formation rates suggested by the SED fits (Table~\ref{tab:sedres}), spanning the range $20-300$ \unit{Gyr^{-1}}.
Our objects all have $>$500 \AA{} equivalent width [\oiii{}] $\lambda \lambda 4959, 5007$ and $\gtrsim$100 \AA{} H$\beta$ emission, placing them securely in the realm of rare $z\sim 0$ extreme emission line galaxies \citep[e.g.][]{Cardamone2009,Izotov2011a}.
While rare at low-redshift, optical equivalent widths of the magnitude observed here have been inferred routinely from IRAC band contamination in systems at $z\gtrsim 7$ \citep[e.g.][]{Labbe2013, Smit2014, Huang2016, Roberts-Borsani2016}.
As for the photometric measurements above, this suggests that the relative intensity of star formation in these systems is comparable to that occurring in UV-selected galaxies at high-redshift.

The gas in these systems is both metal-poor and highly ionized.
The direct-temperature metallicities range from $12+\log_{10}(\mathrm{O/H})\sim 7.8-8.5$, i.e.\ $\sim Z_{\odot}/8 - Z_{\odot}/2$.
In the BPT diagram (Fig.~\ref{fig:bpt}), they lie in the extreme tail of the star-forming sequence towards low [\nii{}]/H$\alpha$ and high [\oiii{}]/H$\beta$.
Our objects show no signs of a deviation from the local SDSS star-forming locus towards higher [\nii{}]/H$\alpha$ as observed at $z\sim 2-3$ \citep[e.g.][]{Steidel2014, Shapley2015, Sanders2016, Kashino2017, Strom2017}; note that this is a natural consequence of their selection using BPT diagram cuts designed to select $z\sim 0$ star-forming galaxies \citep{Shirazi2012}.
The quantity \ott{} $=$ [\oiii{}] $\lambda 4959$ + $\lambda 5007$ / [\oii{}] $\lambda \lambda 3727, 3729$, a proxy for the ionization parameter (or density of ionizing radiation), is also significantly larger than in typical star-forming galaxies nearby.
The \ott{} ratio ranges from 2 to 10 in our sample with a median of 6, whereas the vast majority of nearby SDSS galaxies present \ott{} $\lesssim 1$ \citep[e.g.][]{Sanders2016} at similar \rtt{} ($=$ [\oiii{}] $\lambda 4959$ + $\lambda 5007$ + [\oii{}] $\lambda \lambda 3727, 3729$ / H$\beta$).
These extreme \ott{} values indicate that the gas in these objects is exposed to ionizing radiation from very recently formed stars, and may imply significant Lyman continuum escape \citep[e.g.][]{Jaskot2013,Izotov2016a,Stasinska2015}.

The photometry and optical spectroscopy together reveal systems dominated by recent star formation (sSFR $\sim 100 \unit{Gyr}^{-1}$) with little intrinsic dust reddening ($\mathrm{E(B-V)} \lesssim 0.1$) and extreme optical line emission (\oiii{}] $\lambda \lambda 4959,5007$ EWs $\sim 500-2500$ \AA{}).
Systems with comparable properties observed in the rest-UV at $z\sim 2$ reveal strong UV nebular emission, with \ciii{}] emission reaching EWs of $\sim 15$ \AA{} \citep[e.g.][]{Erb2010,Stark2014}; and at the highest redshifts, similarly strong \ciii{}] and even more extreme \civ{} at $\sim 40$ \AA{} \citep{Stark2015,Stark2015a,Stark2016,Mainali2017}.
Though strong UV nebular emission appears to be common at the highest redshifts, the stellar populations which power it and its dependence on bulk galaxy properties (metallicity, ionization parameter) remain unclear.

\subsection{The UV Spectra}
\label{sec:uvspectra}

\begin{figure*}
	\includegraphics[width=\textwidth]{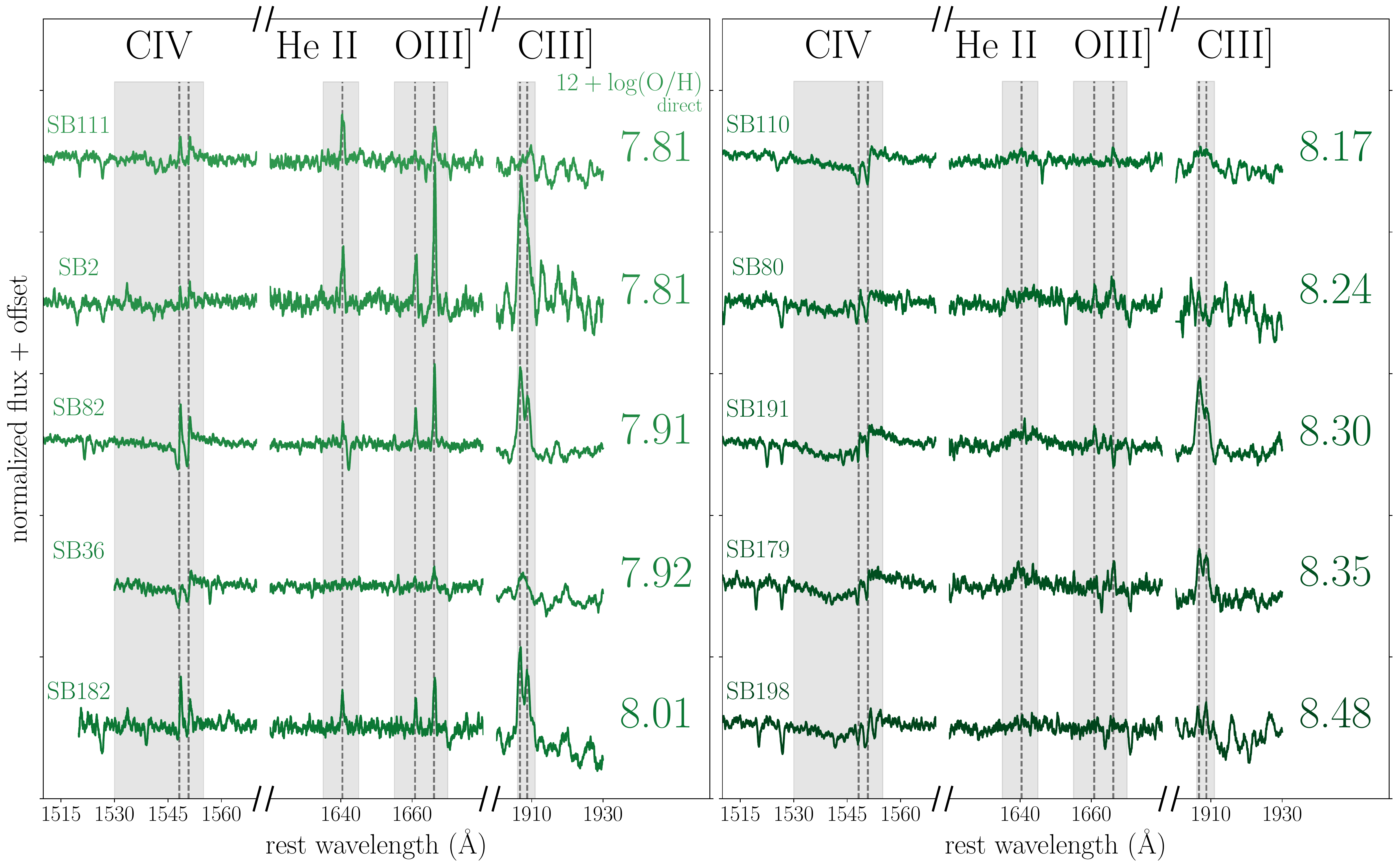}
    \caption{
        The \hstcos{} spectra obtained for our systems, ranked by gas-phase metallicity obtained via the direct method applied to the optical oxygen lines (see Sec.~\ref{sec:optgas}).
        The spectra have been median-normalized and offset vertically for display; and the key metal lines (\civ{}, \heii{}, \oiii{}], \ciii{}]) are highlighted.
        The relatively metal-poor objects (left panel, $12+\log_{10} \mathrm{O/H} \lesssim 7.9$) show extreme emission in these lines, with the \ciii{}] doublet reaching equivalent widths $\sim 15$ \AA{} and nebular \heii{}, \civ{} appearing in four of the five, emission rarely seen in typical star-forming galaxies.
        In contrast, above this metallicity the spectra show relatively weak nebular emission (except \ciii{}], which still reaches equivalent widths $\sim 10$ \AA{} in SB 179 and 191) and are increasingly dominated by stellar wind features in \civ{} and \heii{}.
    }
    \label{fig:majorlines_comp}
\end{figure*}

The \hstcos{} FUV and NUV spectra reveal extreme nebular emission and strong stellar features from the winds of massive stars.
Despite the fact that the systems are uniformly undergoing extreme star formation (sSFR $\gtrsim 20 \unit{Gyr^{-1}}$), there is significant variation in their UV nebular properties.
We plot the spectra in Fig.~\ref{fig:majorlines_comp} with key features highlighted; and present nebular line measurements in Tables~\ref{tab:uvneb_flux} and \ref{tab:uvneb_ew}.

Figure~\ref{fig:majorlines_comp} highlights a clear metallicity trend in the UV spectra.
Above $12+\log\mathrm{O/H} > 8.1$, the average object is characterized by prominent stellar wind features --- deep \civ{} $\lambda\lambda 1548, 1550$ P-Cygni stellar wind features (dominated by massive O stars) and broad \heii{} $\lambda 1640$ emission (produced in the dense, highly-ionized winds of WR stars and very massive OIf supergiants).
The stellar winds of O stars are driven by metal line opacities and observed via a carbon transition, and are thus inherently metallicity-dependent.
The winds of WR stars likely depend similarly on metallicity --- we discuss these winds in more detail in Sec.~\ref{sec:ionsource}.
The most extreme nebular lines (\heii{}, \civ{}) are undetected in this metal-rich subset.
In contrast, the spectra of the more metal-poor objects ($12+\log\mathrm{O/H} \lesssim 8.0$) show weak stellar \civ{} and \heii{} features and are instead dominated by nebular emission in \ciii{}, \civ{}, and \heii{}.

We discuss the UV spectrum of each object in the context of their optical measurements individually below, in order of decreasing gas-phase direct-$T_e$ metallicity.

\begin{table*}
	\centering
	\caption{
        UV nebular emission line flux measurements for \civ, \heii{}, \oiii{}], and \ciii{}] in the \hstcos{} UV spectra.
        We provide $3\sigma$ upper limits when possible.
        The \oiii{}] 1666 line in SB 191 is obscured by MW absorption.
    }
	\label{tab:uvneb_flux}
\begin{tabular}{lcccccc}
\hline
SBID & \civ{} 1548 & \civ{} 1550 & \heii{} 1640 & \oiii{}] 1661 & \oiii{}] 1666 & \ciii{}] 1907, 1909\\ 
 & ($10^{-15}$ ergs/s/cm$^2$)& ($10^{-15}$ ergs/s/cm$^2$)& ($10^{-15}$ ergs/s/cm$^2$)& ($10^{-15}$ ergs/s/cm$^2$)& ($10^{-15}$ ergs/s/cm$^2$)& ($10^{-15}$ ergs/s/cm$^2$)\\ 
\hline
\hline
2 & $0.20 \pm 0.02$ & $0.19 \pm 0.01$ & $1.45 \pm 0.06$ & $1.40 \pm 0.04$ & $3.35 \pm 0.04$ & $11.06 \pm 0.63$\\ 
36 & <$3.3$ & <$3.3$ & <$1.1$ & $0.70 \pm 0.06$ & $2.16 \pm 0.08$ & $7.60 \pm 0.88$\\ 
80 & <$1.6$ & <$1.6$ & <$0.8$ & $0.60 \pm 0.04$ & $1.69 \pm 0.07$ & <$4.1$\\ 
82 & $1.92 \pm 0.05$ & $1.32 \pm 0.05$ & $1.22 \pm 0.05$ & $2.33 \pm 0.05$ & $5.05 \pm 0.05$ & $22.05 \pm 0.40$\\ 
110 & $0.20 \pm 0.05$ & $1.14 \pm 0.09$ & <$1.2$ & <$0.9$ & $1.28 \pm 0.05$ & <$5.2$\\ 
111 & $0.79 \pm 0.04$ & $0.49 \pm 0.03$ & $2.37 \pm 0.05$ & $0.56 \pm 0.06$ & $2.68 \pm 0.07$ & <$3.9$\\ 
179 & <$2.0$ & <$2.0$ & <$0.8$ & <$1.1$ & $1.36 \pm 0.06$ & $6.80 \pm 0.23$\\ 
182 & $1.54 \pm 0.04$ & $0.94 \pm 0.05$ & $1.42 \pm 0.05$ & $0.99 \pm 0.04$ & $2.27 \pm 0.05$ & $12.30 \pm 0.33$\\ 
191 & <$1.0$ & <$1.0$ & <$1.0$ & $1.19 \pm 0.07$ & - & $15.91 \pm 0.39$\\ 
198 & <$3.1$ & <$3.1$ & <$0.8$ & <$1.1$ & <$1.1$ & $3.79 \pm 0.30$\\ 
\hline
\end{tabular}
\end{table*}

\begin{table*}
	\centering
	\caption{
        Equivalent widths of the UV nebular emission lines.
        We provide $3\sigma$ upper limits when possible.
        The \oiii{}] 1666 line in SB 191 is wiped-out by MW absorption.
    }
	\label{tab:uvneb_ew}
\begin{tabular}{lcccccc}
\hline
SBID & \civ{} 1548 & \civ{} 1550 & \heii{} 1640 & \oiii{}] 1661 & \oiii{}] 1666 & \ciii{}] 1907, 1909\\ 
 & (\AA{})& (\AA{})& (\AA{})& (\AA{})& (\AA{})& (\AA{})\\ 
\hline
\hline
2 & $0.25 \pm 0.02$ & $0.22 \pm 0.02$ & $1.70 \pm 0.06$ & $1.94 \pm 0.06$ & $5.05 \pm 0.10$ & $14.86 \pm 1.07$\\ 
36 & $<0.4$ & $<0.4$ & $<0.4$ & $0.26 \pm 0.02$ & $0.84 \pm 0.03$ & $4.98 \pm 0.59$\\ 
80 & $<0.5$ & $<0.4$ & $<0.7$ & $0.55 \pm 0.04$ & $1.68 \pm 0.08$ & $<4.0$\\ 
82 & $0.67 \pm 0.02$ & $0.41 \pm 0.02$ & $0.44 \pm 0.02$ & $0.85 \pm 0.02$ & $1.89 \pm 0.02$ & $12.09 \pm 0.30$\\ 
110 & $0.08 \pm 0.02$ & $0.37 \pm 0.03$ & $<0.4$ & $<0.3$ & $0.48 \pm 0.02$ & $<2.8$\\ 
111 & $0.41 \pm 0.02$ & $0.24 \pm 0.01$ & $1.42 \pm 0.03$ & $0.35 \pm 0.04$ & $1.73 \pm 0.05$ & $<3.3$\\ 
179 & $<0.5$ & $<0.4$ & $<0.6$ & $<0.9$ & $1.17 \pm 0.06$ & $8.71 \pm 0.42$\\ 
182 & $0.96 \pm 0.03$ & $0.55 \pm 0.03$ & $0.94 \pm 0.04$ & $0.70 \pm 0.03$ & $1.73 \pm 0.04$ & $13.35 \pm 0.52$\\ 
191 & $<0.5$ & $<0.4$ & $<0.5$ & $0.64 \pm 0.04$ & - & $11.33 \pm 0.34$\\ 
198 & $<0.7$ & $<0.6$ & $<0.5$ & $<0.8$ & $<0.9$ & $3.38 \pm 0.31$\\ 
\hline
\end{tabular}
\end{table*}

The most metal-rich object in our sample (at $12+\log\mathrm{O/H}=8.48$, or $Z_\odot/2$) is {\bf SB 198}, a star-forming region of stellar mass $10^{4.7} M_\odot$ embedded in a spiral galaxy at 16.5 Mpc.
The UV spectrum of this object is dominated by the \civ{} P-Cygni feature.
For this and all following objects, we quantify the strength of the \civ{} wind feature by integrating the absorption component over 1528--1551 \AA{}, with continuum defined as a linear fit to flux on either side of this interval.
Before integrating this feature, we subtract Gaussian fits to any nebular ($\sigma< 200$ km/s) \civ{} emission and absorption present with continuum defined as a local linear model, and smooth via Fourier filtering to resolution $>200$ km/s to remove residual MW absorption features.
This results in a \civ{} absorption equivalent width of $-5.2 \pm 0.2$ \AA{}.
The \ciii{}] doublet is detected at 3 \AA{} EW, but the \oiii{}] lines are not.
Despite an extremely large sSFR of $150$ \unit{Gyr^{-1}}, the system shows a very low \ott{} of 1.7 and comparatively minimal UV nebular emission.

The next most metal-rich object is {\bf SB 179} at $12+\log\mathrm{O/H}=8.35$.
This is another giant \hii{} region / super star cluster complex embedded in a larger disk system at 25 Mpc, with a correspondingly low mass ($10^{5.2} M_\odot$).
The UV spectrum reveals a prominent \civ{} P-Cygni stellar feature and clear nebular emission from \ciii{}] and \oiii{}].
The \civ{} absorption equivalent width is measured at $-5.6$, comparable to that of SB 198.
However, SB 179 shows prominent broad stellar \heii{} emission as well, as expected given the detection of broad wind emission in the optical SDSS spectra \citep[see Table~\ref{tab:basicprop} and][]{Shirazi2012}.
This wind emission is indicative of a substantial population of Wolf-Rayet stars, which we discuss in more detail in Section~\ref{sec:esispec}.
A two-component fit to \heii{} $\lambda 1640$ identifies purely broad $1600$ km/s FWHM emission with equivalent width $3.1_{-0.7}^{+0.3}$ \AA{}.
Strong \ciii{}] is detected in this system at 8.7 \AA{} equivalent width.

The next most metal-rich object in our sample (at $12+\log\mathrm{O/H}=8.30$) is {\bf SB 191}, a $10^{4.9} M_\odot$ star-forming region in a barred-spiral at 10 Mpc.
SB 191 is marked by very prominent stellar \civ{} and \heii{} features in the FUV, similar to SB 179.
The \civ{} stellar absorption in the spectrum of SB 191 has EW $-6.7\pm 0.1$ \AA{}; and the \heii{} stellar emission fit yields equivalent width $4.3\pm0.4$ \AA{} --- one of the largest values attained in nearby star-forming regions \citep[c.f.][]{Wofford2014,Smith2016}.
Nebular \oiii{}] $\lambda 1661$ is detected at equivalent width 0.7 \AA{}, but the $1666$ component of the doublet is contaminated by an \alii{} MW absorption line.
The \ciii{}] doublet is very prominent, with an equivalent width of $11$ \AA{}.
Neither nebular \heii{} or \civ{} is detected to a 3$\sigma$ upper-limits of $0.5$ \AA{}.
This object has the largest H$\beta$ equivalent width in our sample at 400 \AA{} (sSFR of $130 \unit{Gyr^{-1}}$), and an extremely large \ott{} of 10, further confirming that this object is undergoing a rapid buildup of massive stars.

{\bf SB 80} at $12+\log\mathrm{O/H}=8.24$ is another embedded \hii{} region, this time at 46 Mpc with stellar mass $10^{6.1} M_\odot$.
The UV spectrum shows clear stellar \heii{} ($3.1 \pm 0.3$) and stellar \civ{} absorption ($-4.3 \pm 0.2$).
The only nebular lines detected are the \oiii{}] doublet at 2.2 \AA{} combined EW (with a $3\sigma$ upper limit to \ciii{}] $<4$ \AA{}).
The \ott{} ratio of this system is quite low relative to the rest of the sample at $3.7$, implying a relatively low ionization parameter.

{\bf SB 110} is an isolated compact system (63 Mpc, $10^{6.4} M_\odot$) at $12+\log\mathrm{O/H} = 8.17$.
Stellar \heii{} emission is visible, though significantly less obvious than in SB 191 and 179.
The \civ{} P-Cygni absorption has about the same depth as SB 80, at $-4.4$ \AA{}.
The \ciii{}] doublet is undetected at $\lesssim 3\sigma$ ($<3$ \AA{}), making \oiii{}] 1666 the only confidently-measured nebular line at EW 0.5 \AA{}.

The next system, {\bf SB 182}, presents a substantially different UV spectrum.
This object is at gas-phase metallicity $12+\log \mathrm{O/H} = 8.01$, or $Z/Z_\odot \simeq 1/5$.
It is an isolated compact galaxy with the second-highest mass ($10^{7.3} M_\odot$) of the sample.
At a distance of 191 Mpc, the \hstcos{} aperture radius subtends the largest physical scale probed by our UV spectra: $\sim 1.2 \unit{kpc}$.
The UV stellar features have nearly disappeared (stellar \civ{} absorption EW $> -1.5$ \AA{} and no clear stellar \heii{}), and the \hstcos{} spectra are instead characterized by intense nebular emission.
The \ciii{}] doublet reaches 13 \AA{} and \oiii{}] $\lambda 1666$ nearly 2 \AA{}.
In addition to \ciii{}] and \oiii{}] emission, we now see nebular \heii{} (at EW $0.95$ \AA{}) and nebular \civ{} (doublet combined EW $1.5$ \AA{}).
The presence of nebular \heii{} emission (requiring substantial flux $<228$ \AA{}) and the large ratio \ott{} $= 5.4$ suggest both a large ionization parameter and hard ionizing spectrum.

{\bf SB 36} ($12+\log\mathrm{O/H} = 7.92$) is another isolated system (141 Mpc away) with similar stellar mass ($10^{7.5} M_\odot$) to SB 182.
Its lower H$\beta$ EW (93 \AA{} compared to 150 \AA{} in SB 182, the second-lowest in the sample) presages its weaker UV nebular spectrum.
Both \ciii{}] (EW $\sim 5$ \AA{}) and \oiii{} are detected; but nebular \civ{} and \heii{} are not ($<0.4$ \AA{} EW).

{\bf SB 82} ($12+\log\mathrm{O/H} = 7.91$), a compact galaxy with $10^{6.4} M_\odot$ in stars at 76 Mpc, shows extreme UV emission.
Though there is some hint of a \civ{} P-Cygni feature (equivalent width of absorption $1.5\pm 0.1$ \AA{}), \civ{} nebular emission is clearly present (at combined EW 1.1 \AA{}).
We also detect nebular \heii{} $\lambda 1640$ at EW $0.5$ \AA{} (the nearby MW absorption line does not impact this measurement; we fit it simultaneously with the nebular emission).
The \ciii{}] doublet is measured at 12 \AA{} EW.
The extreme \ott{} ratio of 9 measured for this system is similar to the relatively metal-rich object SB 191, but here both nebular \civ{} and \heii{} are confidently detected alongside the strong \ciii{}] emission.

The first of the two most metal-poor systems, {\bf SB 2} ($12+\log\mathrm{O/H} = 7.81$, $\sim Z_\odot/8$), is comparable to SB 82 and 182 in its UV spectrum.
This BCD component is closer (19 Mpc) and lower in mass ($10^{5.1} M_\odot$) than SB 82 and 182; and presents very strong H$\beta$ (EW of 270 \AA{}) and large \ott{} $= 7.5$.
The \ciii{}] EW achieved is the highest in this sample, at nearly 15 \AA{}.
Both nebular \heii{} and \civ{} are detected, with \heii{} at 1.7 \AA{} and \civ{} at 0.5 \AA{} combined EW.

Finally, {\bf SB 111} is a low-mass ($10^{5.6} M_\odot$) compact galaxy 16 Mpc away and also at gas-phase metallicity $12+\log \mathrm{O/H} = 7.81$.
The UV spectrum shows nebular \heii{} (1.4 \AA{}) and \civ{} (combined 0.65 \AA{}), as well as \oiii{}] (with a strong 1666 component at $2.7$ \AA{}); yet \ciii{}] is undetected ($<4$ \AA{} EW).
The detection of both \civ{} and \heii{} suggest that SB 111 has significant hard ionizing flux beyond $\sim 50$ eV.

Before discussing fits to the full UV spectra (Sec.~\ref{sec:beagle}), we examine the measured nebular emission in \civ{}, \ciii{}], and \oiii{}] in the context of photoionization modeling.
These line provide a nearly direct estimate of the relative abundance of carbon in the ISM \citep[e.g.][]{Garnett1995}, and together with the strong optical lines paint a clearer picture of the state of ionized gas in these objects.
We compare the measured fluxes of nebular \civ{} $\lambda \lambda 1548, 1551$, \oiii{}] $\lambda \lambda 1661, 1666$, and \ciii{}] $\lambda \lambda 1907, 1909$ along with [\oiii{}] $\lambda \lambda 4959, 5007$, [\oiii{}] $\lambda 4363$, [\oii{}] $\lambda 3727$, H$\alpha$, and H$\beta$ to the grid of models described by \citet{Gutkin2016}.
The main adjustable parameters of this model are described in \citet{Gutkin2016} and Section~\ref{sec:beagle} below; in particular, here we allow C/O, the density of the photoionized ISM \nH{}, the upper-mass cutoff of the IMF \mup{}, and the typical ionization parameter of newly-formed \Hii{} regions \Us{} to vary.
We adopt a simplified version of the attenuation model described by \citet{Charlot2000} in this analysis, assuming a slope $\lambda^{-1.3}$ in birth clouds and $\lambda^{-0.7}$ in the ambient ISM.
For simplicity, we adopt a constant star formation model fixed at an age of 100 Myr.
We derive parameter estimates and 68\%-confidence intervals for \Us{}, C/O, and \tauV{} using a $\chi^2$ analysis; the results are displayed in Table~\ref{tab:beagleres}.
These strong emission lines are very well-fit by this model, and prefer gas-phase metallicities and attenuation optical depths in good agreement with those measured above, with median offset from the direct-$T_e$ measurements of $0.2\pm 0.4$ dex.
In addition, this modeling suggests that the gas in these systems is highly-ionized ($-4 < \log \Us{} < -2$) and subsolar in carbon abundance ($-1.4 \lesssim \log(\mathrm{C/O}) \lesssim -0.5$).
Note that while this method of deriving C/O differs from the $T_e$ method applied by \citet{Berg2016}, it is similar to the method of \citet[][found to be consistent with $T_e$]{Perez-Montero2017} and the bulk range of derived abundances is consistent with that found for galaxies of similar metallicity using these methods.
We discuss implications of the derived C/O measurements in more detail in Section~\ref{sec:empiricaltrends} below.

\begin{table}
	\centering
	\caption{
        Parameter estimates derived from comparison of strong optical and nebular \civ{}, \oiii{}], and \ciii{}] UV emission line fluxes to the \citet{Gutkin2016} stellar photoionization model grid.
    }
    \label{tab:beagleres}
\begin{tabular}{lccc}
\hline
SBID & $\log \Us{}$ & $\log(\mathrm{C/O})$ & \tauV{}\\ 
\hline
2 &  $-4.00_{-0.25}^{+0.25}$ & $-1.36_{-0.06}^{+0.06}$ & $0.00_{-0.05}^{+0.05}$ \\
36 &  $-2.75_{-0.17}^{+0.17}$ & $-0.81_{-0.24}^{+0.09}$ & $0.19_{-0.06}^{+0.11}$ \\
80 &  $-2.75_{-0.17}^{+0.17}$ &  $-0.74_{-0.54}^{+0.35}$ & $0.26_{-0.12}^{+0.26}$ \\
82 &  $-2.25_{-0.17}^{+0.17}$ & $-0.52_{-0.04}^{+0.04}$ & $0.53_{-0.18}^{+0.04}$ \\
110 & $-2.75_{-0.17}^{+0.17}$ &  $-0.68_{-0.38}^{+0.17}$  & $0.07_{-0.05}^{+0.08}$ \\
111 & $-2.74_{-0.18}^{+0.18}$ & $-1.02_{-0.07}^{+0.23}$ & $0.12_{-0.08}^{+0.11}$ \\
179 & $-3.25_{-0.17}^{+0.17}$ &  $-1.01_{-0.07}^{+0.23}$ & $0.05_{-0.03}^{+0.03}$ \\
182 & $-2.75_{-0.17}^{+0.17}$ & $-0.75_{-0.07}^{+0.10}$ & $0.22_{-0.08}^{+0.05}$\\
191 & $-2.25_{-0.17}^{+0.17}$ &  $-0.74_{-0.07}^{+0.10}$ & $0.00_{-0.05}^{+0.09}$ \\
198 & $-3.25_{-0.17}^{+0.17}$ &  $-0.80_{-0.25}^{+0.13}$ & $0.13_{-0.12}^{+0.17}$ \\
\hline
\end{tabular}
\end{table}

\subsection{The ESI Spectra}
\label{sec:esispec}

As hot massive stars, Wolf-Rayet (WR) stars may be an important source of the ionizing radiation necessary to power the observed UV nebular emission.
The \hstcos{} spectra reveal strong stellar \heii{} $\lambda 1640$ emission in several systems (Fig.~\ref{fig:majorlines_comp}).
In particular, three objects above $12+\log\mathrm{O/H}>8.2$ have broad $1600$ km/s FWHM \heii{} components with equivalent widths $>2$ \AA{} (SB 80, 191, 179).
To further characterize the WR populations present and ultimately assess their impact on the UV spectra, we also investigate WR emission in the deep optical ESI data.

WR stars are O stars stripped of their outer layers of hydrogen and observationally defined by the presence of broad emission lines from highly-ionized winds \citep[e.g.][]{Crowther2007}.
This includes lines of helium and nitrogen (CNO cycle) as well as carbon and sometimes oxygen (triple-$\alpha$ process) --- nuclear processed material being swept-off the exposed core of a massive star.
The WR class is subdivided primarily into two categories.
WN stars are dominated by lines of helium and nitrogen from hydrogen burning, especially \heii{} $1640, 4696$.
In contrast, WC stars primarily show carbon emission, particularly \civ{} $\lambda 5808$.
The relative numbers of these subtypes can inform understanding of how these stars are produced --- likely some combination of wind-driven mass loss and binary mass transfer at low metallicity \citep[e.g.][]{Conti1976,Maeder1994,Schaerer1998,Eldridge2009}.
Very massive OIf supergiants can also contribute to broad \heii{} emission during their hydrogen-burning main sequence lifetime, as they drive similar dense ionized winds during this time \citep[e.g.][]{Crowther2016,Smith2016}.

The impact of WR and WR-like stars on high-ionization emission remains unclear.
The Lyman continuum flux from WR stars is generally comparable to that of O stars, but the contribution to $\mathrm{He}^+$-ionizing flux may be substantial for the hottest WR stars \citep{Crowther2007}.
Stellar population synthesis models often predict a boost in photons beyond $<228$ \AA{} in the short timespan when WR stars are active \citep[e.g.][]{Schaerer1998,Leitherer1999,Vidal-Garcia2017}, suggesting they may play a large role in powering high-ionization nebular emission.
However, previous observational studies have found nebular \heii{} emission without accompanying WR signatures \citep[e.g.][]{Garnett1991,Brinchmann2008,Shirazi2012}.
Determining the impact these stars have on UV nebular emission is critical for interpreting UV lines at high-redshift.

\begin{figure}
	\includegraphics[width=\columnwidth]{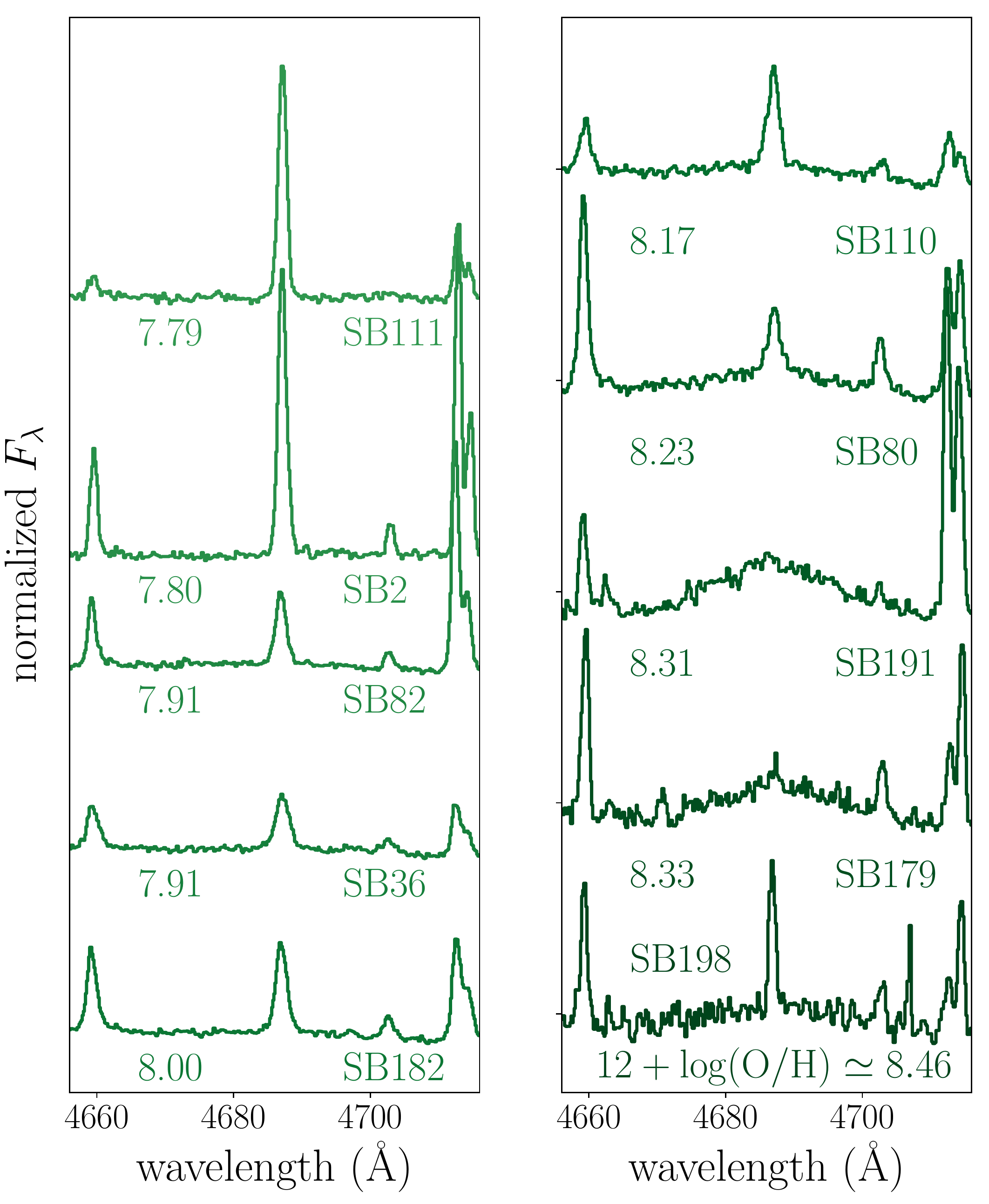}
    \caption{
        Cutouts of Keck/ESI optical spectra centred on the \heii{} $\lambda 4686$ line for all ten objects, ordered by gas-phase direct-$T_e$ metallicity.
        The more metal-poor galaxies in our sample tend to show stronger nebular emission and weaker broad stellar components, which suggests evolution in both the stellar winds and ionizing spectral slope with metallicity.
    }
    \label{fig:esi_heii}
\end{figure}

The resolution of the ESI data allow us to confidently identify broad FWHM $>150$ km/s emission in the \heii{} $\lambda 4686$ \AA{} line.
We plot the ESI spectra centred on this line in Fig.~\ref{fig:esi_heii}.
The results of a simultaneous fit to the broad and narrow components to the lines are displayed in Table~\ref{tab:esi_heii}; since the region around this transition is crowded with other nebular lines, we use median-filtering (with kernel size 10 \AA{}) to better represent the continuum beyond 25 \AA{} from line centre.
As seen for \heii{} $\lambda 1640$ with \hstcos{} (Sec.~\ref{sec:uvspectra}), the ESI data reveal a range of \heii{} $\lambda 4686$ nebular strengths; from undetected to \heii{}$\lambda 4686$/H$\beta$ $\simeq 4\e{-3}$.
Two systems (SB 179 and 191) reveal purely stellar \heii{} emission, mistaken in low-resolution SDSS spectra for nebular.
Broad components are detected only in the five most metal-rich systems, above $12+\log\mathrm{O/H} > 8.1$ ($Z/Z_\odot > 1/4$).
These are uniformly measured at 1200 km/s FWHM, with fluxes relative to H$\beta$ $\sim 10^{-2}$ (Table~\ref{tab:esi_heii}).
We also investigate the dominant line emitted by WC stars, \civ{} $\lambda 5808$ \AA{}.
This line is detected at $\gtrsim 3\sigma$ in four galaxies (80, 110, 179, and 191) with $5-15\%$ the stellar \heii{} $\lambda 4686$ flux.

We can also constrain the number of WR stars relative to O stars present in these systems.
This measurement comes with a significant caveat: since WR winds are metallicity-dependent (discussed further in Sec.~\ref{sec:ionsource}), we must use a scaling relation to estimate the wind line luminosity of WN and WC stars at the gas-phase metallicity of each galaxy.
The resulting number estimates are subject directly to systematic uncertainties in these wind scalings, which are presently calibrated with individual SMC and LMC stars \citep{Crowther2006}.
For simplicity, we assume that only WN stars are present --- the flux of the \civ{} $\lambda 5808$ stellar feature (using the wind luminosity scalings below) implies very small WCE/WNL ratios $< 10\%$.
This is consistent with expectations from population synthesis models \citep[e.g.][]{Eldridge2006} that at these metallicities stripping is insufficient to reach He-burning products.\footnote{However, it is important to note that SDSS samples will likely be biased against purely WC-dominated systems as the dichroic split occurs at $5900-6100$ \AA{}, near \civ{} $\lambda 5808$ for low redshift galaxies.}.
In this simple calculation we do not correct for OIf supergiants, as we are primarily interested in the relative number of massive stars driving dense ionized winds.
Using the simple linear metallicity relation derived by \citet{Lopez-Sanchez2010} from the models of \citet{Crowther2006} to predict the line luminosity of a single WNL star, we estimate the ratio of WNL stars to the equivalent number of O7V stars derived from H$\beta$.
For the five galaxies in which the stellar 4686 line is detected in our ESI spectra, we obtain a ratio of $\mathrm{WR/O} \simeq 0.04-0.1$ (see Table~\ref{tab:esi_heii}).
The remaining systems, at metallicities below $12+\log\mathrm{O/H}<8.1$, appear to harbor smaller proportions of WR stars, $\lesssim 0.05$.
These results are broadly consistent with that derived for larger samples of WR galaxies and with the predictions of binary population synthesis models \citep[e.g.][]{Brinchmann2008,Lopez-Sanchez2010}.

\begin{table}
	\centering
    \caption{\heii{} stellar and nebular fluxes measured from simultaneous two-component fits to the ESI data.
    The WR to O star ratio is estimated as described in Sec.~\ref{sec:esispec}.
    Upper limits represent the 84th-percentile confidence interval.}
	\label{tab:esi_heii}
\begin{tabular}{lccc}
\hline
SBID & \heii{} $\lambda 4686$ / H$\beta$ & \heii{} $\lambda 4686$ / H$\beta$ & N(WR)/N(O)\\ 
 & nebular (x$10^3$) & stellar (x$10^3$) & \\ 
\hline
2 & $13.0 \pm 0.6$ & $< 3.4$ & $< 0.02$\\ 
36 & $14.6 \pm 0.5$ & $< 13.3$ & $< 0.06$\\ 
80 & $3.1 \pm 0.3$ & $12.0 \pm 0.9$ & $0.05$\\ 
82 & $8.5 \pm 0.2$ & $< 6.4$ & $< 0.03$\\ 
110 & $15.3 \pm 0.5$ & $22.6 \pm 1.6$ & $0.09$\\ 
111 & $39.4 \pm 1.3$ & $< 5.9$ & $< 0.03$\\ 
179 & $< 1.4$ & $14.8 \pm 1.1$ & $0.05$\\ 
182 & $12.4 \pm 0.6$ & $< 4.6$ & $< 0.02$\\ 
191 & $< 1.4$ & $13.5 \pm 1.2$ & $0.05$\\ 
198 & $4.3 \pm 0.3$ & $12.0 \pm 1.2$ & $0.04$\\ 
\hline
\end{tabular}
\end{table}

\section{Full UV Spectral Synthesis}
\label{sec:beagle}

Stellar population synthesis coupled with photoionization modeling has become a commonplace tool in the analysis of galaxy spectra.
However, comprehensive tests of this mode of analysis with typical datasets is difficult.
In particular, while much work has been done to quantify the performance of different population synthesis models on photometric data \citep[e.g.][]{Wofford2014} and on integrated-light spectra probing older stellar populations \citep[e.g.][]{Bruzual2003,Schiavon2004,Conroy2009}, tests in the regime of young stellar populations dominated by massive stars and nebular emission are comparatively few.
Recently, joint analysis of optical and UV stellar and nebular features at $z\sim 2$ has become possible \citep{Steidel2016}, revealing significant discrepancies between inferred stellar and nebular abundances; but at these distances, this analysis can only be performed on stacked spectra or a small number of lensed systems.
The data presented here provide a rare opportunity to test these tools and the underlying stellar models by fitting jointly both nebular emission and the wind features of the massive stars present in individual galaxies, covering a range of metallicities and UV spectral properties.

Here we present our initial results obtained by fitting the COS spectra of the 10 galaxies in our sample, while more extensive modeling results of these data will be presented in a companion paper (Chevallard et al.\ 2017, in-prep).
We fit, pixel by pixel, the full \hstcos{} UV spectrum of each object, with the goal of matching both strong stellar and nebular features simultaneously.
To achieve this we use the Bayesian spectral interpretation tool BEAGLE \citep{Chevallard2016}, which incorporates in a flexible and consistent way the production of radiation in galaxies and its transfer through the interstellar and intergalactic media.
Before running the fitting, we remove ISM and MW absorption features and smooth the spectra to a uniform resolution of 2.3 \AA{} FWHM, appropriate for comparison to the stellar templates.
We rely on the full set of models of \citet{Gutkin2016} which combine the latest version of the \citet{Bruzual2003} stellar population synthesis model with the standard photoionization code \cloudy{} \citep{Ferland2013} to compute the emission from stars and the interstellar gas.
In particular, the nebular emission is computed following the prescription of \citealt{Charlot2001}, where the main adjustable parameters of the photoionized gas are the interstellar metallicity, \Zism, the dust-to-metal mass ratio, \xid\ (which characterizes the depletion of metals on to dust grains), and the typical ionization parameter of newly ionized \Hii\ regions, \Us\ (which characterizes the ratio of ionizing-photon to gas densities at the edge of the Stroemgren sphere).
We consider here models with hydrogen density $\nH=100\,\mathrm{cm}^{-3}$ and C/O abundance ratios ranging from 0.1 to 1.0 times the solar ratio [$(\mathrm{C/O})_\odot\approx0.44$].
Finally, we describe attenuation by dust using the 2-component model of \citet{Charlot2000}. 

Following Section 3.3, we adopt a constant star formation history of variable age, where we let the age freely vary in the range $6.0\leq\log(\mathrm{age}/\mathrm{yr})\leq9$, finding negligible differences in the results when adopting a more complex SFH.
We adopt a standard \citet{Chabrier2003} initial mass function and we test two different values for the upper mass cutoff, 100 and 300 $M_\odot$.
We further adopt the same metallicity for stars and star-forming gas ($Z=\Zism$) and assume that all stars in a galaxy have the same metallicity, in the range $-2.2\leq\log(Z/Z_\odot)\leq0.25$.
We let freely vary the dust-to-metal mass ratio and the ionization parameter in the ranges $0.1\leq\xid \leq0.5$ and $-4\leq\log\Us\leq-1$ respectively.
We consider $V$-band dust attenuation optical depths in the range $0\leq\tauV\leq5$ and let the fraction of this arising from dust in the diffuse ISM rather than in giant molecular clouds freely vary in the range $0\leq\mu\leq1$. 

\begin{figure*}
    \includegraphics[width=0.8\textwidth]{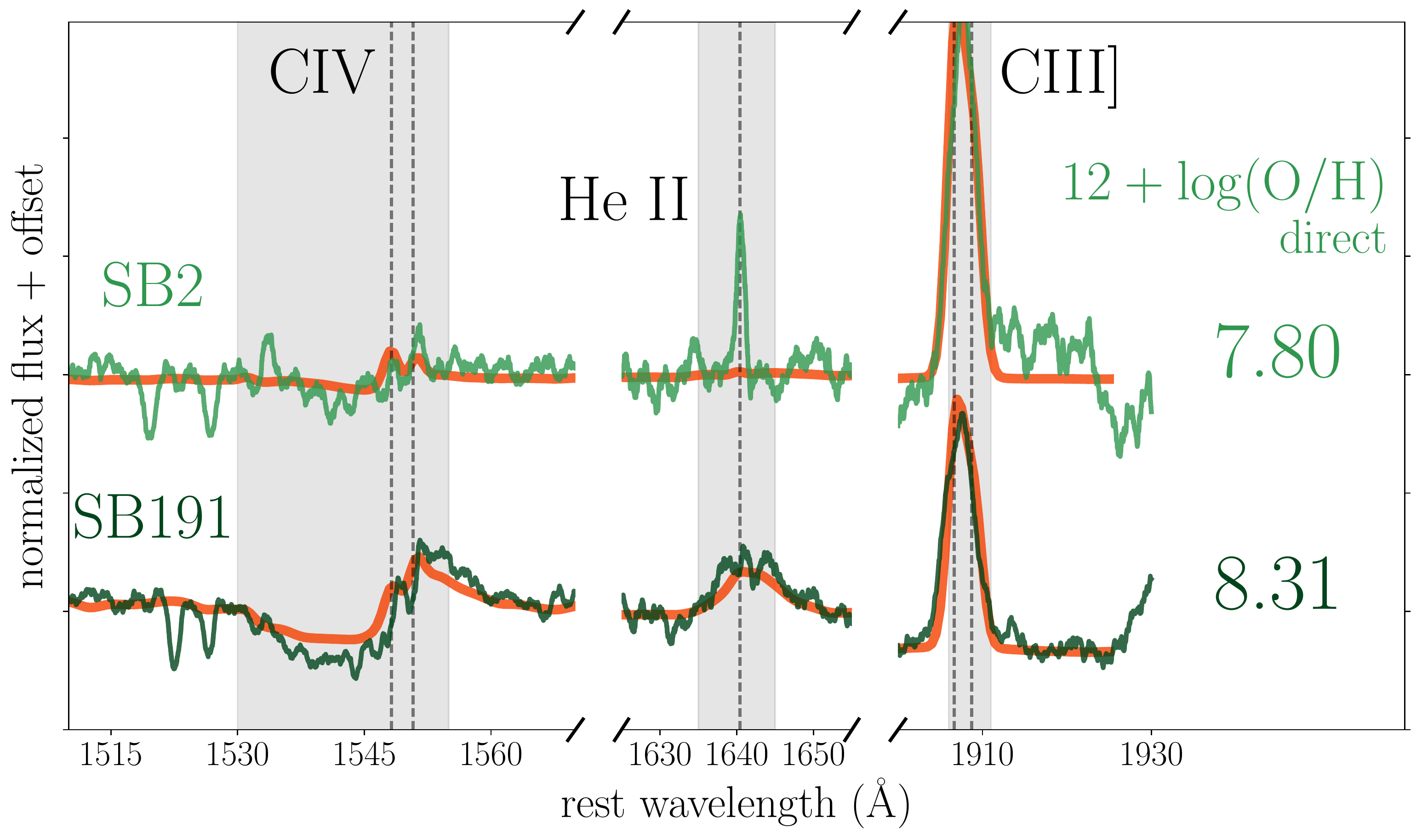}
    \caption{
        \beagle{} fits to the \hstcos{} UV spectra of two representative galaxies from our sample.
        The observed UV spectra are displayed in shades of green, and the maximum a posteriori estimation model is displayed as a thicker orange line (note that ISM/MW absorption lines were removed and the data was smoothed before fitting).
        The models are able to simultaneously reproduce well both stellar and nebular features in both cases, with the exception of nebular \heii{} $\lambda 1640$.
        The extremely strong stellar \heii{} in SB 191 (and SB 179) is matched very well.
        However, the prominent nebular \heii{} in SB 2 (and SB 111, 182) is not reproduced by the fiducial models without invoking extremely low metallicities inconsistent with the optical spectra.
    }
    \label{fig:beagle_uv}
\end{figure*}

In Fig.~\ref{fig:beagle_uv} we show an example of a fit to a relatively metal-rich galaxy (SB 191; direct-$T_e$ $12+\log\mathrm{O/H} = 8.3$) and one of the most metal-poor systems (SB 2; direct-$T_e$ $12+\log\mathrm{O/H} = 7.8$).
The stellar \civ{} and \heii{} and nebular \ciii{}], \oiii{}], and \civ{} are reproduced very well in all cases while the prominent nebular \heii{} emission in SB 2, 182, and 111 is difficult to achieve.
Increasing the maximum allowed stellar mass $M_{up}$ to 300 $M_\odot$ increases the strength of nebular \heii{}, and does allow the emission in SB 111 to be fit; but only for extremely low metallicity $12+\log\mathrm{O/H} \simeq 6.5$ inconsistent with the optical data.
These fits will be explored in quantitative detail in Chevallard et al. (2017, in-prep).
We discuss the discrepant nebular \heii{} empirically below.

\section{Discussion}
\label{sec:discussion}

Spectroscopy of local star-forming galaxies in the UV provides an important empirical baseline for interpreting observations at high redshift and informs our understanding of low-metallicity stellar populations.
Deep observations at both $z\sim 2$ and now $z\sim 6-7$ have revealed emission in \ciii{}] and \civ{} far stronger than in typical star-forming systems at low redshift.
The gas properties and stellar populations necessary and sufficient to power such high-ionization emission remain unclear, especially with limited existing data below $Z_\odot/3$.
In Section~\ref{sec:empiricaltrends} we discuss the empirical properties of the strong UV line emitters and the potential utility of UV nebular emission as diagnostics of physical conditions in the reionization era.
Then we discuss the shape of the ionizing spectrum inferred from the nebular emission in Section~\ref{sec:ionspec}, before addressing directly the possible sources of the necessary flux in Section~\ref{sec:ionsource}.

\subsection{Characterizing Strong UV Line Emitters}
\label{sec:empiricaltrends}

Detections of strong \ciii{}], \civ{}, and other high-ionization lines in the reionization era represent a new opportunity.
Since they originate from species with ionization potentials in the range of extreme-UV photons, they are more sensitive to the most extreme radiation emitted by young metal-poor stellar populations.
This sensitivity also makes these lines challenging to interpret --- without an observational baseline, the results of photoionization modeling will be subject to substantial systematic uncertainties.
To draw robust inferences about reionization-era galaxies from rest-UV nebular emission, we must develop an empirical understanding of the stellar populations and gas conditions which support these lines.
This in turn requires a local sample of extreme UV line emitters to which photoionization modeling and high-redshift observations can be compared.

Nearby star-forming galaxies with high-ionization UV emission comparable to that seen at $z>6$ have proved mostly elusive.
Archival studies focused on selecting UV-bright star-forming galaxies have not yet identified a significant population analogous to that observed at $z\sim 7$.
Individual galaxies at $z\sim 6-7$ have revealed \ciii{}] at rest-frame equivalent widths in-excess of $20$ \AA{} \citep{Stark2015,Stark2017} and \civ{} at $\sim 20-40$ \AA{} \citep{Stark2015a,Mainali2017}.
Only seven (one) star-forming galaxies with secure \ciii{}] equivalent widths $> 10$ \AA{} ($>15$ \AA{}, respectively) have been found by FOS, GHRS, IUE, and COS; and only three with detections of nebular \civ{} \footnote{As discussed in Appendix~\ref{sec:archive_app}, this ignores two star-forming regions in the LMC and SMC for which the FOS aperture subtends a very small physical distance; and Tol 1214-277, where the continuum was undetected by FOS.}.

The ten star-forming galaxies presented here include four new extreme $>10$ \AA{} \ciii{}] emitters, including the second highest \ciii{}] equivalent width detected locally at $15$ \AA{} (SB 2).
In addition, our data constrain both stellar and nebular \civ{} and \heii{}.
Four of our UV spectra reveal clear nebular \civ{} and \heii{} in emission, which \citep[together with three strong emitters from the sample presented by][]{Berg2016} allows us to conduct the first thorough analysis of these extreme UV lines in nearby star-forming galaxies.
The detection rate of UV nebular emission in this sample selected to show \heii{} emission in the optical is extremely high, with \oiii{}] detected in all but one; yet even in this sample we see a wide range of UV nebular properties.

\begin{figure}
	\includegraphics[width=\columnwidth]{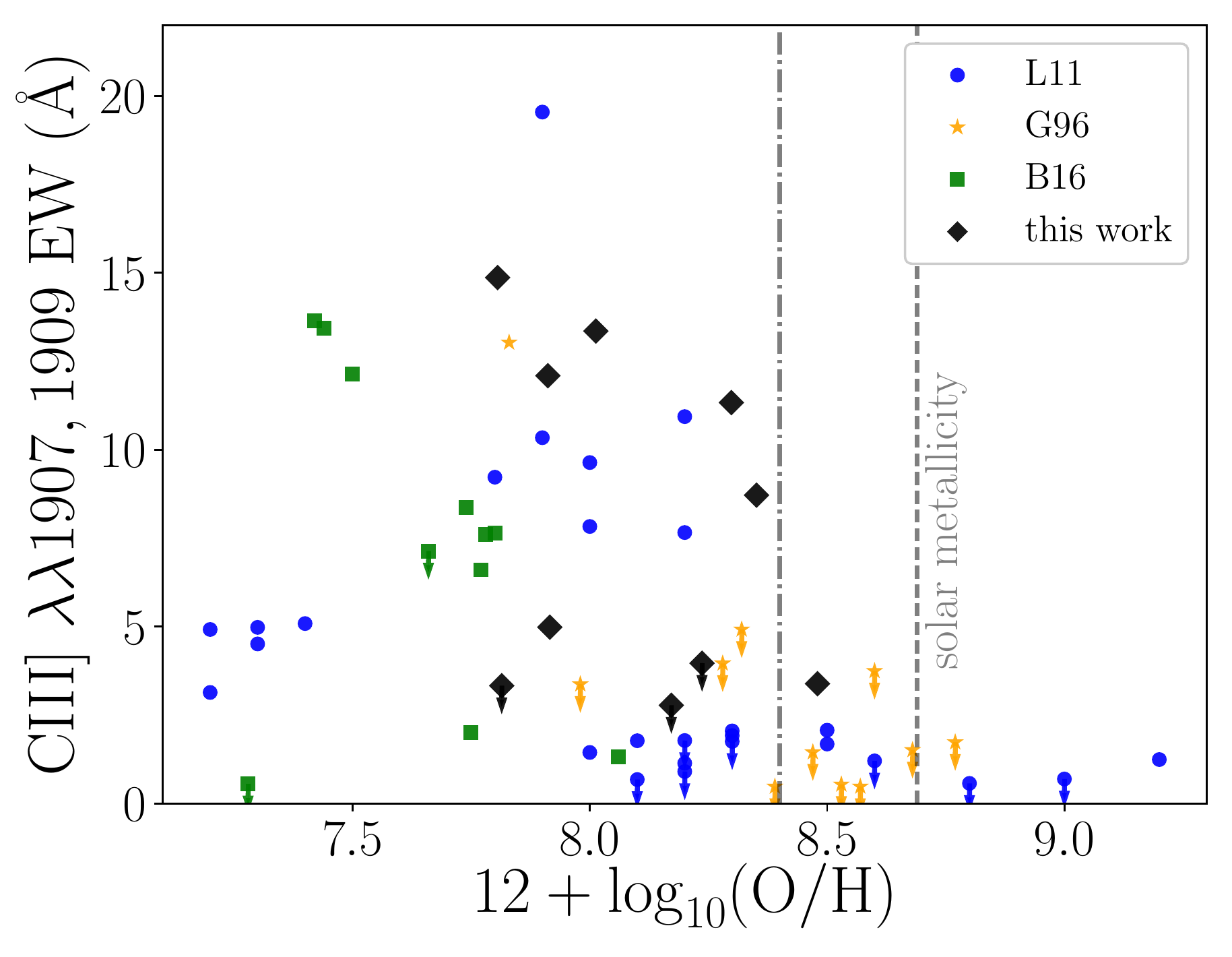}
    \caption{
        Equivalent width of the combined \ciii{}] $\lambda \lambda 1907, 1909$ semi-forbidden doublet as a function of gas-phase metallicity.
        In addition to the \hstcos{} data presented in this paper, we plot archival data from nearby galaxies catalogued by \citet[][L11]{Leitherer2011}, \citet[][G96]{Giavalisco1996}, and \citet[][B16]{Berg2016} --- see Appendix~\ref{sec:archive_app} for more details.
        These data suggest an empirical metallicity threshold for \ciii{}]: above $12+\log\mathrm{O/H}\sim 8.4$ (marked by the dash-dotted line), \ciii{}] equivalent widths do not exceed $\sim 5$ \AA{}.
        Below this threshold, extremely high equivalent widths $\gtrsim 15$ \AA{} are achieved (comparable to that observed in the reionization era: \citealt{Stark2015, Stark2017}), though not uniformly.
    }
    \label{fig:ciii_metal}
\end{figure}

We expect metallicity to play a role in modulating UV nebular emission.
Metal-poor gas cools less efficiently, leading to higher electron temperatures and stronger collisionally-excited emission; and lower metallicity stars evolve to hotter effective temperatures with weaker winds, yielding harder emergent stellar spectra \citep[e.g.][]{Schaller1992,Schaerer2003}.
Previous authors have found unclear trends between \ciii{}] equivalent width and metallicity, focusing on the large scatter at fixed metallicity \citep[e.g.][]{Bayliss2014,Rigby2015}.
In Figure~\ref{fig:ciii_metal}, we plot the \ciii{}] doublet equivalent widths versus gas-phase metallicity for our sample as well as for archival local data gathered from IUE, GHRS, FOS, and COS (see Appendix~\ref{sec:archive_app}).
There is a striking transition with metallicity evident in these local galaxies.
Above $12+\log\mathrm{O/H}\gtrsim 8.4$, \ciii{}] equivalent widths do not exceed $\sim 5$ \AA{}, and only reach a median of $0.9$ \AA{}.
Below this metallicity ($Z/Z_\odot \lesssim 1/2$), the median increases by a factor of five (to $5.0$ \AA{}), and individual systems display extremely strong emission $\sim 15-20$ \AA{}, approaching reionization-era values.
In our sample, we see a hint of evolution below half-solar metallicity as well, with stronger median equivalent width at and below metallicities $12+\log\mathrm{O/H}\lesssim 8.0$ ($\sim 12$ \AA{}) than above ($\sim 3.4$ \AA{}).
We observe a similar trend in \oiii{}] $\lambda \lambda 1661,1666$ in our sample, with galaxies below $Z/Z_\odot \lesssim 1/5$ reaching median doublet equivalent width 2.4 \AA{}, three times higher than at higher metallicities (0.8 \AA{}).
Evidently, metallicity has a substantial effect on UV nebular emission; suppressing strong \ciii{}] and \oiii{}] entirely above a half-solar, and enhancing it at yet lower metallicities (at least down to $Z/Z_\odot \sim 1/8$).

Our data reveal a similar but lower threshold for nebular \civ{} and \heii{} production.
As is clear in Fig.~\ref{fig:majorlines_comp}, we detect nebular \civ{} and \heii{} in four of the five most metal-poor objects, but none above $12+\log\mathrm{O/H}>8.0$ ($Z/Z_\odot > 1/5$).
The three detections of likely nebular \civ{} and \heii{} from \citet{Berg2016} all occur in systems below a fifth-solar, and the highest equivalent width in \civ{} (11 \AA{} combined) is attained at $12+\log\mathrm{O/H} \simeq 7.44$.
This implies that the ionizing flux (and in the case of \civ{}, high electron temperatures) necessary to power these lines is not present above a fifth solar --- see Sec.~\ref{sec:ionspec} for further discussion.
We are currently limited by small number statistics in this extremely metal-poor regime.
However, the present detections suggest that very strong $>10$ \AA{} \civ{} emission may require even lower metallicities, $Z/Z_\odot \lesssim 1/10$.

\begin{figure}
	\includegraphics[width=\columnwidth]{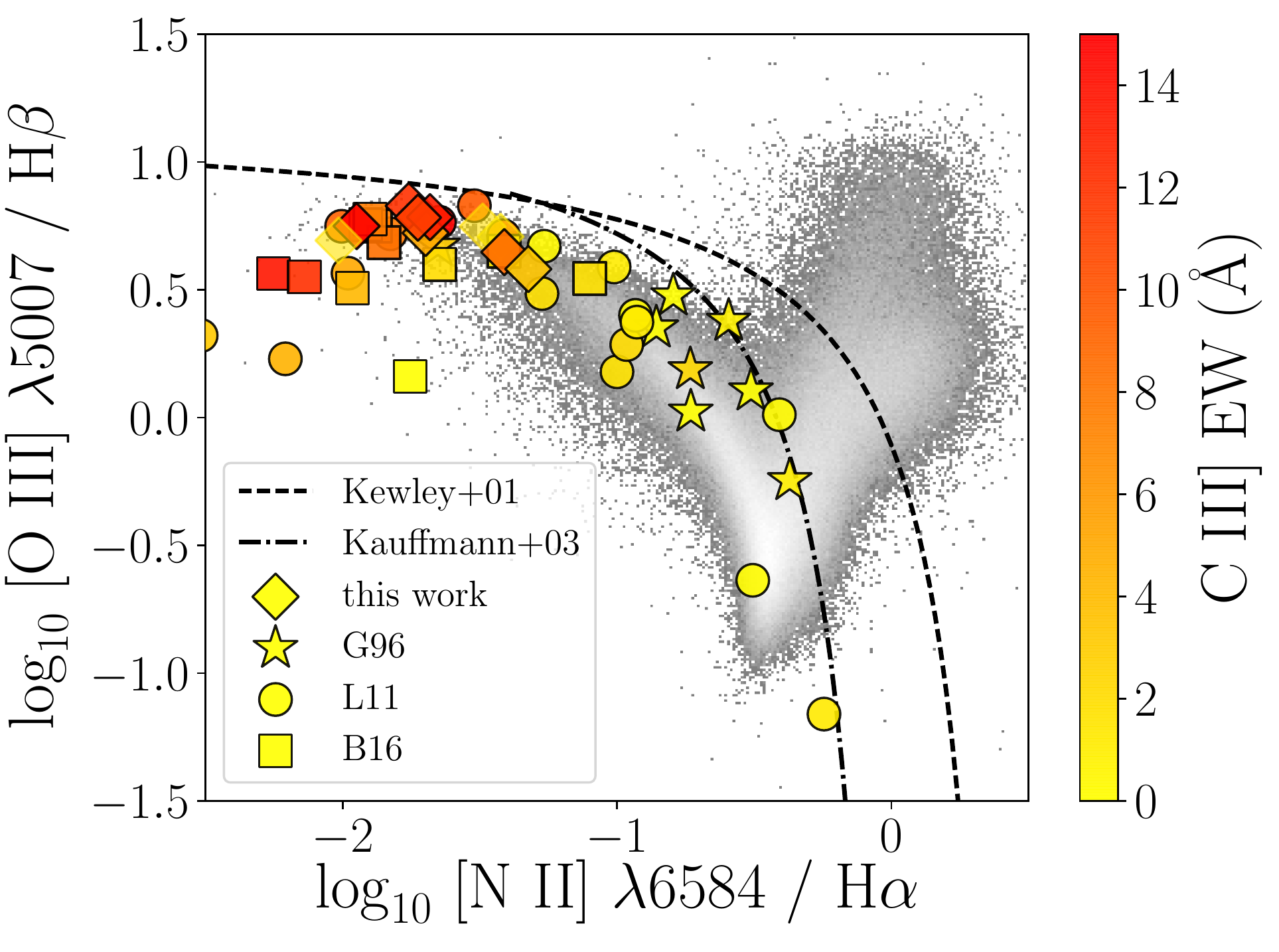}
    \caption{
        The BPT diagram with colour indicating the measured \ciii{}] EW for our sample (circles) and local predominantly star-forming galaxies from the literature observed with IUE \citep[][G96]{Giavalisco1996}, FOS \citep[][L11]{Leitherer2011}, and COS \citep[][B16]{Berg2016}.
        The greyscale log-histogram in the background represents all local SDSS galaxies with $\mathrm{S/N}>3$ in the relevant lines.
        It is apparent that on-average, the highest \ciii{}] EWs occur in systems in the extreme upper left, where [\oiii{}]/H$\beta$ is maximized and [\nii{}]/H$\alpha$ is minimized, corresponding to highly-ionized, lower-metallicity systems.
    }
    \label{fig:bpt_uv}
\end{figure}

\begin{figure}
	\includegraphics[width=\columnwidth]{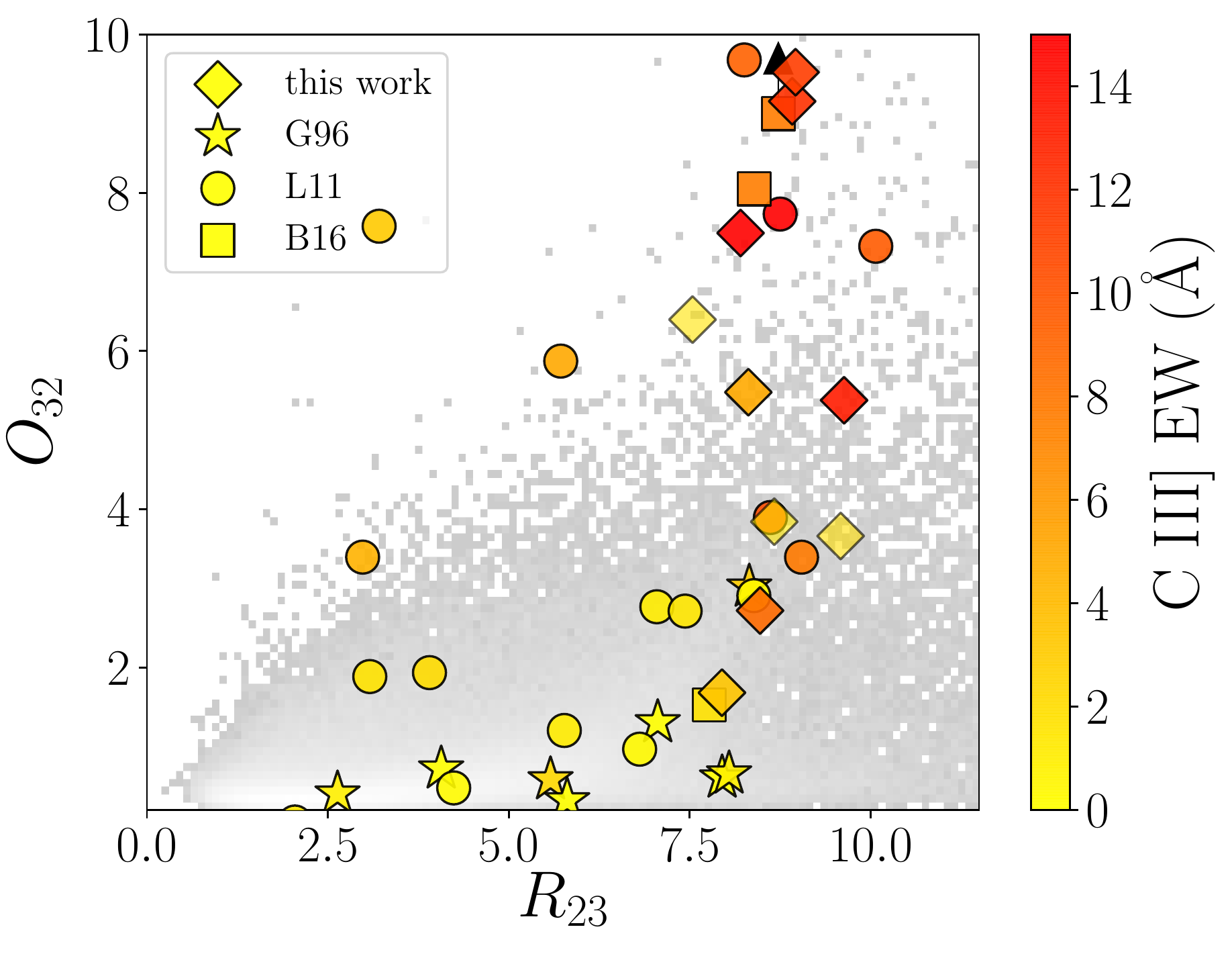}
    \caption{
        The \ott{}-\rtt{} diagram, with symbols as-in Fig.~\ref{fig:bpt_uv}.
        Like the BPT diagram, these line ratios performs surprisingly well at separating the most extreme \ciii{}] EWs; these occur in the upper right where both the ionization parameter (\ott{}) and collisional excitation (due to inefficient cooling) are maximized.
    }
    \label{fig:o32_uv}
\end{figure}

The strength of \ciii{}] emission is clearly not a function of only metallicity.
The BPT diagram probes both the gas composition and the incident ionizing spectrum, which allows for a more nuanced separation of objects than metallicity alone.
In Fig.~\ref{fig:bpt_uv}, we plot our objects alongside the archival sample (Appendix~\ref{sec:archive_app}) on the BPT diagram, coloured according to observed restframe \ciii{}] equivalent width.
The \rtt{}-\ott{} diagram separates galaxies in a similar way, but with greater sensitivity to the ionization state of the gas uncoupled from abundances.
The upper right of this diagram corresponds to high ionization parameter (high \ott{}) and moderately metal-poor gas at high temperature (high \rtt{}, peaking around $12+\log\mathrm{O/H} \sim 8.0-8.5$; e.g.\ \citealt{Kobulnicky1999,Dopita2000}).
We plot these line ratios for our data and the comparison sample in Fig.~\ref{fig:o32_uv}.
In both diagrams, we see that strong \ciii{}]-emitters systematically populate the regions corresponding to highly-ionized metal-poor gas.
The objects with \ciii{}] equivalent widths above $>10$ \AA{} have median $\log\mathrm{[\nii{}] 6584}/\mathrm{H\alpha} = -1.8$ and $\log\mathrm{[\oiii{}] 5007}/\mathrm{H\beta} > 0.55$ (median $0.75$), placing them securely in the extreme upper-left tail of the star-forming sequence in the BPT diagram.
Similarly, these systems present median $\mathrm{\rtt{}} = 8.8$ and $\mathrm{\ott{}} > 3.8$ (median $7.8$), indicative of a substantially more ionized medium than typical in nearby galaxies (c.f.\ the SDSS greyscale histogram in Figures~\ref{fig:bpt_uv} and \ref{fig:o32_uv}).

\begin{figure}
	\includegraphics[width=\columnwidth]{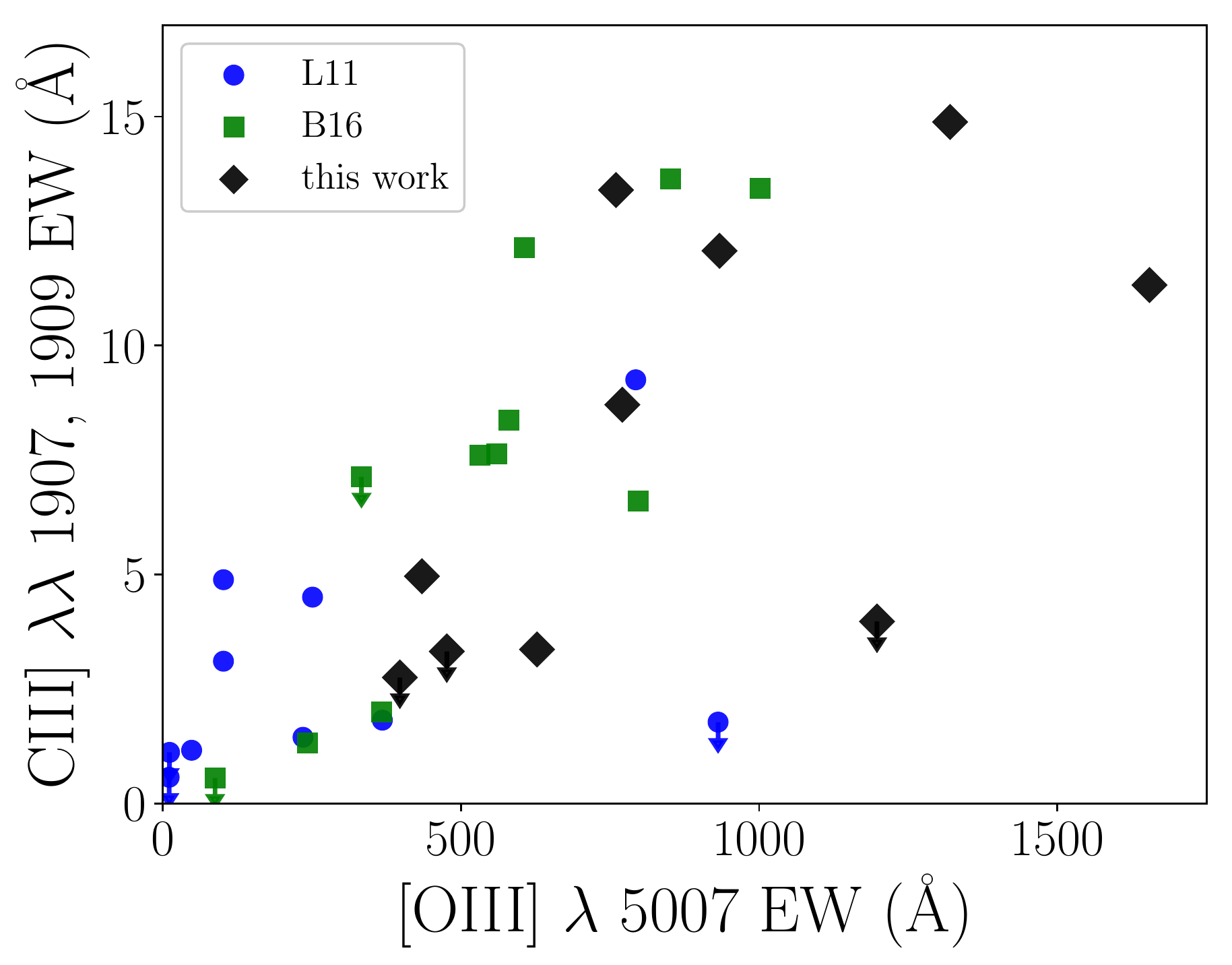}
    \caption{
        Equivalent width of the combined \ciii{}] $\lambda \lambda 1907, 1909$ doublet plotted against [\oiii{}] $\lambda 5007$ equivalent width measured in matched SDSS spectra.
        In addition to the \hstcos{} data presented in this paper, we plot archival data from nearby galaxies catalogued by \citet[][L11]{Leitherer2011} and \citet[][B16]{Berg2016}.
        This plot reveals a correlation between the equivalent widths of these lines, and implies that selecting galaxies based on optical [\oiii{}] equivalent width may be an efficient way to find strong UV line emitters.
    }
    \label{fig:ciii_oiii5007}
\end{figure}

The equivalent width of [\oiii{}] emission in the optical also appears to correlate strongly with \ciii{}] equivalent width in these local galaxies.
In Figure~\ref{fig:ciii_oiii5007} we plot \ciii{}] doublet equivalent widths against the equivalent width of [\oiii{}] $\lambda 5007$.
The [\oiii{}] $\lambda 5007$ EWs are measured from SDSS spectra only to enforce aperture consistency --- thus this plot only includes systems for which an SDSS spectrum with pointing matched to within 2\arcsec{} of the COS/FOS pointing was available.
In these local metal-poor galaxies, \ciii{}] emission at $>5$ \AA{} only occurs in galaxies with extreme optical line emission, with [\oiii{}] $\lambda 5007$ equivalent width $\gtrsim 500$ \AA{}.
In addition, this plot suggests a rising trend beyond even this cutoff, with galaxies above $\gtrsim 750$ \AA{} in 5007 displaying typical \ciii{}] equivalent widths $\sim 10-15$ \AA{}.
Equivalent widths of this magnitude are extremely rare in the local universe out to $z\sim 1$, found only in low-mass systems with very high sSFR \citep[e.g.][]{Cardamone2009,Amorin2015}.

These empirical correlations paint a clearer picture of the factors which govern UV metal line production.
Exciting the UV lines requires the presence of massive hot stars capable of providing the necessary ionizing flux, and is thus associated with very recent star formation.
Above half-solar metallicity, some combination of efficient gas cooling and inefficient production of hard $\gtrsim 25$ eV photons prevents \ciii{}] from reaching equivalent widths $> 5$ \AA{}.
As a result, strong ($>10$ \AA{}) \ciii{}] emitters have optical line signatures indicative of metal-poor, highly-ionized gas.
This is supported by detailed photoionization modeling focusing on \ciii{}], which show a similar metallicity threshold \citep{Jaskot2016}.
The high electron temperatures and ionizing flux $\gtrsim 50$ eV necessary to power nebular \civ{} and \heii{} appears to require even more metal-poor stellar populations and gas, below a fifth-solar metallicity.

Understanding in detail how various physical parameters influence UV line equivalent widths is critical for predicting and understanding the spectra of distant populations.
Two additional factors complicate interpretation of \ciii{}] in particular.
First, the strength of \ciii{}] relative to \oiii{}] is related directly to the C/O ratio since both have similar ionization and excitation potentials.
Due to some form of (pseudo)secondary production and release of carbon, C/O is found to correlate with O/H such that low-metallicity systems have systematically low C/O \citep[e.g.][]{Garnett1995,Berg2016}.
As expected for such metal-poor galaxies, photoionization modeling of our galaxies indicates systematically sub-solar C/O, ranging from $-0.5 \geq \log\mathrm{C/O} \geq -1.4$ (Sec.~\ref{sec:uvspectra}).
In addition, the strength of \ciii{}] is further modulated by the ionization state of carbon: if sufficient flux is available above 47.9 eV, carbon may be triply-ionized in quantities sufficient to weaken \ciii{}] emission.

The scatter in \ciii{}] equivalent width at fixed metallicity observed in our sample is well explained in this context, as we illustrate for the most prominent outliers.
First consider SB 111, the lowest metallicity system (see Fig.~\ref{fig:majorlines_comp} and Table~\ref{tab:uvneb_ew}).
The \ciii{}] doublet is undetected, but nebular \oiii{}] is present as well as \civ{}, consistent with both a low C/O ratio ($\log\mathrm{C/O} = -1$; see Table~\ref{tab:beagleres}) and highly-ionized carbon.
Second, SB 36 presents a drastically different UV spectrum from SB 82 despite being at the same gas-phase metallicity.
In SB 82 we see prominent \ciii{}] (12 \AA{}) as well as nebular \heii{} and \civ{}; whereas \ciii{}] in SB 36 is relatively weak ($\sim 5$ \AA{}) and the only other UV line detected is \oiii{}].
Though both have very high sSFR, SB 82 has a higher [\oiii{}] $\lambda \lambda 4959,5007$ equivalent width (1300 \AA{} compared to 600 \AA{}) and stronger H$\beta$ (180 \AA{} versus 90 \AA{}).
This suggests that the very recent star formation history in SB 82 has produced a somewhat more dominant population of massive stars in this galaxy, reflected in the substantially different UV nebular spectra.

In contrast with the trends described above, two galaxies in our sample with relatively metal-rich gas (SB 191 and 179, $Z/Z_\odot \sim 1/3$) present large \ciii{}] equivalent widths ($>5$ \AA{}).
In addition to their high [\oiii{}] equivalent widths (SB 191 has the highest in our sample, with [\oiii{}] $\lambda \lambda 4959,5007$ together reaching 2400 \AA{}), these two systems have prominent broad stellar wind features in the UV (\heii{} emission and \civ{} P-Cygni; see Fig.~\ref{fig:majorlines_comp}).
As we will discuss further in Sec.~\ref{sec:ionsource}, this suggests that Wolf-Rayet stars at moderately low metallicity ($\sim Z_\odot/3$) are capable of powering strong $\gtrsim 10$ \AA{} \ciii{}] emission.
These systems present the highest \ciii{}] equivalent widths at their metallicities, very close to the empirical cutoff discussed above (see Fig.~\ref{fig:ciii_metal}).
Such extreme stellar \heii{} $\lambda 1640$ equivalent widths ($>3$ \AA{}) are rare in the local universe \citep{Wofford2014}, and the range of \ciii{}] equivalent widths systems in this state can power is presently unclear.

The empirical correlations we have discussed above have significant implications for reionization-era galaxies observed in deep rest-UV spectra.
Consider EGS-zs8-1, an extremely bright $3\times L_{\mathrm{UV}}^*$ galaxy selected by IRAC broad-band colour indicative of high equivalent width [\oiii{}]+H$\beta$ and confirmed to lie at $z=7.73$ via Ly$\alpha$ \citep{Oesch2015,Roberts-Borsani2016}.
A deep Keck/MOSFIRE spectrum of this object revealed strong \ciii{}] emission at 22 \AA{} equivalent width \citep{Stark2017}.
At $z= 1-3$, typical massive $\sim L^*$ star-forming galaxies emit in \ciii{}] at the $\sim 1-2$ \AA{} level, with the most extreme Ly$\alpha$ emitters ($W_{0, \mathrm{Ly}\alpha}> 20$ \AA{}) reaching \ciii{}] $\sim 5$ \AA{} \citep[e.g.][]{Shapley2003,Du2017}.
Photoionization modeling and the bulk of local star-forming galaxies suggest that \ciii{}] equivalent widths $\gtrsim 15$ \AA{} are only attained below $Z/Z_\odot \lesssim 1/5$ (Fig.~\ref{fig:ciii_metal}).
Indeed, stellar population synthesis modeling applied to EGS-zs8-1 indicate that this system requires a very low metallicity of $Z/Z_\odot = 0.11\pm 0.05$.
However, the surprising detection of \ciii{}] at $\sim 10$ \AA{} alongside extreme stellar \heii{} emission and strong [\oiii{}] at $Z/Z_\odot \sim 1/3$ presented here opens the possibility that short-lived massive stars at higher metallicity may also be capable of powering comparable emission in systems with particular high sSFR.
As we discuss further in Sec.~\ref{sec:ionsource}, uncertainties in the evolutionary channels producing these stars and in their emergent ionizing spectra have significant consequences for physical quantities like metallicities inferred from photoionization modeling of UV lines.

While \ciii{}] emission approaching reionization-era detections has been found in local systems, no nearby star-forming galaxies have yet been found with \civ{} comparable to the $\sim 20-40$ \AA{} emission found at $z>6$.
There are now two detections of nebular \civ{} in the reionization era \citep{Stark2015a,Mainali2017}.
These systems are gravitationally lensed and lower mass ($\sim 10^8 M_\odot$) than bright unlensed systems like EGS-zs8-1.
On the basis of the trends discussed above and in Sec.~\ref{sec:ionspec}, and one detection of $>10$ \AA{} \civ{} at yet lower metallicities, this emission may be associated with extremely low-metallicity gas and stars below a tenth solar metallicity.
\hstcos{} programs targeting very metal poor ($Z/Z_\odot < 1/10$) galaxies (e.g.\ GO:13788, PI: Wofford; GO:14679, PI: Stark) will be essential to establishing this, and identifying the stellar populations and conditions capable of powering this extreme \civ{} emission.

We have uncovered a variety of optical indicators which distinguish strong UV line emitters.
The high prevalence of extreme [\oiii{}]+H$\beta$ emission inferred from IRAC excesses at $z>6$ \citep[e.g.][]{Labbe2013,Smit2014} suggests that these rest-UV lines are far more common in the reionization era than at lower redshift, which has already been borne-out in the first deep spectroscopy of galaxies in this era.
These indicators may be of significant utility in locating UV line emitters at $z\sim 2$ as well, where increasingly large samples of rest-optical spectra are being assembled \citep[e.g.][]{Steidel2014, Shapley2015, Kashino2017}.
For the highest redshifts $z>12$ where the rest-optical is inaccessible to \jwst{}, the rest-UV lines may be our only tools.
At all redshifts, high-ionization UV lines provide an extremely sensitive probe of the ionizing spectrum which powers them.

\subsection{The Ionizing Spectrum}
\label{sec:ionspec}

The sudden appearance of high-ionization nebular emission with decreasing metallicity discussed above suggests a significant transition in the ionizing spectrum.
Triply-ionizing carbon to allow nebular \civ{} emission requires 47.9 eV photons, and nebular \heii{} $\lambda 1640$ and $\lambda 4686$ are recombination lines powered by photons beyond the $\mathrm{He^+}$ ionizing edge at 54.4 eV.
Direct observation of the extreme-UV ($\sim 10-100$ eV) output of massive O stars is essentially impossible at any metallicity due to the weak emergent flux and heavy attenuation by nascent gas towards these stars.
High-ionization nebular emission thus potentially provides one of the only windows into the ionizing spectrum of metal-poor young stellar populations.

\begin{figure}
	\includegraphics[width=\columnwidth]{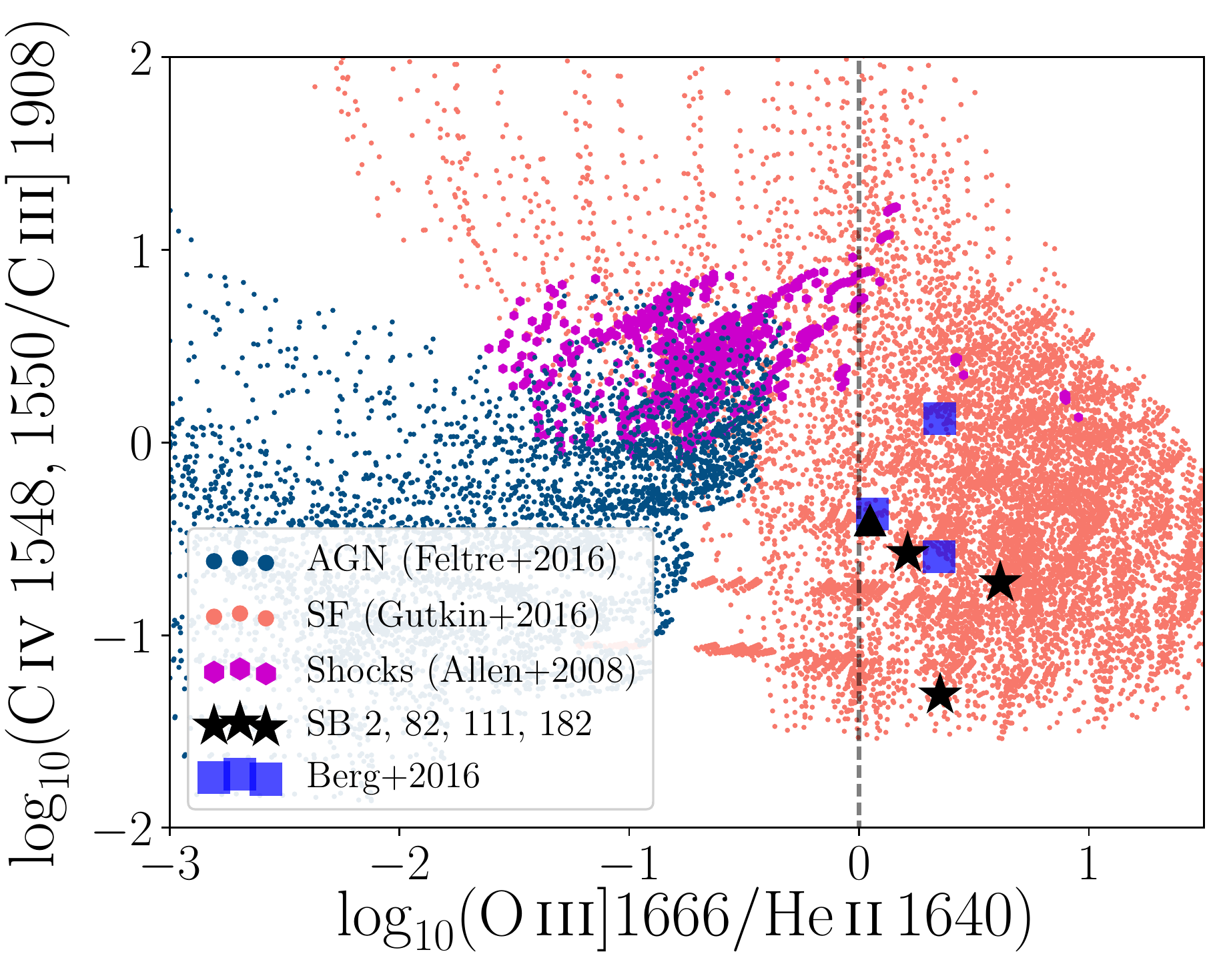}
    \caption{
        A line diagnostic diagram incorporating all of the high-ionization UV lines.
        The background circles are photoionization models powered by AGN \citep[dark blue;][]{Feltre2016}, stellar \citep[salmon;][]{Gutkin2016}, and shock \citep[fuchsia, shock and precursor;][]{Allen2008} spectra with fixed solar C/O abundance.
        The four metal-poor systems presented in this paper with detections of both nebular \heii{} and \civ{} are plotted as black stars (or, in the case of SB 111, as a lower-limit caret corresponding to the \ciii{}] upper-limit).
        Three similar systems from \citet{Berg2016} with COS detections of all four lines are plotted as blue squares.
        The ratios of \oiii{}]/\heii{} are inconsistent with any AGN models --- all have more flux in \oiii{}] 1666 than \heii{} 1640.
    }
    \label{fig:feltre_oiiiheiicivciii}
\end{figure}

The ratios of high-ionization lines provide powerful constraints on the shape of the ionizing continuum.
Photoionization modeling suggests that UV line diagnostic diagrams are useful tools for differentiating stellar from AGN ionization and potentially for inferring metallicity \citep{Feltre2016,Gutkin2016}, especially at low metallicities where traditional optical line diagnostics can fail to separate AGN from highly-ionized star-forming regions \citep[e.g.][]{Groves2006}.
However, empirical constraints on the precise diagnostic line space populated by star-forming galaxies are lacking due to the rarity of multiple UV line detections.
In Figure~\ref{fig:feltre_oiiiheiicivciii}, we plot the ratios \civ{}/\ciii{}] and \oiii{} 1666 / \heii{} for the four systems from our sample with detected \civ{} and \heii{} alongside photoionization models from \citet{Feltre2016} for AGN and \citet{Gutkin2016} for star-forming galaxies.
In addition, we plot photoionization predictions for fast radiative shocks (see Sec.~\ref{sec:shocks}) from \citet[][shock plus precursor]{Allen2008}, which span a range of shock velocities 100 -- 1000 km/s and metallicities down to $30\%$ solar.
We also include measurements of the three extreme galaxies from \citet{Berg2016} with detected nebular \civ{} and \heii{}.
The position of our observations on this diagram fall within the locus of star-forming models.
The hard break in the stellar spectrum beyond the $\mathrm{He}^+$ ionizing edge, absent from the flatter power-law spectra yielded by AGN and radiative shocks, produces a clear separation in \oiii{}]/\heii{} \citep[][]{Feltre2016, Mainali2017}.
All of our observations lie to the star-forming side of this division, with \oiii{}] $\lambda 1666$ comparable or much stronger than \heii{} $\lambda 1640$.
Similar bulk agreement with the stellar photoionization predictions is achieved in a plot of \civ{}/\heii{} against \oiii{}]/\heii{}.
The disjoint position of our objects with respect to the AGN and shock models strongly implies that the nebular \civ{} and \heii{} we observe are primarily powered by stellar photoionization.

\begin{figure}
	\includegraphics[width=\columnwidth]{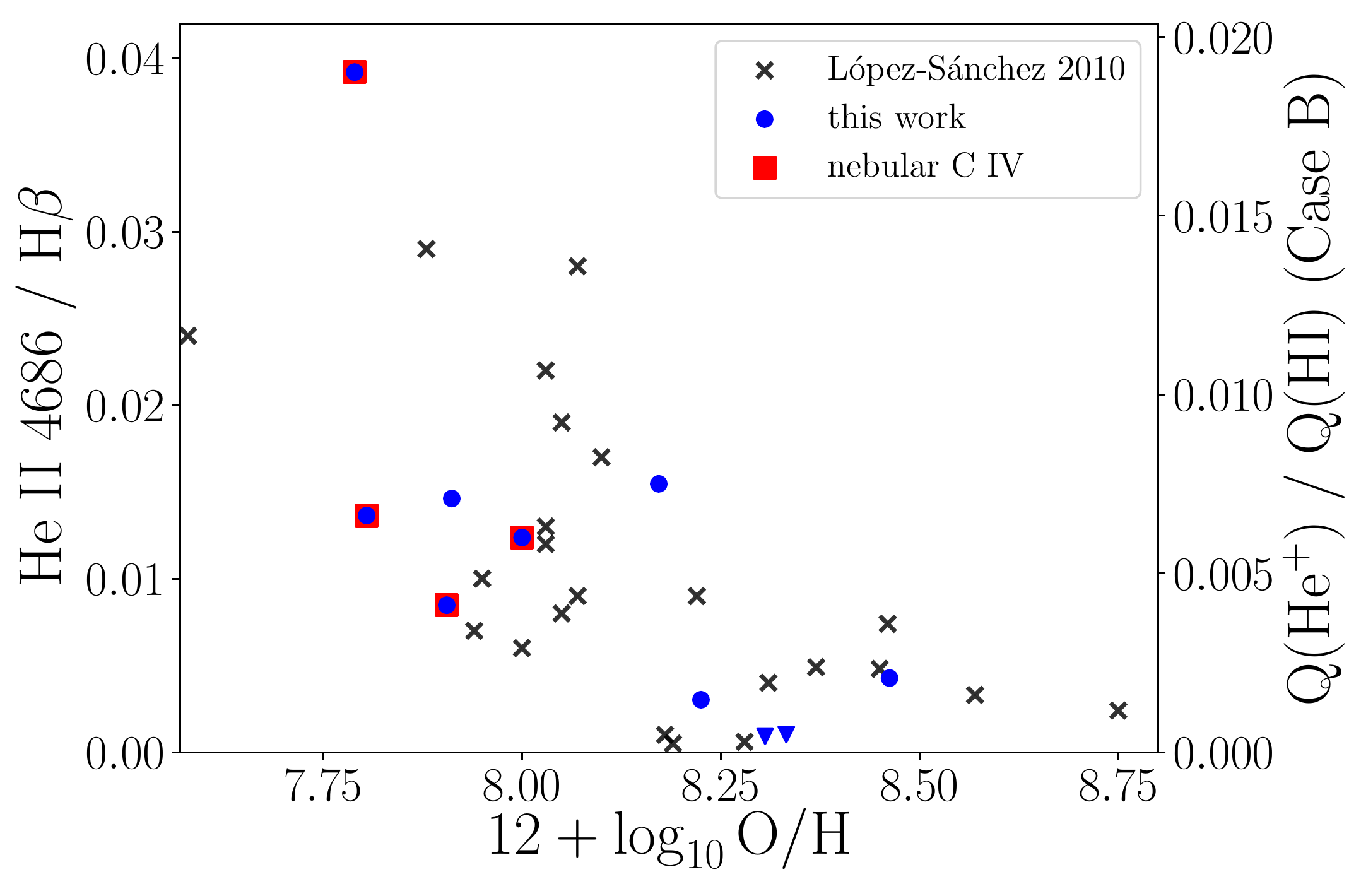}
    \caption{
        Measured nebular \heii{} 4686 / H$\beta$ values (a proxy for the ratio of $\mathrm{He^+}$-ionizing to H-ionizing photons) versus gas-phase metallicity.
        Both our galaxies (blue circles, backed by red squares if nebular \civ{} was detected) and the Wolf-Rayet galaxies presented by \citet{Lopez-Sanchez2010} are presented.
        The right axis indicates the corresponding $\mathrm{Q(He^+)/Q(H)}$ assuming Case B recombination for both species.
        Both observational samples show a sharp transition with metallicity, with substantially larger ratios and harder spectra below $12+\log\mathrm{O/H}\sim 8$.
    }
    \label{fig:heiihbeta_metal}
\end{figure}

The deep ESI spectra we obtained also provide constraints on the shape of the ionizing spectrum between $10-50$ eV.
The ratio of flux between the nebular \heii{} and H recombination lines is closely related to the ratio of $\mathrm{He^+}$-ionizing ($<228$ \AA{}) to H-ionizing photons ($<912$ \AA{}).
The optical \heii{} $\lambda 4686$ and H$\beta$ lines are ideal for avoiding the uncertainties introduced by dust and aperture corrections.
Our ESI spectra easily resolve the nebular and broad components of the \heii{} line (see Fig.~\ref{fig:esi_heii}) and thus allow a robust simultaneous fit to both, to which H$\beta$ in the same spectrum can be compared (Sec.~\ref{sec:esispec} and Table~\ref{tab:esi_heii}).
The (dust-corrected) ratio of nebular \heii{}/H$\beta$ is plotted as a function of gas-phase metallicity in Fig.~\ref{fig:heiihbeta_metal}.
In addition to the sample presented here, we plot the published measurements of this ratio for the local Wolf-Rayet galaxies analyzed by \citet{Lopez-Sanchez2010}.
The axis on the right displays the ratio $\mathrm{Q(He^+)/Q(H)}$ assuming ideal Case B recombination proceeds for both H and $\mathrm{He^+}$ (and $10^4$ K, $10^2$ \unit{cm^{-3}}; \citealt{Hummer1987,Draine2011}).
The Case B predictions should be treated with care as collisional excitation and low ionization parameters can result in deviations from this assumption \citep[e.g.][]{Stasinska2001,Raiter2010}, but demonstrate clearly the effect of changing the ionizing spectrum slope at the $\mathrm{He^+}$ ionizing edge.

The nebular \heii{}/H$\beta$ ratio reveals a sharp transition in the ionizing spectrum with decreasing metallicity.
The maximum value of \heii{}/H$\beta$ increases as the gas-phase metallicity decreases, with a marked upturn in this envelope around $12+\log\mathrm{O/H} \sim 8.0 - 8.2$.
Systems below $12+\log\mathrm{O/H}< 8.2$ ($< Z_\odot/3$) reach relative nebular \heii{} fluxes nearly ten times higher than those above this metallicity.
This implies an order of magnitude increase in $\mathrm{Q(He^+)}/\mathrm{Q(HI)}$ as the metallicity of a young stellar population is decreased from $Z_\odot$ to $Z_\odot/10$.
A similar trend towards higher \heii{}/H$\beta$ below $12+\log\mathrm{O/H}<8.0$ was found in the full sample of SDSS spectra analyzed by \citet{Brinchmann2008} and \citet{Shirazi2012}, though we find a far more pronounced transition.
This may be due in part to confusion of strong stellar \heii{} for nebular emission in SDSS --- our higher-resolution ESI spectra revealed that both SB 179 and 191 have purely stellar emission in this line.
This increase in \heii{}/H$\beta$ mirrors the transition we noted in the UV spectra, with the only nebular \civ{} detections occurring below $\lesssim Z_\odot/5$.
A trend towards more highly-ionized gas is also seen in the flux ratio of \heii{} $\lambda 4686$ / \hei{} $\lambda 4713$ visible in Fig.~\ref{fig:esi_heii}, which increases from $\lesssim 1$ in the four most metal-rich objects to $\gtrsim 2$ (up to 8 in SB 111) below $12+\log\mathrm{O/H} < 8.2$.

Our selection method complicates interpretation of Fig.~\ref{fig:heiihbeta_metal}, but does not easily explain the observed transition.
Both our sample and that of \citet{Lopez-Sanchez2008} were selected to show emission at \heii{} $\lambda 4686$ \citep[in the latter case, broad stellar rather than nebular;][]{Schaerer1999}.
We are thus biased towards higher \heii{}/H$\beta$ than a \heii{}-blind sample.
There is no clear reason why we would be biased against finding any objects in the upper-right of Fig.~\ref{fig:heiihbeta_metal} --- that is, with high nebular \heii{}/H$\beta$ and $12+\log\mathrm{O/H}>8.2$.
It is possible that strong nebular and stellar \heii{} were partially blended for some moderately metal-poor objects in SDSS, yielding very high \heii{}/H$\beta$ values and resulting in an AGN classification in SB2012.
However, classification as an AGN in the SB2012 scheme required the identification of AGN-like features such as broad Balmer lines, strong \nev{} $\lambda 3426$, \feii{} emission, or an unusual [\oiii{}] $\lambda 4363$/H$\gamma$ ratio in the SDSS data as well.
Regardless, future work targeting galaxies without considering \heii{} emission in selection will enable a more complete study of this emission as a function of metallicity (e.g.\ HST GO:14679, PI: Stark).

There is significant tension between these results and commonly-used stellar population synthesis models. 
\citet{Shirazi2012} found that both \texttt{Starburst 99} \citep{Leitherer1999,Leitherer2010} and \texttt{BPASS} \citep{Eldridge2009} can reproduce nebular \heii{}/H$\beta$ ratios as high as 1-10\% for instantaneous bursts of star formation, but only for a short period of time approximately 3 Myr after the burst; and only above $Z/Z_\odot > 1/5$.
This modeling suggests that strong nebular \heii{} should only occur when Wolf-Rayet (WR) stars are active, and that large WR populations should power strong \heii{}/H$\beta$ $\gtrsim 1\%$.
This is in contrast to the observations, where nebular \heii{} is commonly seen in systems without WR features at low metallicity \citep[Figure~\ref{fig:esi_heii}; see also][]{Brinchmann2008,Shirazi2012}.
At the relatively high-metallicity end ($12+\log\mathrm{O/H}>8.1$) where strong WR features are seen, our high-resolution view of \heii{} reveals that nebular \heii{} is consistently weaker than predicted (\heii{} $\lambda 4686$/H$\beta< 1\%$).
As discussed in Sec.~\ref{sec:beagle}, fits to the UV spectra presented here indicate that the prominent nebular \heii{} $\lambda 1640$ emission visible in SB 111, 2, and 182 cannot be fit with the models of \citet{Gutkin2016} without simultaneously invoking $M_{up}=300 M_\odot$ and metallicity substantially lower than the optically-derived gas-phase metallicity.
If we are to accurately interpret high-ionization emission in distant galaxies, we must understand the origin of these discrepancies.

\subsection{Sources of Ionizing Radiation}
\label{sec:ionsource}

Accurate predictions for extreme ultraviolet (EUV) ionizing flux from star-forming galaxies are critical for modeling and interpreting the high-ionization UV nebular lines that this flux powers.
As discussed above, current stellar population synthesis models do not reproduce the strong metallicity dependence observed in \heii{}/H$\beta$ near $Z/Z_\odot \sim 1/4$ (Fig.~\ref{fig:heiihbeta_metal}).
Our data reveals a sharp transition in this ratio consistent with an order of magnitude increase in $\mathrm{Q(He^+)/Q(HI)}$ as gas-phase metallicity is decreased from $Z_\odot$ to $Z_\odot/5$ (Sec.~\ref{sec:ionspec}).
Identifying the physical origin of this transition will help direct adjustments or additions to these models.

Our deep moderate-resolution \hstcos{} and ESI data allow us to put direct constraints on the massive stellar populations and other ionizing sources potentially present in these systems.
By virtue of their selection using \heii{} diagnostics \citep[see Sec.~\ref{sec:selection} and][]{Shirazi2012} and as inferred from sensitive UV line diagnostics (Fig.~\ref{fig:feltre_oiiiheiicivciii}), an AGN contribution to photoionization is highly unlikely in these systems.
This leaves three primary ionizing sources which could contribute to the high-ionization emission lines, which we discuss in-turn: stars, fast radiative shocks, and X-ray binaries.

\subsubsection{Stars}

The EUV flux output from young stellar populations which powers high-ionization nebular emission lines is very uncertain and intimately linked to stellar winds.
Both OB stars and their hydrogen-stripped relatives Wolf-Rayet (WR) stars drive winds through metal line absorption in the EUV, which diminish in strength as stellar metallicity is decreased \citep[e.g.][]{Castor1975,Kudritzki1987,Crowther2002,Vink2005,Crowther2006}
The amount of EUV flux which passes through these winds to power nebular emission is decreased substantially by high-density winds and difficult to calibrate directly, especially beyond the $\mathrm{He^+}$ ionizing edge \citep[e.g.][]{Smith2002,Crowther2006,Smith2014}.
As metallicity decreases and winds weaken, WR stars will be produced less efficiently by wind-driven mass loss.
However, mass transfer in binaries can efficiently remove the outer layers of a donor star, producing a hot stripped star without the need for high wind mass loss \citep[e.g.][]{Maeder1994,Eldridge2008,Gotberg2017,Smith2017}.
In addition, the mass gainer in a binary system may be sufficiently spun-up by angular momentum transport to become fully-mixed during its hydrogen burning phase, leading to quasi-chemically homogeneous evolution and similarly high effective temperatures \citep[][]{Maeder1987,Yoon2005,Eldridge2012,deMink2013}.
If sufficient metals are available, these stars will reveal themselves in broad wind emission lines such as \heii{} like canonical WR stars; but at low metallicities, these sources of hard ionizing radiation will be effectively invisible except for their impact on nebular emission.
The efficiency of mass transfer and the effects of rotation are both likely enhanced at lower metallicities, where weaker stellar winds allow massive stars to retain their mass longer \citep[e.g.][]{Maeder2000,Eldridge2009,Szecsi2015}.
Our moderate-resolution \hstcos{} and ESI spectra allow us to measure stellar wind features directly, and thus ascertain whether the wind properties of the massive stars present correlate with the nebular emission we detect.

The UV spectra reveal a significant weakening in stellar wind lines with declining gas-phase metallicity (Fig.~\ref{fig:majorlines_comp}).
The five systems above $12+\log\mathrm{O/H}\gtrsim 8.2$ have \civ{} P-Cygni absorption deeper than $-4$ \AA{}; whereas the feature is uniformly weaker than this for the lower-metallicity systems.
Quantitatively, the trend is consistent with other local galaxies.
The empirical relation derived by \citet{Crowther2006a} predicts a gas-phase oxygen abundance of $12+\log\mathrm{O/H}=8.2$ for a P-Cygni absorption depth of $-4$ \AA{}, in good agreement with our direct-$T_e$ gas-phase metallicities.
The largely monotonic decrease in the strength of this feature with decreasing gas-phase oxygen abundance is strong evidence that the stellar metallicity and thus stellar wind strengths track the gas-phase metallicity, as expected.
This confirms that the systems with the strongest observed high-ionization UV emission harbor populations of very metal-poor stars in addition to hot metal-poor gas.

\begin{figure}
	\includegraphics[width=\columnwidth]{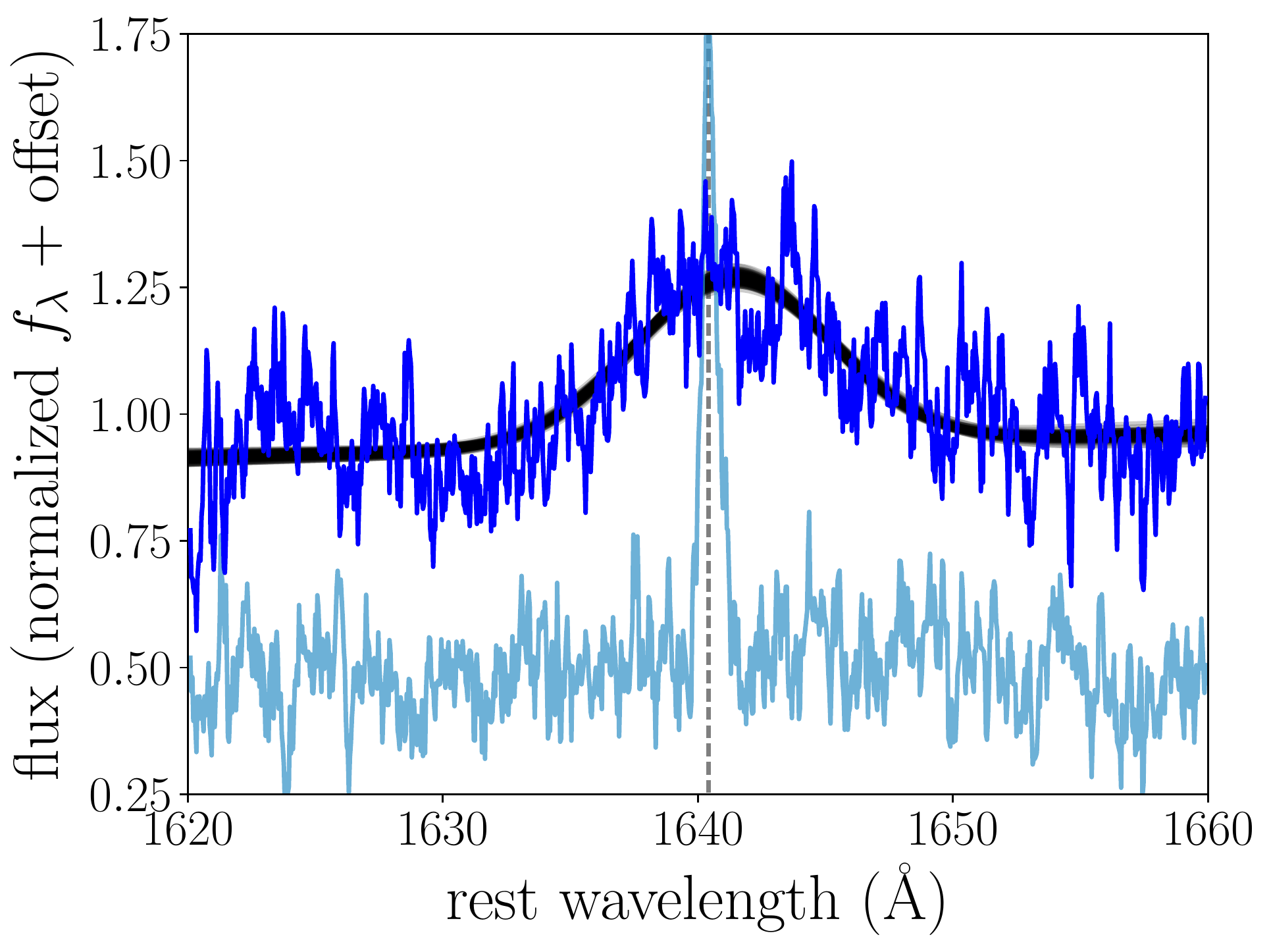}
    \caption{
        Stacked \hstcos{} spectra centred on \heii{} 1640 for the five most metal-rich galaxies (SB 110, 179, 191, 80, 198) and the five metal-poor systems (SB 111, 2, 36, 82, 182; transparent).
        Samples from the MCMC Gaussian fit posterior to the metal-rich stack are overlaid in black.
        The moderately metal-poor stack ($12+\log\mathrm{O/H}\gtrsim 8.1$) reveals purely broad stellar emission at extremely large equivalent width (3.4 \AA{}).
        The low-metallicity stack, in contrast, is dominated by nebular emission with no obvious stellar wind contribution.
    }
    \label{fig:heii_stack}
\end{figure}

The five relatively high-metallicity systems show surprisingly strong WR star signatures.
The stacked G160M spectra of the systems above $12+\log\mathrm{O/H}>8.1$ reveals very strong broad \heii{} $\lambda 1640$ emission, with equivalent width $3.4 \pm 0.2$ \AA{} and resolution-corrected FWHM $1600$ km/s --- easily distinguished from the purely nebular emission seen in the low-metallicity stack (Fig.~\ref{fig:heii_stack}).
While the width of this feature is consistent with typical late WN-type (WNL) stars \citep[WN6-WN8:][]{Chandar2004}, the strength is larger than expected by common population synthesis predictions, which do not typically exceed 2 \AA{} below solar metallicity \citep{Brinchmann2008a}.
Incorporating quasi-homogeneous evolution for mass gainers in binary population synthesis prescriptions enhances this wind line by producing more very hot stars capable of driving ionized winds at moderately low metallicity \citep{Eldridge2012}.
In addition, IMF variations that allow for more very massive stars will boost this line, and may be required to explain even stronger stellar \heii{} emission in nearby star clusters \citep[e.g.][]{Wofford2014}.
The fits presented in Section~\ref{sec:beagle} are able to qualitatively reproduce the extreme stellar \heii{} in SB 179 and 191 (Fig.~\ref{fig:beagle_uv}) with the updated single-star models of Charlot \& Bruzual (2017, in-prep) which incorporate an updated treatment of WR atmospheres.
In any case, the prominence of broad \heii{} emission in these systems means that massive stars at these metallicities are driving dense winds.
These winds require the absorption of EUV ionizing radiation which would otherwise escape to power nebular emission.

Strong broad \heii{} emission in our sample is accompanied by nebular emission in \ciii{}] and \oiii{}], but weak emission in higher-ionization lines.
As discussed above, nebular \civ{} and \heii{} are undetected in the UV above $12+\log\mathrm{O/H}>8.1$ where broad \heii{} emission is detected.
This is consistent with strong absorption of EUV flux in the dense winds we see in broad emission.
However, at $Z/Z_\odot > 1/5$, two of the three systems with extreme stellar \heii{} equivalent widths $>2$ \AA{} show very strong \ciii{}] emission, at 9 and 11 \AA{} (SB 179 and 191, respectively; Fig.~\ref{fig:majorlines_comp}).
This is in striking contrast to the systems at similar metallicity but with lower equivalent width stellar \heii{} $\lambda 1640$  (SB 198 and 110), where \ciii{}] is $<5$ \AA{}.
While moderately metal-poor massive stars do not appear to  be efficient producers of $\mathrm{He^+}$-ionizing photons, this result and the archival analysis of \citet{Rigby2015} suggests they may be an important source of the less extreme radiation necessary to produce \ciii{}].
If this alternative channel (extreme $Z/Z_\odot\sim 1/2$ massive star populations) can power emission comparable to the $22$ \AA{} equivalent width \ciii{}] detected in a massive $>L_\star$ reionization-era galaxy \citep{Stark2017}, extremely metal-poor $Z/Z_\odot < 1/10$ gas and stars may not need to be invoked (see Sec.~\ref{sec:empiricaltrends}).
However, the efficacy of this channel is presently unclear --- both SB 80 and the cluster NGC 3125-A1 analyzed by \citet{Wofford2014} show $>3$ \AA{} stellar \heii{} $\lambda 1640$ but undetected \ciii{}] emission, whereas NGC 5253 contains a cluster with $>3$ \AA{} stellar \heii{} and 8 \AA{} \ciii{}] emission \citep{Smith2016}.
As \citet{Smith2016} highlight, what we interpret in this paper as WR emission in \heii{} may in some cases be produced largely by very massive H-burning stars rather than canonical He-burning WR stars --- but since they drive similarly dense winds, these OIf supergiants likely present similar EUV spectra.
These systems with strong stellar \heii{} emission must have undergone very recent star formation to produce the observed massive stars.
However, as discussed above, significant uncertainties in stellar evolutionary channels result in uncertain predictions for the ionizing spectrum and nebular emission from these stellar populations.
Clearly, more data is required to quantify the range of \ciii{}] equivalent widths these stars can power.

The strongest nebular emission in our sample appears below $12+\log\mathrm{O/H}<8.1$, where stellar wind signatures vanish.
Nebular \civ{} and \heii{}, and the highest \ciii{}] equivalent widths in our sample ($15$ \AA{}), are found in those systems where broad stellar \heii{} emission is undetected (Fig.~\ref{fig:majorlines_comp}); and the ESI data reveal \heii{}/H$\beta \gtrsim 0.01$ in these systems, systematically higher than for the more metal-rich objects (Fig.~\ref{fig:heiihbeta_metal}).
The number of WR stars relative to O stars appears to decrease below this threshold (Table~\ref{tab:esi_heii}), as previously observed and predicted by population synthesis \citep[e.g.][]{Brinchmann2008}.
Though visible WR stars become less common at low metallicities, models predict that lower metallicity stellar populations will still contain hot stripped cores produced via binary mass transfer and rotation effects \citep[e.g.][]{Gotberg2017}.
Due to declining wind densities, all of these stars are predicted to show both weaker wind lines and commensurately stronger EUV flux, especially beyond the $\mathrm{He^+}$-ionizing edge --- just as we infer from their nebular emission.

Our results suggest that hot stars likely dominate production of the observed nebular emission.
However, we have shown that the transition to strong nebular \heii{} and \civ{} may be sharper than predicted from population synthesis (Sec.~\ref{sec:ionspec}), consistent with dense stellar winds blocking more $>50$ eV ionizing flux than predicted at higher metallicities.
This suggests that these emission lines are potentially more useful probes of metallicity than we might have expected theoretically.
Deep spectra of nearby integrated stellar populations provide one of the only windows onto the uncertain emergent ionizing spectra of metal-poor stars, with significant implications for nebular emission modeling at $z>6$.
Further work is needed to ensure that population synthesis models are able to fully reproduce the nebular and stellar features in extreme systems such as these.

\subsubsection{Fast radiative shocks}
\label{sec:shocks}

Fast radiative shocks produced by supernovae explosions or stellar winds can provide significant EUV flux \citep[e.g.][]{Allen2008}, but their ability to explain the nebular \heii{} emission in these systems is unclear.
Shocks are not expected to produce the strong metallicity dependence observed in the \heii{}/H$\beta$ ratio (Fig.~\ref{fig:heiihbeta_metal}).
At fixed shock velocity, the resulting \heii{}/H$\beta$ ratio typically increases with increasing metallicity, opposite the observed trend (see \citealt{Izotov2012}, using the models of \citealt{Allen2008}).
There is no clear reason that these shocks should become more prevalent with decreasing metallicity.
In addition, UV line ratios are expected to discriminate effectively between shock and stellar photoionization \citep[e.g.][]{Villar-Martin1997,Allen1998, Jaskot2016}.
Figure~\ref{fig:feltre_oiiiheiicivciii} reveals that the observed UV line ratios are inconsistent with pure shock photoionization, but note that current model grids do not extend below SMC metallicities \citep{Allen2008}.

In local star-forming galaxies, shocks have been invoked primarily to explain [\nev{}] $\lambda 3426$ emission \citep[e.g.][]{Thuan2005,Izotov2012}.
The ionization potential necessary to quadruply-ionize neon is $97.1$ eV, requiring photons nearly twice as energetic as for doubly-ionizing helium and potentially a harder ionizing spectrum such as that provided by radiative shocks.
Our MMT spectra did not reach the blue-end depth necessary to detect this line at a few percent H$\beta$.
However, SB 111 has a previously reported detection of [\nev{}] emission at the $\sim 2 \sigma$ level \citep[J1230+1202 in][]{Izotov2012}.
The ratio of [\nev{}]/\heii{} (0.2) is the lowest in the sample explored by \citeauthor{Izotov2012}.
Though the nebular line ratios produced by shocks are predicted to be quite different from those expected by stellar photoionization, \citet{Izotov2012} found that they may explain the rare [\nev{}] emission if the shock contribution to the strong lines is small and for shock velocities in the narrow range of 300-500 km/s.

With the spectral resolution provided by ESI and \hstcos{}, we can check directly for evidence of broad shock contributions to the optical lines.
First, we do not see any evidence for the FWHM$\sim 1000-2000$ km/s broad emission in H$\beta$ and [\oiii{}] observed in some [\nev{}] emitters and attributed to adiabatic shock propagation \citep{Chevalier1977}.
Though \heii{} may be produced in the compressed pre-shock region at smaller widths, we would expect to see some broad contribution to \heii{} and strong optical lines such as H$\beta$ and [\oiii{}] at widths near the shock velocities \citep[$300-500$ km/s:][]{Izotov2012}.
The \heii{} $\lambda 4686$ \AA{} lines show no evidence for contributions besides the $\sigma<65$ km/s nebular component and the very broad $>1500$ km/s stellar one (see Fig.~\ref{fig:esi_heii}).
We detect broad components to H$\beta$ and [\oiii{}] $\lambda 4959$ in some systems, at $\sim 0.5-2$\% of the narrow flux and with $\sigma \lesssim 300$ km/s; but these broad components show no clear correlation with the observed \heii{}/H$\beta$ ratio, and the velocity is lower than the $300-500$ km/s expected from modeling of [\nev{}] emitters \citep{Izotov2012}.
We conclude that fast radiative shocks are unlikely to contribute substantially to the high-ionization nebular emission in these systems.

\subsubsection{X-ray binaries}

Another possible source of hard ionizing radiation in star-forming galaxies are high-mass X-ray binaries (HMXBs), systems in which a compact object (neutron star or stellar-mass black hole) is accreting material from a massive O or B companion.
Recently, detailed study of metal-poor star-forming galaxies nearby has revealed a strong metallicity dependence in the observed hard X-ray luminosity.
At fixed SFR, lower metallicity systems appear to be significantly more X-ray luminous \citep{Prestwich2013,Brorby2014,Brorby2016}.
This is likely due to some combination of weaker stellar mass loss and thus more efficient Roche lobe overflow; and more efficient mass transfer due to weaker X-ray heating of metal-poor gas \citep[e.g.][]{Thuan2004,Linden2010}.
However, archival data do not show a one to one correspondence between \heii{} (or [\nev{}]) and X-ray point source detections, leading other authors to conclude that HMXBs are not primarily responsible for this emission \citep[e.g.][]{Thuan2004,Thuan2005,Shirazi2012}.
In addition, adding HMXBs to stellar population synthesis models will generally yield a flatter ionizing continuum in the EUV \citep[e.g.][]{Power2013}, resulting in high-ionization line ratios divergent from the stellar locus in Fig.~\ref{fig:feltre_oiiiheiicivciii}.

Archival Chandra data is available for the most extreme nebular \heii{}-emitter in our sample, SB 111.
A Chandra X-ray point source was reported as associated with SB 111 ([RC2] A1228+12 by \citet{Brorby2014}.
We reprocessed this data following the same approach, and found that this detection is $>11''$ ($>0.9$ kpc at 16 Mpc) offset from the centre of the SDSS optical fiber.
No point source is detected within the $\sim 3''$ spectroscopic aperture to a limiting luminosity of $L_{X} \sim 3\e{38}$ erg/s in the 0.3-8 keV band; the reported point source cannot be responsible for the observed \heii{} emission.
More data is required to characterize the X-ray binary content of metal-poor galaxies such as these, and may additionally help constrain binary population synthesis predictions in a novel way.

\section{Summary}

We present \hstcos{} UV spectra of ten nearby galaxies with signatures of very recent star formation.
The medium-resolution gratings easily disentangle nebular emission, stellar wind features, and interstellar absorption; and enable measurement of the full suite of nebular emission lines emerging in high-$z$ rest-UV studies: \ciii{}], \oiii{}], \civ{}, and \heii{} (see Figure~\ref{fig:majorlines_comp}).
The systems are all vigorously forming stars (with sSFR of-order 100 $\mathrm{Gyr}^{-1}$), yet show a remarkable diversity of UV spectral properties.
Emission in the \ciii{}] semi-forbidden doublet is detected in seven objects, with equivalent widths reaching extremely high values in some cases ($\sim 10-15$ \AA{}).
Systems above $Z_\odot/5$ ($12+\log \mathrm{O/H} \gtrsim 8.0$) are dominated by stellar features, presenting strong P-Cygni absorption at \civ{} formed in the winds of massive O-stars and broad $\sim 1600$ km/s emission in \heii{} indicative of hot ionized winds from WR or very massive O stars.
Below $Z_\odot/5$, these wind features disappear and are replaced by prominent nebular emission in both \heii{} and \civ{}.

We investigate the variation in \ciii{}] equivalent width in detail.
In combination with archival local samples, our data support a metallicity threshold for \ciii{}] production in star-forming galaxies, with equivalent widths $>5$ \AA{} achieved only below $12+\log\mathrm{O/H} \lesssim 8.4$ ($Z/Z_\odot \lesssim 1/2$; Fig.~\ref{fig:ciii_metal}).
Below this threshold, some objects reach equivalent widths comparable to those seen in reionization-era systems, while in others the doublet is undetected.
This variation appears to be well-explained by variation in C/O and specific star formation rate.
The hot stars and inefficient ISM cooling found below $Z_\odot/2$ appears to be required to power high-EW \ciii{}], but emission is only observed in systems with a population of very recently-formed massive stars (sSFR $\gtrsim 10$ $\mathrm{Gyr}^{-1}$) which can provide the necessary ionizing flux.
We find that galaxies with \ciii{}] above $>5$ \AA{} are associated with high [\oiii{}] $\lambda 5007$ EW $\gtrsim 500$ \AA{}, $\mathrm{[\oiii{}]}\lambda 5007/\mathrm{H}\beta \gtrsim 3$, and $\mathrm{\ott{}} \gtrsim 4$ (Section~\ref{sec:empiricaltrends}).
In this context, the high \ciii{}] detection rate thus far at $z>6$ \citep{Stark2015,Zitrin2015,Stark2017,Ding2017} is entirely consistent with the increasingly large [\oiii{}]-H$\beta$ EWs inferred from IRAC excesses \citep[e.g.][]{Smit2014,Smit2015}.
Both imply that the galaxies with UV metal line detections at $z>6$ are metal-poor (confidently $Z/Z_\odot < 1/2$) and undergoing rapid star formation.
We also find strong $\sim 10$ \AA{} \ciii{}] emission alongside prominent WR signatures at $Z/Z_\odot \sim 1/3$.
Massive stars produced in very recent bursts of star formation may be able to power extreme \ciii{}] at higher metallicity than predicted by population synthesis models, though the typical nebular output of such populations remains unclear at present.

The appearance of nebular \civ{} and \heii{} below $Z/Z_\odot \lesssim 1/5$ implies a lower metallicity cutoff for emission in these lines.
Indeed, we observe a sharp increase in the ratio of \heii{}/H$\beta$ at this metallicity, consistent with an order of magnitude increase in the hardness of the ionizing spectrum beyond the $\mathrm{He^+}$-ionizing edge ($Q(\mathrm{He^+})/Q(\mathrm{H})$) below $12+\log\mathrm{O/H} \sim 8.0$ (Figure~\ref{fig:heiihbeta_metal}).
This transition is more rapid than predicted by stellar population synthesis models, suggesting stellar wind densities and evolutionary pathways yielding hot stripped stars may change more quickly than anticipated over this metallicity range.
Our results suggest that nebular \heii{} and \civ{} may prove to be useful probes of metallicity in distant galaxies; but none of our systems exceed $\sim 2$ \AA{} equivalent width in either.
A previous detection of \civ{} at combined doublet EW $11$ \AA{} in a nearby star-forming system at $12+\log\mathrm{O/H}\sim 7.44$ \citep{Berg2016} provides a hint that such extreme \civ{} may only be produced below $Z/Z_\odot < 1/10$.
The detection of \civ{} at $> 20$ \AA{} at $z\sim 6-7$ \citep{Stark2015a,Mainali2017} thus may indeed require extremely metal-poor gas and stars.

The \hstcos{} and ESI spectra acquired for our targets enable constraints to be placed on the origins of this ionizing flux.
The clean transition from primarily stellar to purely nebular \heii{} with decreasing metallicity (Fig.~\ref{fig:esi_heii}) is naturally explained by the weakening of O and WR star winds, which leads to both easier escape of $>50$ eV flux and to more efficient production of hot EUV-bright stars via binary mass transfer and rotation (Section~\ref{sec:ionsource}).
Massive stellar evolution is poorly constrained by observations in this metallicity regime, and thus high-ionization nebular emission in metal-poor dwarf galaxies is a useful window into the ionizing spectrum of these stars.
While we cannot firmly rule out a contribution from shocks or X-ray binaries, these sources appear unlikely to provide the bulk of the EUV flux in these galaxies.
We find no clear evidence for fast radiative shocks in high-resolution Keck/ESI spectra, and the most extreme \heii{}-emitter has archival Chandra data which puts a stringent upper limit on the presence of high-mass X-ray binaries.

With the growing number of high-ionization UV line detections in the reionization era, understanding how to interpret these features is becoming increasingly important.
Observations at $z\sim 0$ are critical for understanding both what stellar populations power this emission and how to translate rest-UV observations at high-$z$ into meaningful physical constraints.
Our results confirm that these lines will be useful not only for redshift measurement, but as empirical probes of metallicity and star formation in the very early universe.

\section*{Acknowledgements}

We thank the referee for their timely and helpful report.
Based on observations made with the NASA/ESA Hubble Space Telescope, obtained from the data archive at the Space Telescope Science Institute.
Support for program \#14168 was provided by NASA through a grant from the Space Telescope Science Institute, which is operated by the Association of Universities for Research in Astronomy, Inc., under NASA contract NAS 5-26555.
Observations reported here were obtained at the MMT Observatory, a joint facility of the University of Arizona and the Smithsonian Institution.
The scientific results reported in this article are based in part on observations made by the Chandra X-ray Observatory and published previously in cited articles.
Some of the data presented herein were obtained at the W.M. Keck Observatory, which is operated as a scientific partnership among the California Institute of Technology, the University of California and the National Aeronautics and Space Administration.
The Observatory was made possible by the generous financial support of the W.M. Keck Foundation. 
The authors wish to recognize and acknowledge the very significant cultural role and reverence that the summit of Mauna Kea has always had within the indigenous Hawaiian community.
We are most fortunate to have the opportunity to conduct observations from this mountain.

DPS acknowledges support from the National Science Foundation through the grant AST-1410155.
AV, JC, and SC acknowledge support from the European Research Council via an Advanced Grant under grant agreement no. 321323-NEOGAL.
TJ acknowledges support from NASA through Hubble Fellowship grant HST-HF2-51359.001-A awarded by the Space Telescope Science Institute.
AF acknowledges support from the ERC via an Advanced Grant under grant agreement no.\ 339659-MUSICOS.

This research made use of Astropy, a community-developed core Python package for Astronomy \citep{AstropyCollaboration2013}.

\bibliographystyle{mnras}
\bibliography{uvgals}

\appendix

\section{Distances and Identifiers}
\label{app:distances}

{\bf SB 2:}
(SDSS J094401.87-003832.1, MCG+00-25-010, CGCG 007-025, SDSS Plate-MJD-Fiber 266-51630-100) An extended BCD with at least four bright components in a linear configuration in SDSS imaging.
The SDSS and COS spectral apertures were centred on the brightest central object.
We compute a distance of 18.7 Mpc.
At this distance, the \hstcos{} aperture radius corresponds to $\sim 113$ pc.

{\bf SB 36:}
(SDSS J102429.25+052450.9, Plate-MJD-Fiber 575-52319-521) An isolated BCD with a cometary shape.
We find a distance of 141 Mpc, well into the Hubble flow; at which the \hstcos{} aperture radius corresponds to $\sim 850$ pc.

{\bf SB 80:}
(SDSS J094256.74+092816.2, Plate-MJD-Fiber 1305-52757-269) A bright \hii{} region embedded in the warped disk of UGC 5189.
We find a distance of 46.5 Mpc from the infall-corrected redshift, in good agreement with the similarly-derived result of 46.4 Mpc \citep{Baillard2011}.
The \hstcos{} aperture radius subtends 280 pc at this distance.

{\bf SB 82:}
(SDSS J115528.34+573951.9, Mrk 193 , Plate-MJD-Fiber 1313-52790-423) An isolated BCD.
We compute a distance of $75.6$ Mpc, corresponding to a \hstcos{} aperture physical radius of $\sim 460$ pc.

{\bf SB 110:}
(SDSS J094252.78+354726.0, Plate-MJD-Fiber 1594-52992-563) An isolated BCD with only a hint of extended emission beyond the bright \hii{} region.
We find a flow-corrected distance of 63.0 Mpc, at which distance the \hstcos{} aperture radius corresponds to $\sim 380$ pc.

{\bf SB 111:}
(SDSS J123048.60+120242.8, LEDA 41360, Plate-MJD-Fiber 1615-53166-120) An isolated BCD.
This cometary galaxy is dominated by a bright \hii{} region where the SDSS spectral aperture was placed, with a much dimmer diffuse component extending to the west.
It was assigned membership to the Virgo cluster by \citet{Giovanelli2007}, following the group assignment procedure described by \citet{Springob2007} and the peculiar velocity model of \citet{Tonry2000}.
Thus, we assume the distance of the Virgo cluster, 16.5 Mpc.
At this separation, the \hstcos{} $1.25''$ radius aperture probes a physical scale of-order $\sim 100$ pc.

{\bf SB  179:}
(SDSS J112914.16+203452.0, LEDA 35380, Plate-MJD-Fiber 2500-54178-84) A blue \hii{} region / cluster at the southwest end of the disturbed disk galaxy IC 700.
The distance of IC 700 has been estimated from the Tully-Fisher relation (typical errors of-order 20\%) to be $\sim 21.5$ Mpc \citep{Bitsakis2011}.
Our local flow estimate provides a distance of 24.6 Mpc, in reasonable agreement with the Tully-Fisher result.
We adopt the 24.6 Mpc measurement, at which distance the \hstcos{} aperture radius subtends $\sim 150$ pc.

{\bf SB 182:}
(SDSS J114827.33+254611.7, LEDA 36857, Plate-MJD-Fiber 2510-53877-560) A BCD with a neighbouring object of similar brightness to the northwest (likely a star based on SDSS photometry).
The redshift of this object suggests it is well into the Hubble flow at a distance of 191 Mpc, where the \hstcos{} aperture probes a region of radius $\sim 1.2$ kpc.

{\bf SB 191:}
(SDSS J121518.58+203826.6, Mrk 1315, LEDA 39187, KUG 1212+209B, Plate-MJD-Fiber 2610-54476-421) A prominent \hii{} region embedded in the southeast arm of the barred-spiral NGC 4204.
A precise distance measurement is not available to this source; the Updated Nearby Galaxy Catalog provides an estimate of $8.0$ Mpc based on surface fluctuations \citep{Karachentsev2013}.
The Extragalactic Distance Database \citep{Tully2009} states a preferred distance estimate of 10 Mpc based on a numerical action kinematic model.
Our local flow model suggests there is some degeneracy between this position and one consistent with the Virgo cluster --- HyperLeda \citep{Makarov2014} estimates a distance of 14.3 Mpc from the flow-corrected velocity.
We adopt a distance estimate of 10 Mpc for this work.
At 10 Mpc, the \hstcos{} aperture radius corresponds to $\sim 60$ pc.

{\bf SB 198:}
(SDSS J122225.79+043404.7, SDSSCGB 20312.2, Plate-MJD-Fiber 2880-54509-277) A bright \hii{} region embedded in the face-on spiral galaxy NGC 4301. 
Its velocity and position suggest that NGC 4301 is in the Virgo cluster \citep[see also][]{GildePaz2007}.
Thus, we adopt a distance of 16.5 Mpc.
At this distance, the \hstcos{} aperture radius corresponds to $\sim 100$ pc.

\section{Archival UV Targets}
\label{sec:archive_app}

In order to explore the factors which control the UV line emission (and in particular that of \ciii{}]), we compare our high-sSFR sample to a larger set of local systems with UV and optical measurements in Section~\ref{sec:discussion}.
We collect archival UV data from MAST for star-forming galaxies catalogues by \citet[][FOS and GHRS; hereafter L11]{Leitherer2011}, \citet[][IUE; G96]{Giavalisco1996}, and \citet[][COS; B16]{Berg2016}.
We fit \ciii{}] using our line-fitting software (see Section~\ref{sec:linefit}) and adopt the gas-phase metallicity measurements from these authors.
We also gather optical measurements from the literature and from SDSS for these objects where available, attempting to match the UV aperture pointings.
The objects and the sources of their optical line measurements are summarized in Table~\ref{tab:archival_data}.

In addition to the objects explicitly included in the FOS/GHRS atlas assembled by \citet{Leitherer2011}, we also examine the other systems with FOS data presented by \citet{Garnett1995} which were excluded.
These are C1543+091, 30 Doradus, and SMC-N88A.
The latter two are individual star-forming regions located in the LMC and SMC, respectively.
Both present extremely high-equivalent width \ciii{}] emission in FOS spectra (29 and 43 \AA{})
Since the FOS 1$''$ circular aperture used subtends less than a parsec at the distance of the LMC/SMC, we attribute this to aperture effects --- these spectra do not include the full continuum of the ionizing sources and are thus not representative of integrated galaxy spectra, so we ignore them.

Tol 1214-277 presents clear \ciii{}] emission but the continuum is undetected, leading to a very large uncertainty in the equivalent width and an unphysically large best fit ($>30$ \AA{}).
We ignore it in the above analysis.

One galaxy in the \citet{Leitherer2011} sample shows potentially nebular \civ{} emission. 
We measure an equivalent width $\sim 25$ \AA{} in the G190H spectrum of IC 3639 \citep{Leitherer2011}, but this galaxy harbors a Compton-thick AGN \citep[e.g.][]{Boorman2016} which we assume is responsible for this line.

Ultraviolet spectra covering \ciii{}] in nearby star-forming systems has also been taken with the Space Telescope Imaging Spectrograph (STIS) onboard HST (GO: 12472, PI: Leitherer).
Measurements of \ciii{}] and metallicity for this sample were reported by \citet{Rigby2015}; the highest \ciii{}] equivalent width measured therein was 8.3 \AA{}.
As these measurements have not yet been published in detail, we do not include them in our analysis; but note that they confirm the general trend displayed in Figure~\ref{fig:ciii_metal}.

Finally, note that the science apertures of these instruments vary in size dramatically.
The spatial scales probed range from the $0''.2$-wide slit of STIS to the $10''\times 20''$ aperture of IUE (with FOS, GHRS, and COS in the $1''-3''$ regime).
As seen in the case of FOS spectra of clusters in the LMC and SMC (above), very small apertures at low-redshift can significantly affect measured equivalent widths.
While we exclude these extreme outliers from our study, a full analysis of the effect of aperture and distance differences is beyond the scope of this paper.

\begin{table*}
	\centering
	\caption{
        Archival nearby star-forming UV targets drawn from \citet[][B16]{Berg2016}, \citet[][G96]{Giavalisco1996}, and \citet[][L11]{Leitherer2011}.
        Metallicities are as-derived by the respective authors.
        The optical line ratio measurements are drawn from the sources referenced in the last column and dust-corrected as necessary using the Balmer decrement.
        All SDSS, \ciii{}], and [\oiii{}] $\lambda 5007$ equivalent width measurements were made with the line-fitting technique described in Sec.~\ref{sec:linefit}.
    }
	\label{tab:archival_data}
\begin{tabular}{lcccccc}
\hline
Name & $12+\log\mathrm{O/H}$ & \ciii{}] & [\oiii{}] 5007 & \ott{} & [\oiii{}] 5007 / H$\beta$ & optical\\ 
(UV Atlas) & (atlas) & $W_0$ \AA{} & $W_0$ \AA{} (SDSS) &  &  & source\\ 
\hline
SDSSJ025426 (B16)& 8.06 & $1.3 \pm 0.4$ & $242 \pm 13$& --- & 4.47 & SDSS\\ 
SDSSJ082555 (B16)& 7.42 & $13.6 \pm 2.4$ & $851 \pm 125$& --- & 3.66 & SDSS\\ 
SDSSJ085103 (B16)& 7.66 & $<7.1$ & $333 \pm 19$& --- & 3.21 & SDSS\\ 
SDSSJ095137 (B16)& 8.20 & $<53.7$ & $53 \pm 2$ & 1.5 & 3.50 & SDSS\\ 
SDSSJ104457 (B16)& 7.44 & $13.4 \pm 1.0$ & $1002 \pm 235$& --- & 4.52 & SDSS\\ 
SDSSJ115441 (B16)& 7.75 & $2.0 \pm 0.7$ & $367 \pm 36$& --- & 3.97 & SDSS\\ 
SDSSJ120122 (B16)& 7.50 & $12.1 \pm 0.8$ & $606 \pm 129$& --- & 3.55 & SDSS\\ 
SDSSJ122436 (B16)& 7.78 & $7.6 \pm 1.1$ & $532 \pm 65$ & 8.1 & 5.61 & SDSS\\ 
SDSSJ122622 (B16)& 7.77 & $6.6 \pm 1.6$ & $798 \pm 86$& --- & 5.69 & SDSS\\ 
SDSSJ124159 (B16)& 7.74 & $8.4 \pm 1.1$ & $581 \pm 157$& --- & 4.84 & SDSS\\ 
SDSSJ124827 (B16)& 7.80 & $7.6 \pm 1.1$ & $560 \pm 48$ & 10.7 & 5.97 & SDSS\\ 
SDSSJ141454 (B16)& 7.28 & $<0.6$ & $88 \pm 8$& --- & 1.45 & SDSS\\ 
Haro 15 (G96)& 8.57 & $<0.5$& --- & 0.6 & 2.25 & \citet{Atek2014}\\ 
IC 0214 (G96)& 8.68 & $<1.5$& --- & 0.7 & 1.27 & \citet{Atek2014}\\ 
Mrk 26 (G96)& 8.60 & $<3.7$& --- & 0.6 & 1.54 & \citet{Moustakas2006}\\ 
Mrk 347 (G96)& 8.53 & $<0.5$& --- & 0.3 & 1.05 & \citet{Atek2014}\\ 
Mrk 496 (G96)& 8.77 & $<1.7$& --- & 0.4 & 0.57 & \citet{Atek2014}\\ 
Mrk 499 (G96)& 8.47 & $<1.4$& --- & 0.7 & 2.39 & \citet{Atek2014}\\ 
Mrk 66 (G96)& 8.39 & $<0.5$& --- & 1.3 & 3.00 & \citet{Atek2014}\\ 
Pox 120 (G96)& 7.83 & $13.0 \pm 1.3$& ---& ---& --- & ---\\ 
Pox 124 (G96)& 8.28 & $<4.0$& ---& ---& --- & ---\\ 
Tol 1924-416 (G96)& 8.32 & $<4.9$& --- & 3.0 & 4.69 & \citet{Atek2014}\\ 
Tol 41 (G96)& 7.98 & $<3.4$& ---& ---& --- & ---\\ 
C1543+091 (L11)& 7.80 & $9.2 \pm 0.6$ & $792 \pm 89$ & 9.7 & 5.60 & SDSS\\ 
IZw18 (L11)& 7.20 & $4.9 \pm 0.9$ & $102 \pm 4$& --- & 1.20 & SDSS\\ 
IZw18-NW HIIR (L11)& 7.20 & $3.1 \pm 0.6$ & $102 \pm 5$ & 7.6 & 2.09 & \citet{Kehrig2016}\\ 
IZw18-SE HIIR (L11)& 7.30 & $4.5 \pm 0.6$ & $251 \pm 18$ & 3.4 & 1.70 & \citet{Kehrig2016}\\ 
Mrk 71 (L11)& 7.90 & $19.5 \pm 1.0$& --- & 7.7 & 5.81 & \citet{Moustakas2006}\\ 
NGC 1569 (L11)& 8.20 & $<0.9$& --- & 2.9 & 4.68 & \citet{Moustakas2006}\\ 
NGC 2403-vs38 (L11)& 8.50 & $1.7 \pm 0.4$& --- & 1.9 & 1.51 & \citet{Esteban2009}\\ 
NGC 2403-vs44 (L11)& 8.50 & $2.1 \pm 0.4$& --- & 1.9 & 1.93 & \citet{Esteban2009}\\ 
NGC 2403-vs9 (L11)& 8.10 & $1.8 \pm 0.4$ & $367 \pm 34$& --- & 3.06 & SDSS\\ 
NGC 3690 (L11)& 8.80 & $<0.6$ & $10.8 \pm 0.4$ & 0.5 & 1.03 & \citet{Moustakas2006}\\ 
NGC 4214 (L11)& 8.10 & $<0.7$& --- & 1.0 & 2.51 & \citet{Moustakas2006}\\ 
NGC 4670 (L11)& 8.20 & $1.1 \pm 0.3$ & $48.5 \pm 0.9$ & 1.2 & 2.36 & \citet{Moustakas2006}\\ 
NGC 4861 (L11)& 7.90 & $10.3 \pm 1.2$& --- & 3.9 & 5.14 & \citet{Moustakas2006}\\ 
NGC 5055 (L11)& 9.00 & $<0.7$& ---& ---& --- & ---\\ 
NGC 5253-HIIR-1 (L11)& 8.20 & $7.7 \pm 0.8$& ---& ---& --- & ---\\ 
NGC 5253-HIIR-2 (L11)& 8.20 & $10.9 \pm 1.8$& ---& ---& --- & ---\\ 
NGC 5253-UV1 (L11)& 8.30 & $1.9 \pm 0.6$& ---& ---& --- & ---\\ 
NGC 5253-UV2 (L11)& 8.30 & $<1.8$& ---& ---& --- & ---\\ 
NGC 5253-UV3 (L11)& 8.30 & $2.1 \pm 0.8$& ---& ---& --- & ---\\ 
NGC 5457-NGC 5455 (L11)& 8.20 & $<1.8$ & $934 \pm 171$ & 2.8 & 3.88 & \citet{Croxall2016}\\ 
NGC 5457-NGC 5471 (L11)& 8.00 & $9.6 \pm 3.1$& --- & 7.3 & 6.74 & \citet{Croxall2016}\\ 
NGC 5457-Searle5 (L11)& 8.60 & $<1.2$ & $11.5 \pm 0.5$& --- & 0.23 & SDSS\\ 
NGC 7552 (L11)& 9.20 & $1.2 \pm 0.4$& --- & 0.0 & 0.07 & \citet{Storchi-Bergmann1995}\\ 
SBS 0335-052 (L11)& 7.30 & $5.0 \pm 0.3$& ---& ---& --- & ---\\ 
SBS 1415+437 (L11)& 7.40 & $5.1 \pm 0.9$& --- & 5.9 & 3.67 & \citet{Thuan1999}\\ 
Tol 1345-420 (L11)& 8.00 & $7.8 \pm 1.4$& --- & 3.4 & 5.23 & \citet{Dessauges-Zavadsky2000}\\ 
UM 469 (L11)& 8.00 & $1.4 \pm 0.3$ & $235 \pm 14$ & 2.7 & 4.08 & \citet{Kniazev2004}\\ 
\hline
\end{tabular}
\end{table*}

\end{document}